\documentclass[
 pra,
 aps,
 reprint,
 superscriptaddress,
 citeautoscript,
 floatfix,
]{revtex4-2}

\usepackage[]{standalone}
\usepackage{anyfontsize}
\usepackage{import}
\usepackage{amssymb} 
\usepackage{amsmath}
\allowdisplaybreaks 
\usepackage{cmap}
\usepackage[utf8]{inputenc}
\usepackage[T1]{fontenc}
\usepackage{graphicx}
\usepackage[hyperfootnotes=false,breaklinks=true]{hyperref}
\usepackage{setspace}
\usepackage{multirow, makecell}
\hypersetup{
 bookmarksopen=true,
 bookmarksopenlevel=1,
 colorlinks=true,
 linkcolor=blue,
 anchorcolor=blue,
 citecolor=blue,
 filecolor=blue,
 urlcolor=blue,
 pdfpagemode=UseOutlines,
 pdfstartview={XYZ null null 1},
}
\usepackage{amsmath}
\allowdisplaybreaks 
\usepackage{amssymb}
\usepackage{amstext}
\usepackage{amsfonts}
\usepackage{amsthm}
\usepackage{mathtools}

\usepackage{bm}
\usepackage{braket}
\usepackage{physics}
\usepackage{enumitem}
\usepackage[
 capitalise,
]{cleveref}
\crefname{table}{Table}{Tables}%
\crefname{section}{Sec.}{Sects.}%
\Crefname{section}{Section}{Sections}%
\Crefname{table}{Table}{Tables}%

\usepackage{dsfont}
\usepackage[dvipsnames, table]{xcolor}

\usepackage[caption=false]{subfig}
\usepackage{overpic}

\definecolor{verylightgray}{gray}{0.97}

\newcommand{\pdag}{{\protect\phantom{\dagger}}}

\newcommand{\perm}{T} 

\newcommand{\mycomment}[1]{}  

\graphicspath{{.}{figs/}{submissions/_figs/}}

\usepackage{relsize}
\usepackage{cancel}

\usepackage[
 draft, 
 commentmarkup=uwave,
 authormarkup=superscript, 
 authormarkupposition=left,
]{changes}
\newcommand{\addAuthor}[2]{\definechangesauthor[name={#1}, color=#2]{#1}}

\newcommand{\AuthorColorList}{%
  EFD/Green,
  Yan/cyan,%
  Yuan/orange,%
  Luke/Plum,%
  Kevin/Blue,%
  SMG/red}

\addAuthor{EFD}{Green}
\addAuthor{Yan}{cyan}
\addAuthor{Yuan}{orange}
\addAuthor{Luke}{Plum}
\addAuthor{Kevin}{Blue}
\addAuthor{SMG}{red}
\usepackage{pgffor}
\foreach \id/\clr in \AuthorColorList
{%
 \expandafter\xdef\csname \id Add\endcsname##1{\noexpand\added[id=\id]{##1}}
 \expandafter\xdef\csname \id Delete\endcsname##1{\noexpand\deleted[id=\id]{##1}}
 \expandafter\xdef\csname \id Comment\endcsname##1{\noexpand\comment[id=\id]{##1}}
 \expandafter\xdef\csname \id Replace\endcsname##1##2{\noexpand\replaced[id=\id]{##2}{ ##1}}
}

\usepackage{subfiles}
\usepackage{etoolbox}

\theoremstyle{plain}
\newtheorem{thm}{Theorem}

\newtheorem{lem}[thm]{Lemma}

\theoremstyle{definition}

\theoremstyle{remark}

\crefname{lemma}{Lemma}{Lemmas}
\Crefname{lemma}{Lemma}{Lemmas}
\crefname{thm}{Theorem}{Theorems}
\Crefname{thm}{Theorem}{Theorems}

\usepackage{nameref}
\makeatletter
\newcommand{\myref}[1]{\cref{#1}\mynameref{#1}{\csname r@#1\endcsname}}
\newcommand{\Myref}[1]{\Cref{#1}\mynameref{#1}{\csname r@#1\endcsname}}

\def\mynameref#1#2{%
  \begingroup
    \edef\@mytxt{#2}%
    \edef\@mytst{\expandafter\@thirdoffive\@mytxt}%
    \ifx\@mytst\empty\else
    \space(\nameref{#1})\fi
  \endgroup
}
\makeatother

\makeatletter
\def\blfootnote{\xdef\@thefnmark{}\@footnotetext}
\makeatother

\usepackage{orcidlink}

  \mycomment{
\definecolor{qmclr}{RGB}{225,241,225}
\definecolor{qbclr}{RGB}{229,201,253}
\tikzset{%
  level 0/.style = {draw=none},
  level 1/.style = {dash pattern=on 2pt off 6pt},
  level 2/.style = {dash pattern=on 4pt off 4pt},
  level 3/.style = {dash pattern=on 6pt off 2pt},
  level 4/.style = {solid},
  bond/.style args = {#1,#2}{draw=#1,level #2, line width=2.5pt},
}%
\tikzset{%
  pics/squareSWAPs/.style args={#1,#2,#3,#4|#5,#6,#7,#8|#9}{%
  code = {%
\graph [nodes={draw, circle, inner sep=1.5pt, font=\sffamily\small}, counterclockwise,
 radius=0.5cm, n=4, phase=45,
 edges={thick},
] {
  subgraph C_n [name=qumode, nodes={draw, fill=qmclr, rectangle, rounded corners=1pt, minimum size=0.3cm}, V={#5,#6,#7,#8}]
  <->[shorten <=0pt, shorten >=0pt]
  subgraph I_n [name=qubit, nodes={draw, fill=qbclr}, V={0,1,2,3},radius=0.25cm]
};

\draw[bond={clrA,#1}] (qubit 0) -- (qubit 1);
\draw[bond={clrB,#2}] (qubit 1) -- (qubit 2);
\draw[bond={clrA,#3}] (qubit 2) -- (qubit 3);
\draw[bond={clrB,#4}] (qubit 3) -- (qubit 0);

\node [inner sep=0cm, outer sep=0cm,
] at (0,0) {\textbf{#9}};
  }
  }
}%
  }

\begin{document}

\clearpage \newpage \twocolumngrid
\pagenumbering{arabic} 

\blfootnote{This manuscript has been authored by UT-Battelle, LLC, under Contract No.\ DE-AC0500OR22725 with the U.S.\ Department of Energy. The United States Government retains, and the publisher, by accepting the article for publication, acknowledges that the United States Government retains a nonexclusive, paid-up, irrevocable, worldwide license to publish or reproduce the published form of this manuscript, or allow others to do so, for the United States Government purposes. The Department of Energy will provide public access to these results of federally sponsored research in accordance with the DOE Public Access Plan.}

\title{Co-Designing Spectral Transformation Oracles with Hybrid Oscillator-Qubit Quantum Processors: From Algorithms to Compilation}

\author{Luke Bell\,\orcidlink{0000-0002-9440-0077}}
\email{lukebell@ucsb.edu}
\affiliation{Department of Physics, Yale University, New Haven, Connecticut 06520, USA}
\affiliation{Department of Physics, University of California Santa Barbara, Santa Barbara, California
93106, USA}

\author{Yan Wang\,\orcidlink{0000-0002-6545-6434}}
\email{wangy2@ornl.gov}
\affiliation{\mbox{Computational Sciences and Engineering Division, 
Oak Ridge National Laboratory, 
Oak Ridge, Tennessee 37831, USA}}

\author{Kevin C. Smith\,\orcidlink{0000-0002-2397-1518}}
\altaffiliation{Present address: IBM Quantum, Cambridge, MA}
\affiliation{Brookhaven National Laboratory, Upton, New York 11973, USA}
\affiliation{Yale Quantum Institute, Yale University, New Haven, Connecticut 06520-8263, USA}
\affiliation{Department of Physics, Yale University, New Haven, Connecticut 06520, USA}

\author{Yuan Liu\,\orcidlink{0000-0003-1468-942X}}
\affiliation{\mbox{Department of Electrical and Computer Engineering, North Carolina State University, Raleigh, North Carolina 27606, USA}}
\affiliation{Department of Computer Science, North Carolina State University, Raleigh, North Carolina 27606, USA}
\affiliation{Department of Physics, North Carolina State University, Raleigh, North Carolina 27606, USA}

\author{Eugene Dumitrescu\,\orcidlink{0000-0001-5851-9567}}
\email{dumitrescuef@ornl.gov}
\affiliation{\mbox{Computational Sciences and Engineering Division, 
Oak Ridge National Laboratory, 
Oak Ridge, Tennessee 37831, USA}}

\author{S.~M. Girvin\,\orcidlink{0000-0002-6470-5494}}
\affiliation{Department of Physics, Yale University, New Haven, Connecticut 06520, USA}
\affiliation{Yale Quantum Institute, Yale University, New Haven, Connecticut 06520-8263, USA}

\begin{abstract}
We co-design a family of quantum eigenvalue transformation oracles that can be efficiently implemented on hybrid discrete/continuous-variable (qubit/qumode) hardware. To illustrate the oracle's representation-theoretic power and near-term experimental accessibility, we encode a Gaussian imaginary time evolution spectral filter. As a result, we define a continuous linear combination of unitaries block-encoding. Due to the ancillary qumode's infinite-dimensional nature, continuous variable qumodes constitute a powerful compilation tool for encoding continuous spectral functions without discretization errors while minimizing resource requirements. We then focus on the ubiquitous task of preparing eigenstates in quantum spin models. For completeness, we provide an end-to-end compilation which expresses high-level oracles in terms of an experimentally realizable instruction set architecture in both 1D and 2D. Finally, we examine the leading-order effects of physical errors and highlight open research directions. Our algorithms scale linearly with the spatial extent of the target system and are applicable to both near-term and large-scale quantum processors. 
\end{abstract}

\date{\today}

\maketitle

\tableofcontents

\section{Introduction} \label{sec:intro}

Can algorithmic and physical co-design overcome current challenges to establish a near-term quantum computational advantage? From a computer science perspective, to unleash exponential quantum speedups, one wishes to compile opaque quantum oracles exemplified by singular value transformations~\cite{Gilyen2019,Martyn2021} and eigenvalue filters~\cite{Irmejs2024}. On the other hand, from experimental and co-design perspectives, one wishes to maximize a physical platform's computational reach by judiciously leveraging physical features. For example, expanding the compilation space from qubit to qudit on mixed-radix superconducting hardware was shown to minimize the total effective gate cost~\cite{Litteken2023}. 

This work takes advantage of quantum harmonic oscillators (also called qumodes~\cite{Menicucci2008,Liu2024}), which are a manifestly continuous quantum computational type, to co-design a novel family of quantum eigenvalue transformation oracles. Qumodes are ubiquitous and, working as an ancillary degree of freedom, already implicitly leveraged to compile two-body and multi-qubit entangling gates in superconducting and trapped ion machines~\cite{sorensen1999quantum, Martinez2016, katz2023demonstration}. Furthermore, by using instruction sets based on natively encoded qumodes, recent works have optimized the quantum simulation of simple lattice gauge theories~\cite{Davoudi2021, Crane2024}. Qumodes are fundamentally different from finite-dimensional qudits due to their \textit{infinite} Hilbert space dimension. It has therefore been conjectured that qumode quantum registries could radically expand computational capabilities~\cite{Liu2016e}. For example, using a continuous variable ancilla, Ref.~\onlinecite{zhang2021cvthermoQS} proposed an opaque quantum algorithm to prepare quantum spin systems at finite temperatures. However, end-to-end compilation strategies for arbitrary many-body interacting models that lower abstract spectral filters all the way down to the hardware layer have yet to be realized. 

We therefore focus on general architectures that comprise a lattice of continuous variable (CV) oscillators and an associated lattice of discrete variable (DV) qubits. An example of a physical hardware realization, with all necessary native gates and the connectivity used in our co-design, is provided by the popular lattice of superconducting cavities and transmon qubits~\cite{Teoh2023,Chou2024,Liu2024}. The individual quantum operations necessary for our approach---qubit-conditioned oscillator displacements \cite{EickbuschECD} followed by high-fidelity measurement of the boson number---were efficiently performed in recent experiments using boson sampling to simulate Franck-Condon factors in triatomic molecules \cite{Wang2020FCFs} and simulating molecular dynamics near a conical intersection~\cite{WangConicalIntersection}. 

Within this setting, we propose a constructive algorithm for CV-controlled time-evolution oracles for filtering the spectral weight of spin-$\frac{1}{2}$ (qubit) systems interacting under a target Hamiltonian. Utilizing a concrete instruction set architecture (ISA)~\cite{Liu2024}, we provide existence theorems as well as a practical end-to-end compilation of paradigmatic singular-value transformation oracles which spectrally filter quantum spin chains. 

Our protocol emphasizes the algorithmic utility of the oscillators in compilations which now concretely represent a powerful tool for resource-efficient quantum signal processing. This goes beyond previous work where continuous variables were used to efficiently and naturally simulate $U(1)$ gauge fields. To show how our formalism efficiently compiles subroutines that usually lack efficient (or even explicit) implementations, we provide concrete algorithms for eigenstate preparation based on Gaussian filtering. As a result of compiling spectral-filtering continuous transformations via CV ancillary degrees of freedom, our proposed quantum imaginary time evolution (QITE) with the squared Hamiltonian as the generator simultaneously minimizes hardware resource requirements, saturates optimality bounds, and is embarrassingly parallelizable given multiple ancillary qumodes.

\section{Main Results and Outline}
This section provides an outline of this work and summarizes our main results. Readers are encouraged to first read this section as a high-level guide to our work and then proceed to the sections they are most interested in. To provide a unified theoretical framework, while also giving concrete examples, we i) provide a high-level picture and the intuition for the power and utility of CV/DV quantum computation, ii) describe a gate-set which enables hardware compilation, iii) present one detailed paradigmatic example (this can be used as a template to compile to other generalized tasks or for software automation), and iv) discuss physical errors. 

Before delving into the technical details, we present an informal, high-level outline of the scheme. For pedagogical purposes, let us consider for now only a single qumode which can interact with a spin chain via a series of hybrid qubit-qumode gates. Express the input state of the spin chain as a superposition of eigenenergy basis states of the target qubit Hamiltonian $\hat{H}$ (or the scaled-shifted $\hat{H}'$), and let the qumode be in the vacuum (ground) state,
\begin{align}
    \ket{\Psi_\text{in}}
 &= \sum_n \psi_n \ket{E_n} \otimes \ket{0}. \label{eq:Intro1}
\end{align}
The key primitive of our protocol is a parameterized circuit (denoted as a unitary operator $D_{\hat{H}'}(\alpha)$ and to be described in detail later) which displaces the qumode coherent state away from the origin by a distance proportional to each of the energy components comprising the spin chain wavefunction. Ideally, the protocol maps the initial state to 
\begin{align}
    \ket{\Psi_\text{out}} 
 &= D_{\hat{H}'}(\alpha) \ket{\Psi_\text{in}} 
  = \sum_n \psi_n \ket{E_n} \otimes \ket{\alpha_n},
    \label{eq:Intro2}
\end{align}
where $\ket{\alpha_n}$ is a qumode coherent state with amplitude proportional to the $n$th energy eigenvalue of the spin system $\alpha_n = \alpha (E_n-E_\text{s})$. The constant parameter $\alpha$ will depend on the details of the circuit, and the energy shift (or offset) $E_\text{s}$ will be set by the user as a target energy for filtered states. The final step is to measure the photon number in the qumode. Post-selecting on obtaining zero, the measurement disentangles the spin chain from the oscillator. This leaves the oscillator in the final state
\begin{align}
    \ket{\Psi_\mathrm{final}}
  & \propto \sum_n \psi_n \ket{E_n}\! \ip{0}{\alpha_n},
    \label{eq:Intro2b}
\end{align}
where we have ignored the normalization factor. Upon measuring the qumode to be in the vacuum state, Gaussian QITE energy filtering arises from the overlap $\ip{0}{\alpha_n} = {\exp}(-|\alpha_n|^2/2) = {\exp}(-\alpha^2 |E_n - E_\text{s}|^2/2)$.

The components of this protocol are introduced in detail in \cref{sec:Methods} where we rigorously formulate the spectral filter, provide further intuition, and prove the equivalence of the CV-DV protocol with Gaussian QITE. To do so, we first describe the unitary $D_{\hat{H}'}^{(j)}(\alpha)$ and the initialization and measurement of the $j$th auxiliary qumode. \Cref{thm:cv-dv phase} highlights CV-DV phase-kickback, which is the key to qubit-controlled-displacements that are proportional to the spin system's energetics. This result is then leveraged to perform a CV/DV encoding of continuous linear combinations of unitaries. Our fundamental theorem of CV-DV co-design (\cref{thm:codesign}) proves $\ev{D_{\hat{H}'}^{(j)}(\alpha)}{0} = e^{-\frac{1}{2} \hat{H}'^2\alpha^2}$, i.e., an equivalence between Gaussian QITE filters $e^{-\frac{1}{2} \hat{H}'^2\alpha^2}$ and a block-encoded CV/DV circuit $\ev{D_{\hat{H}'}^{(j)}(\alpha)}{0}$. 

Since success probability is a grand challenge that limits the realization of qubitization-based and block-encoded algorithms, the remainder of \cref{sec:Methods} provides alternatives to the typical block-encoding analysis, which relies on success probabilities or amplitude amplification. As part of this, we describe the intriguing generalized filtering block encodings which are realized by measuring non-vacuum states. For photon number-resolving measurements we uncover Gauss-Hermite spectral filter transformations [\cref{eq:photon_POVM}], while homodyne measurement leads to complex-time evolutions [\cref{eq:Ux0-beta}]. Both are potentially useful in their own right, but we leave further analysis as an open research topic. We then proceed to our main goal of detailing the compilation of the previously opaque spectral filter to a specific hardware instruction set. 

To achieve our main goal, \cref{sec:hardware} then provides a review of CV-DV terminology, existing hardware architectures, and a concrete gate set (summarized in \cref{tab:gates}). While our methodology is generally applicable to a variety of platforms that control matter qubits using continuous degrees of freedom (e.g., trapped ions~\cite{Araz2025}), for the sake of concreteness, we explicitly compile to superconducting qubits coupled to high-$Q$ resonators (considered for example in Refs.~\cite{Liu2024, Crane2024}) and leave compilation to additional platforms as open research. We imagine that such compilation can be automated in existing software workflows, and we note that the CV-DV gate sets considered in this work are implemented in Bosonic Qiskit~\cite{BosonicQiskit}.

\Cref{sec:app-heisenberg-model} selects the Heisenberg model spin Hamiltonian as a paradigmatic example%
~\footnote{In addition to being widely used in describing many magnetic materials, this model and its variants also find broad applications in many-body localizations, spin dynamics, and even quantitative finance~\cite{Ciceri2025}, and is thus often used as a benchmark for algorithmic advantage from quantum computing~\cite{Childs2018}.}
to demonstrate how an end-to-end compilation (i.e., mapping a specific but opaque oracle to hardware-level gates) is performed. By judiciously constructing SWAP networks, that compile $k$-local ($k \geq 2$) interaction Hamiltonians between neighboring and non-neighboring lattice sites, the compilation techniques for this algorithm can be straightforwardly generalized (following the logic in \cref{fig:compilation}) to more exotic spin models. For example, the Lipkin-Meshkov-Glick model will incur an additional polynomial $N^2$ overhead due to the all-to-all, but 2-local, spin-spin interactions. In addition, leveraging locality-preserving fermion-to-spin mappings~\cite{Verstraete_2005, Whitfield2016, Jiang2019, Derby2021}, our algorithm could also spectrally filter fermionic systems such as the (geometrically 2-local) Fermi-Hubbard model, (4-local) quantum chemistry problems, and (typically 4- or 8-local) Sachdev--Ye--Kitaev models. 

Compilation to hardware requires constructively approximating the time evolution under a target Hamiltonian. As such, \cref{thm:LTHS} generalizes \cref{thm:codesign} to the approximate case and provides bounds on the precision errors that arise via Trotter-synthesizing a spectral filter. For the Heisenberg model, we first compile the individual local Hamiltonian terms (this compilation can again be generalized to arbitrary $k$-local Hamiltonians) and then, as illustrated by \cref{fig:compilation}, approximate the global Hamiltonian dynamics. This results in an approximated spectral filter $\hat{R}_{\hat{H}'}(\alpha)$, such that $\norm{\hat{R}_{\hat{H}}(\alpha) - \hat{P}_{\hat{H}}(\alpha)} \leq \varepsilon$ where $\hat{P}_{\hat{H}'}(\alpha)$ is the ideal filter and $\varepsilon$ is a precision chosen by the programmer. To the best of our knowledge, this constitutes the first rigorous error bound analysis, without assuming a finite Fock space cutoff, and compilation of an opaque continuous energy-filtering oracle to hardware-level instructions. 

While a single oscillator is sufficient to compile the spectral filter, we generalize this result and describe how additional oscillators, which are present in superconducting and trapped-ion hardware architectures, can be leveraged in parallel to achieve a quadratic speedup in the filtering process. Specifically, \cref{thm:M-oscillator} proves the oscillator resources-time trade-off in the exact filter. The key insight is that, as a parallelized compilation, the circuit depth of the SWAP network used in a multi-oscillator compilation is the same depth as in the original compilation. \cref{thm:parallel} then combines parallelization with the $\epsilon$-approximate filter to give precision bounds for the quadratically sped-up scaling that can be achieved in realistic multi-oscillator hardware. \cref{fig:visualCircuits} and \cref{fig:6siteCircuits} provide explicit examples of how SWAP networks are utilized in multi-oscillator compilation. 

To complement our analytical results, we verify and validate our findings with numerical simulations in \cref{sec:Numerics}. Importantly, this result details the regimes at which numerical errors are well behaved (meaning vanishingly small $\epsilon$) and describes the Floquet case. Here one shifts perspective to see that Trotterization filters into the Trotter-Floquet eigenstates which, for small Trotter-steps, are many-body localized to the original time-independent eigenstates~\cite{Heyl2019}. Since the Floquet eigenstates share symmetries with the original eigenstates, this regime is proposed as a bootstrapping technique to first project into the appropriate symmetry sectors before later applying either more refined energetic filters (small $\epsilon$ case), defined by our algorithm, or subgroup symmetry projectors. \cref{tab:H4_eigs} in \cref{sec:AnalyticalResults} provides details of the example we have chosen, but we conjecture that this holds for more general models. Finally, in \cref{sec:2D}, the algorithm is compiled to a 2D hardware layout, re-emphasizing our filter's generality with respect to a target model's dimensionality.

After a detailed accounting for algorithmic errors, \cref{sec:physical_error} then analyzes the effects of leading-order physical errors. Specifically, we consider the effects of photon loss. In addition, we outline randomized benchmarking-inspired protocols to provide routes to experimentally disentangle algorithmic and physical errors. Lastly, we conclude in \cref{sec:conclusion} by discussing the next steps required to experimentally realize our algorithms, as well as the new algorithmic and compilation research avenues which our work introduces.

\section{Methods}
\label{sec:Methods}

We now discuss the components needed to realize the spectral filter. In later sections we will see how to compile it using an experimentally available instruction set. Specifically, \Cref{ssec:spectral-projection-hs-trans} outlines spectral filters and how they are, via Hubbard-Stratonovic transformation, represented via continuous linear superpositions of time evolutions. \Cref{ssec:block-encoding-oracle} then details how the complex-time evolution oracle is physically encoded via a controlled displacement and measurement in a hybrid CV-DV system. Considering both continuous and discrete CV measurements, \Cref{ssec:oracular-int-transform} further generalizes the spectral and singular-value filters to those with Hamiltonian-Hermite polynomials or additional real-time evolutions.

\subsection{Gaussian Imaginary Time Evolution}
\label{ssec:spectral-projection-hs-trans}

Given a target Hamiltonian $\hat{H}$, denote its eigenenergy-eigenvector pairs, in the order of ascending energy, as $\qty{\qty(E_n, \ket{E_n})}_{n=0}^N$. A spectrum-shifted Hamiltonian $\hat{H}' = \hat{H} - E_\text{s}$, has the same eigenstates as $\hat{H}$. When shifting by the ground state energy, $E_\text{s} \coloneqq E_0$, the generator $\hat{H}'$ is positive semidefinite $\hat{H}' \geq 0$, meaning that its spectrum is non-negative. For the remainder of the article for simplicity, we assume a closed ($E_{\text{s}} \in \mathbb{R}$) principal system. Define a Gaussian QITE propagator: 
\begin{align}
   \hat{P}_{\hat{H}'}(\tau) 
 &= e^{-\frac{1}{2} \hat{H}'^2\tau^2} = \sum_{n} e^{-\frac{1}{2}(E_n - E_\text{s})^2 \tau^2} \op{E_n} \label{eq:filter_general_shift}\\
 &= \op{E_0} + 
    \sum_{n\neq 0} e^{-\frac{1}{2}(E_n - E_0)^2 \tau^2} \op{E_n}.
 \label{eq:filter_GSE_shift}
\end{align}
\Cref{eq:filter_general_shift} indicates $\hat{P}_{\hat{H}'}(\tau)$ filters a quantum state by Gaussian weights, centered around $E_\text{s}$, on its components in terms of the Hamiltonian eigenbasis (common basis for both $\hat{H}$ and $\hat{H}'$). In \cref{eq:filter_GSE_shift} $E_\text{s} \coloneqq E_0$. If $E_n > E_0$ for all $n \neq 0$, meaning the shift is with respect to a nondegenerate ground state~\footnote{The method can be applied to degenerate ground states with some small adaptation}). In the same limit but with $E_\text{s} \coloneqq E_n$, we obtain an eigenstate projector $\op{E_n}$, while for arbitrary $E_\text{s}$ it results in an energy filter near $E_\text{s}$. For targeting a ground state, recent works investigated iteratively learning $E_0$ when a spectral gap exists but the energy itself is a-priori unknown~\cite{Lin2020}. We note that spectral filtering and eigenstate preparation are related, but distinct, tasks. In the limit that the spectral filter's variance is much smaller than the spectral gap between the desired and its neighboring eigenstates spectral, as in the limit $\tau \to \infty$, filtering becomes eigenstate preparation with, e.g., $\hat{P}_{\hat{H}'}(\tau\rightarrow\infty)=\op{E_0}$ being a ground state projector. Since eigenstate preparation is in the worst case QMA-hard (for a very small spectral gap) but coarse spectral filtering need not be so challenging, our protocol offers an new pathways to explore the complexity of this task in a controlled but approximate setting. 

Instead of naively compiling \cref{eq:filter_general_shift} in terms of an exponential of the unwieldy generator $\hat{H}'^2$, we simplify compilation with the Hubbard-Stratonovich (HS) integral transformation. Begin with the Gaussian integral identity $\mathds{1} = \sqrt{a/\pi} \int dp\, {\exp}[-a(p \mathds{1}+\hat{K})^2]$, where $\Re a \geq 0$ for $a \neq 0$, $\hat{K}$ is any operator, and $\mathds{1}$ is the corresponding identity operator. Multiplying both sides of the Gaussian integral identity with $e^{a\hat{K}^2}$ results in
\begin{align}
     e^{a\hat{K}^2}
  &= \sqrt{\frac{a}{\pi}} \int_{-\infty}^{+\infty} dp\, e^{-a p^2} e^{-2 a\hat{K} p}. \label{eq:HS_transform}
\end{align}

The next step is to interpret \cref{eq:HS_transform} as a HS integral transformation encoding a Hamiltonian-generated QITE propagator, as in \cref{eq:filter_GSE_shift}, in terms of a \textit{continuous} linear combination of unitaries. To see this, select $\hat{K} =i\hat{H}^{'} \tau / 2$ and define tensor product space time evolution operators as 
\begin{equation}
    \label{eq:tensor_product_evolution}
    U^{(\hat{B})}_{\hat{Q}}(t) \equiv e^{-i2t \hat{B} \otimes \hat{Q}}.
\end{equation}
Then, letting $\mathcal{N}(\mu,\sigma)$ denote a normal distribution and setting $a=2$ (see \cref{sec:bosonic_convention}), we arrive at
\begin{align}
  \label{eq:filter_integral_transform}
 \hat{P}_{\hat{H}'}(\tau) &= \sqrt{\frac{2}{\pi}} \int_{-\infty}^{+\infty} dp\, 
    e^{- 2p^2}e^{-2i \tau \hat{H}' p}  \\ 
  &=   \int_{-\infty}^{+\infty} dp\, \mathcal{N}(0,1/2) U_{\hat{H}'}^{(p)}(\tau). 
\end{align}
The next section outlines how compilation in terms of CV/DV real-time evolution unitaries and bosonic measurements is more resource-efficient than factorizing the QITE propagator directly in the qubit's Hilbert space.

\begin{figure}
  \centering
  \includegraphics[width=1.0\linewidth]{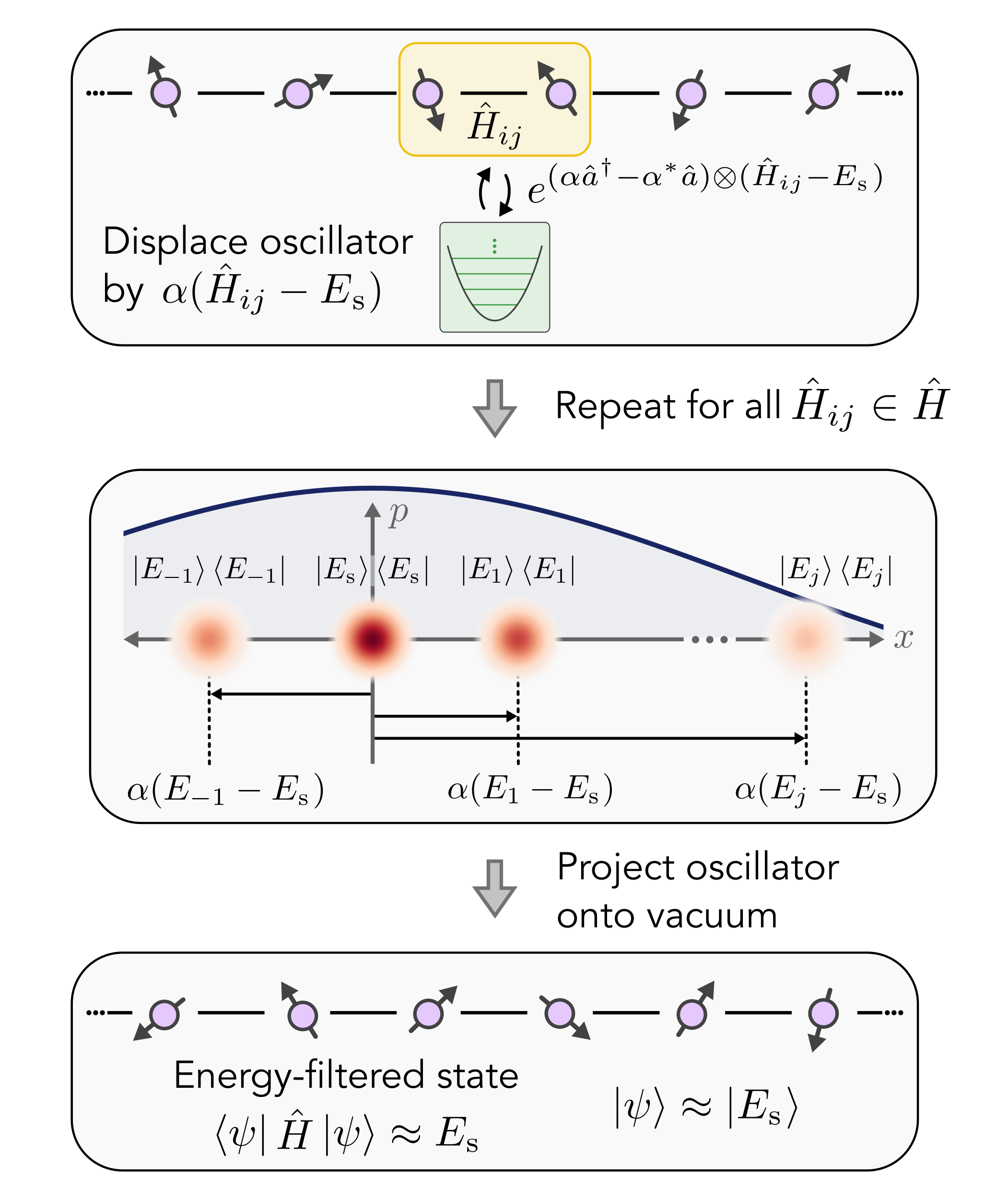}
\caption{Top panel: Schematic illustration of a displacement $\alpha$ controlled on a local interaction $H_{ij}$. Middle panel: The spin spectrum decays as a result of spectrum-conditioned oscillator translations and measurement which results in the bottom panel's approximate ground state.}
  \label{fig:high-level}
\end{figure}

\subsection{Block Encoding Eigenvalue Transformation Oracles}
\label{ssec:block-encoding-oracle}

Block-encoding is a powerful tool for encoding general purpose, non-unitary operations~\cite{Gilyen2019,Martyn2021} within quantum computations. An efficient realization of block-encodings is crucial for achieving general purpose linear-algebraic advantages, e.g., in quantum singular value transform-based algorithms. It has previously been shown that DV quantum resources (ancillary qubits) can be used to construct efficient block-encodings of both DV \cite{camps2022fable,kuklinski2024s,camps2024explicit} and CV operators \cite{sinanan2023single,Rossi23,Liu2024}. By ``bosonizing'' the continuous linear combination of unitaries oracle, we simultaneously remove discretization errors and minimize the ancillary resources required by effectively carrying out the Gaussian integral exactly rather than approximating it by a Riemann sum. Further, the non-trivial \texttt{Prepare} and \texttt{Select} (LCU) or iterate circuits (qubitization) are dramatically simplified and replaced by a single CV-DV unitary. The remainder of this section introduces the components of this protocol. Later, in Sec.~\ref{sec:hardware}, we detail a particular hybrid architecture which consists of i) a qubit-sublattice and ii) a qumode-sublattice and discuss the hardware-level synthesis of the protocol. The \textit{native} interactions between the pair of sub-lattices are denoted by pairs of arrows in the top panel of \cref{fig:high-level}.

\subsubsection{Constructing Block-Encodings using Continuous Linear Combination of Unitaries}

In DV-only methodologies, the continuous integral in ~\cref{eq:filter_integral_transform} is necessarily expressed as a Riemann summation over $d$ terms. For example, in the linear combinations of unitaries (LCU) algorithm, a $\log(d)$ sized qubit ancilla bank is required to encode the discretized Gaussian integrals weights $\mathcal{N}(0,1/2)$. Following this discretization strategy in the context of the LCU algorithm~\cite{Childs2012, Dalzell2025}, an $\varepsilon$-approximation of $\hat{P}_\tau$ was recently proposed~\cite{Keen2021}. Doing so involved storing the gaussian weights as amplitudes on multiple ancillary qubits (\texttt{PREP}), which was followed by ancillary-qubit-controlled real time evolutions (\texttt{SELECT}). This method matched the asymptotically optimal precision errors and query complexity bounds (see Table I of \cite{Keen2021}) from Refs.~\cite{Ge2019, Lin2020} which operate within a DV qubitization-based setting, for which the algorithmic optimality is proved~\cite{Lin2020}. Qubitization, described further below, makes $d$ calls to an iterate which applies the desired functional (real time evolution in this case). In contrast, as will be explained below, our method is continuous and alleviates the need for discretization. To achieve this, we encode \cref{eq:filter_integral_transform}'s continuous distribution over real time evolutions directly onto the modulus squared (probabilities) of a CV oscillator's wavefunction. In addition to improving accuracy by removing discretization errors, our method matches resource minimization seen in qubitization while dramatically simplifies the compilation and reduces the gate depth. 

Note that in our algorithm, the block-encoding is essentially a controlled time-evolution of the target many-body system. The cost for block-encoding will be similar in our case as compared to typical DV ancilla, because every term in the target many-body Hamiltonian $H$ has to be entangled with \emph{either} an ancillary qubit (in the typical DV ancilla case) \emph{or} the ancillary qumode (in our case). This means the number of single-qubit controlled displacement in our case should roughly be similar to the number of CNOT gates when using DV ancilla to construct the block-encoding.
In qubitization methods, recall that the overall gate cost for performing a polynomial transformation is proportional to the product between the polynomial transform degree $d$ and the gate cost for constructing the block-encoding. This is because the block-encoding has to be queried roughly $d$ times to realize a degree-$d$ polynomial transformation of the block-encoding (see Ref.~\cite{Ge2019, Lin2020}). In contrast, in our case, only a single call to the block-encoded time-evolution is needed, because the polynomial transformation is realized by the continuous variable wave function $\psi_0(x)=\ip{x}{0}$ of the qumode's initial state $\ket{0}$. Thus, highlighting the benefit of naturally encoding continuous functions without breaking them up into a discrete polynomial basis, the computational complexity is reduced from $\mathcal{O}(d)$ to $\mathcal{O}(1)$. Finally, in addition to reducing errors and resource requirements as compared to qubitization, compilation is also conceptually simpler than LCU because we do not need the separate \texttt{Prepare} and \texttt{Select} operations. In contrast, as described below, our protocol only requires a qubit-qumode interaction. This simplification enables us to shed layers of abstraction and deliver a concrete end-to-end compilation of the time evolution.  

Lastly, in addition to block-encoding, near-term $e^{-\tau H}$, QITE strategies have also been investigated. Ref.~\cite{Motta2020} and follow-on works suggest finding the shortest depth unitary minimizing the $\varepsilon$-error from a target QITE. However, applying this approach to genuine many-body systems generically suffers from exponential resource costs stemming from either the original tomography or later derivative-based variational methods~\cite{Getelina2023}. The guiding rationale of this work is that such issues can be avoided by judiciously pairing algorithmic tasks with hardware that is best suited to accomplish them. 

To realize \cref{eq:filter_integral_transform} as a circuit, we must specify the quantum mechanical operations that are required. The first component is a unitary relying on a controlled interaction between the ancilla qumode and principal system. The second component is the ability to initialize the ancilla CV system into the vacuum state $|n=0\rangle$, as well as to measure the projector onto the $\op{n=0}$ state. Let us now examine how such initializations, unitaries, and measurements serve to define families of nonunitary, block-encoded operators spectrally filtering the principal system.

To begin, \cref{eq:filter_integral_transform} can be encoded by a \emph{single} ancillary qumode. This is because the CV, first quantized dimensionless position $\hat{x} \coloneqq \frac{1}{2}(a+a^\dagger)$ and momentum $\hat{p} \coloneqq \frac{i}{2} (a^\dagger -a)$ operators obey the canonical commutation relation (CCR) $[\hat{x}, \hat{p}] = (i/2) \mathds{1}$ ~\footnote{The CCR indicates that unbounded CV operators cannot reside in a finite dimensional Hilbert space~\cite{[][{, Example 1.}] Gieres2000, [][{, p95.}] Weyl1931} like qubit systems. Suppose such a trace were well defined. Then $\Tr([\hat{x}, \hat{p}]) = \Tr(\hat{x}\hat{p}) - \Tr(\hat{p}\hat{x}) = 0$ by the cyclic property, which contradicts the CCR.} and have continuous eigen-spectrum $x,p \in \mathbb{R} = (-\infty, \infty)$ ~\footnote{The eigenvectors corresponding to these eigenvalues live in a rigged (meaning equipped) Hilbert space so the spectrum need not be quantized (or countable).}. The operators $\hat{a}^\dagger \text{ and }\hat{a}$ are the bosonic creation and annihilation operators (see \cref{sec:bosonic_convention} for more details) for a photon of a superconducting circuit cavity's electromagnetic mode or a phonon of a mechanical vibration mode. Representing \cref{eq:filter_integral_transform} on a single qumode will enable us to later parallelize QITE and reduce the required resources. 

Before entangling with DV qubits, recall the $j^\text{th}$ qumode's (unconditional) displacement gate \cite{Liu2024}: 
\begin{align}
\label{eq:displacement}
     D^{(j)}(\alpha)
  &= e^{\alpha \hat{a}^\dagger_j - \alpha^* \hat{a}_j}
   \equiv e^{\widetilde{\alpha}_j} \\
  &= e^{\alpha \hat{a}^\dagger_j} e^{- \alpha^* \hat{a}_j} e^{- |\alpha|^2/2} \label{eq:BCH_disentangle} \\
  &= e^{2i\left[ (-\Re \alpha) \hat{p}_j + (\Im \alpha) \hat{x}_j \right] } \notag \\
  &= e^{-2i(\Re \alpha) \hat{p}_j} e^{2i(\Im \alpha) \hat{x}_j} e^{i (\Re \alpha) (\Im \alpha)} \notag \\
  &= D^{(\hat{p}_j)}(\Re\alpha) D^{(\hat{x}_j)}(i\Im \alpha) e^{i (\Re\alpha) (\Im\alpha)}. \label{eq:D_product}
\end{align}

Coherent states are then defined as phase-space displacements $\ket{\alpha} \equiv D(\alpha)\ket{0}$. While usually defined in terms of the anti-Hermitian generator $-2 i \alpha \hat{p}_j \equiv \alpha \hat{a}^{\dagger}_j - \alpha^{*} \hat{a}^{\pdag}_j =  -2i[(\Re \alpha) \hat{p}_j - (\Im \alpha) \hat{x}_j]$, it is also useful to examine the Hermitian generators $2 (\Re \alpha) \hat{p}_j  $ and $- 2 (\Im \alpha) \hat{x}_j$. The former (latter), defined in terms of $\hat{p}_j$ ($\hat{x}_j$), generates spatial (momentum) translations displacing the oscillator's position (momentum) $\ev{\hat{x}_j}{\alpha} = \Re \alpha$ ($\ev{\hat{p}_j}{\alpha} = \Im \alpha$). For simplicity, and without loss of generality, we take $\alpha \in \mathbb{R}$ such that $D^{(j)}(\alpha \in \mathbb{R}) = D^{(p_j)}(\alpha) = e^{-2i \alpha \hat{p}_j}$, for the remainder of this work.

\subsubsection{Controlled Displacements and Measurement}

CV bosonic oscillators (indexed with superscript $j \in B$) and DV qubits (indexed with subscript $k \in Q$) are coupled in the hybrid architecture under consideration. The native entangling gate, displacing the oscillator conditioned on a qubit's $\sigma_k^Z:=Z_k$ degree of freedom, is
\begin{align}
\label{eq:cD}
     D^{(j)}_{Z_k}(\alpha) 
  &= e^{-2 i\alpha Z_k \otimes \hat{p}_j}.
\end{align}

From a qubit perspective, \cref{eq:cD} block-encodes a position boost $e^{-2i \alpha \hat{p}_B} = \bra{0}_Q D_{Z_Q}^{(B)}(\alpha) \ket{0}_Q$ in the upper-left block of a qubit's SU(2). Exchanging the role of qubit and the oscillator, \cref{eq:cD} block-encodes a qubit operator within the vacuum transition amplitude (VTA) of the oscillator subspace, i.e., $\bra{0}_B D_{Z_Q}^{(B)}(\alpha) \ket{0}_B = \op{0}_Q \otimes \bra{0}_B e^{2i\alpha\lambda_0 \hat{p}} \ket{0}_B + \op{1}_Q \otimes \bra{0}_B e^{-2i\alpha\lambda_1 \hat{p}} \ket{0}_B = e^{-\alpha^2 / 2} (\op{0}_Q + \op{1}_Q)$. In this case both qubit levels $\{\ket{0}_k, \ket{1}_k\}$ are equally projected since their energies are $\lambda_{0,1} = \pm 1$ and equal in absolute distance from $E_s=0$.

We are now in a position to generalize \cref{eq:cD}, with a qubit-controlled translation, to a many-body-\textit{spectrum}-controlled translation. A precise example is presented later in \cref{sec:app-heisenberg-model}. By promoting single qubit $\mathfrak{su}(2)$ generators to many-body Hamiltonians, as $Z_{k} \to \hat{H}'$, we see:
\begin{lem}[CV-DV phase kick-back] \label{thm:cv-dv phase}
An oscillator position displacement weighted by the spectrum of a spin chain $\hat{H}'$ is, by interchanging the role of spin chain and oscillator (see \cref{fig:high-level}), equivalent to a real-time dynamics of $\hat{H}'$ with the evolution time weighted by the oscillator's momentum:
\begin{align}
  \label{eq:time-position-displacement}
  D_{\hat{H}'}^{(j)}( \alpha \in \mathbb{R})
  \coloneqq e^{-2 i \alpha  \hat{H}' \otimes \hat{p}_j}
  \eqqcolon U_{\hat{H}'}^{(\hat{p}_j)}(\alpha).
\end{align}
\end{lem}
We will use the left hand side expression when discussing hardware compilation and the right hand side expression in an algorithmic context. This leads to our first result:
\begin{thm}[CV-DV-HS Transformation Oracle]
\label{thm:codesign}
The Eigenvalue Transformation Oracle in \cref{eq:filter_general_shift} can be block-encoded as a Vacuum Transition Amplitude (VTA):
\begin{align}
  \langle 0| D_{\hat{H}'}^{(j)}( \alpha) |0\rangle &= \int dp\, \langle 0| p\rangle \langle p|D_{\hat{H}'}^{(j)}( \alpha) |0\rangle,\\
  &=\int dp\, \langle 0| p\rangle\langle p|0\rangle\, e^{-i 2\alpha p \hat{H}' } \\
  &=\hat{P}_{\hat{H}'}(\alpha).
  \label{eq:Ux00SMG}
\end{align} 
Proof: Insert the momentum-space resolution of identity to derive the first line, act in the momentum basis in the second line, and use $\langle 0|p\rangle\langle p|0\rangle= (2/\pi)^{1/2} e^{-2p^2}$ (\cref{sec:bosonic_convention}) to encode the continuous normal distribution $\mathcal{N}(0,1/2)$ in the last step.
\end{thm}
\cref{eq:Ux00SMG} offers a bosonically block-encoded representation of \cref{eq:filter_GSE_shift}. In this interpretation, the final oscillator state after the controlled displacement is a function of $E_n$ as in the middle panel of \cref{fig:high-level}. Projecting the ancillary qumode onto the vacuum state then reveals a diagonal qubit ensemble $\rho_Q = \sum_n p_n \op{E_n}$ with probability $p_n = e^{-\frac{1}{2} (E_n - E_\text{s})^2 \alpha^2} / \sum_n e^{-\frac{1}{2} (E_n - E_\text{s})^2 \alpha^2} $. With $\alpha$ and $E_\text{s}$ parametrically tuning the filter strength and target energy, this corresponds to the Gibbs ensemble for Hamiltonian $\hat{H}'^2$ and the inverse temperature $\beta = \alpha^2/2$. 

\subsection{Generalized Integral Transformations Oracles}
\label{ssec:oracular-int-transform}

For completeness, we now highlight the extended family of block encodings appearing as vacuum to non-vacuum transition amplitudes. Recall that in a DV block encoding one prepares initial DV ancilla state ($\ket{0}$) and, depending on the final measured state ($\ket{j}$), has subsequently applied a (possibly non-unitary) $\hat{A}_{0,j}$. Deterministic quantum algorithms~\footnote{Ref.~\onlinecite{Chen2020} being an inspirational \textit{exception} to this trend} typically consider only the $j=0$ sector. Consider first the positive operator-valued measure (POVM) $\qty{ \op{m} }$ associated with a microwave photon detector measuring the final state of the oscillator in the $m$-photon Fock state basis $\ket{m}$ with $m \in \{0,1,\cdots\} \equiv \mathbb{N}_0$. 

Using $H_m (p)$ to denote the $m$th Hermite polynomial and $m$-photon wavefunction $\psi_m(p) = \ip{p}{m}$ given at the end of \cref{sec:bosonic_convention}, we have 
\begin{align}
    \hat{P}_{\hat{H}'}&(m, \tau) 
     =  \int_{-\infty}^{+\infty} dp \ip{m}{p}U_{\hat{H}'}^{(p)}(\tau) \ip{p}{0} \nonumber \\
   &= \frac{i^m}{\sqrt{\pi 2^{m-1} m!}} \int_{-\infty}^{+\infty} dp\, e^{-2 p^2}  H_m (\sqrt{2} p) e^{-i\hat{H}' \tau 2p} \notag\\
   &= \frac{(\tau \hat{H}')^m}{\sqrt{m!}} e^{-\frac{\tau^2}{2}\hat{H}'^2}.
   \label{eq:photon_POVM}
\end{align}
An additional simplified proof of \cref{eq:photon_POVM} using operator calculus~\cite{Feynman1951} appears in \cref{eq:photon_U_m}.

The integral in the last step can be viewed as \emph{Hermite integral transform} which reduces to the HS transform when $m=0$. The corresponding Hermite transform pair of the time evolution oracle $e^{-it\hat{H}}$ can be looked up in a standard mathematical table. In \cref{eq:filter_integral_transform}, the number dependent prefactor $\braket{m}{z} = \frac{z^m}{\sqrt{m!}}\braket{0}{z}$ simplified to unity.
For $m=0$, \cref{eq:photon_POVM}  reduces \cref{eq:filter_general_shift} and for $\tau=0$ this expression reduces to the Fock-basis delta function $\ip{m}{0}=\delta_{m,0}$. For $\tau>0$ we complete the square and integrate. Changing to variable $p'=p+i H'\tau$, and using $H_m (p)$ to denote the $m$th Hermite polynomial, gives us
\begin{align}
    \hat{P}_{\hat{H}'}(m, \tau) 
 & = \frac{ e^{-\frac{1}{2}\hat{H}'^2\tau^2} }{\sqrt{2\pi}} \int_{-\infty + i\hat{H}'\tau}^{+\infty + i\hat{H}'\tau} dp'  e^{-\frac{p'^2}{2}} H_m (p-i H'\tau)
\end{align}
Next use $H_m(p-iHt) = \sum^{m}_{k=0}\binom{m}{k} H_k(p) (-2iH'\tau)^{m-k}$. Note that the contribution from $k=0$ is just a phase shift of the original projector. In this way we can bound the \textit{additive} error by considering the expansion over $k \geq 1$.

The calculation above shows that detecting a photon number $m$ acts as a symmetrized Hamiltonian polynomial. Complementary to Ref.~\cite{Bespalova2021} which investigated \textit{inverse} Hamiltonian polynomials, factors of $(\tau \hat{H}')^m$ can also be investigated as an algorithmic resource, a research direction left for future work. Also note that \cref{eq:photon_POVM}'s Hermite basis~\footnote{Strictly speaking, the Rodrigues formula in \cref{eq:photon_POVM} defines the complex It\^{o}-Hermite polynomial basis $H_{m,n}(z,z^*)$.} complements the Chebyshev basis use in quantum signal processing~\cite{Low2019}.

Alternatively, consider projecting the oscillator to a final CV coherent state $\ket{\beta}=D(\beta)\ket{0}$. A weighted summation of \cref{eq:photon_POVM} leads to
\begin{align}
     \hat{P}_{\hat{H}'}(\beta \in \mathbb{C}, \alpha) &= \langle \beta | U_{\hat{H}'}^{(B)}(\alpha) | 0\rangle_B \\
  &=e^{-i {\alpha \beta \hat{H}'}} P_{\hat{H}'}(\alpha) \\
  &= \sum_n e^{-\frac{(\alpha E_n+i \beta)^2}{2}}  \ket{E_n} \bra{E_n}.
  \label{eq:Ux0-beta}
\end{align}
which reduces to \cref{eq:filter_general_shift} when $\beta = 0$. The above projection onto Gaussian coherent states corresponds to the POVM $\qty{ \frac{1}{\pi} \op{\beta} }$ that can be performed with a heterodyne detector~\cite{Chabaud2019, Jackson2022, Jackson2023}. Note the non-orthogonality $\ip{\alpha}{\beta} \neq \delta_{\alpha,\beta}$ for complex $\alpha$ and $\beta$. This POVM satisfies the overcomplete resolution of the identity $\frac{1}{\pi} \int_{\mathbb{C}} d \beta^2 \op{\beta}= \mathds{1}$ where $d\beta^2 = d\mathfrak{R}(\beta) d\mathfrak{I}(\beta)$.

\begin{figure}
  \centering
  \includegraphics[width=1.0\linewidth]{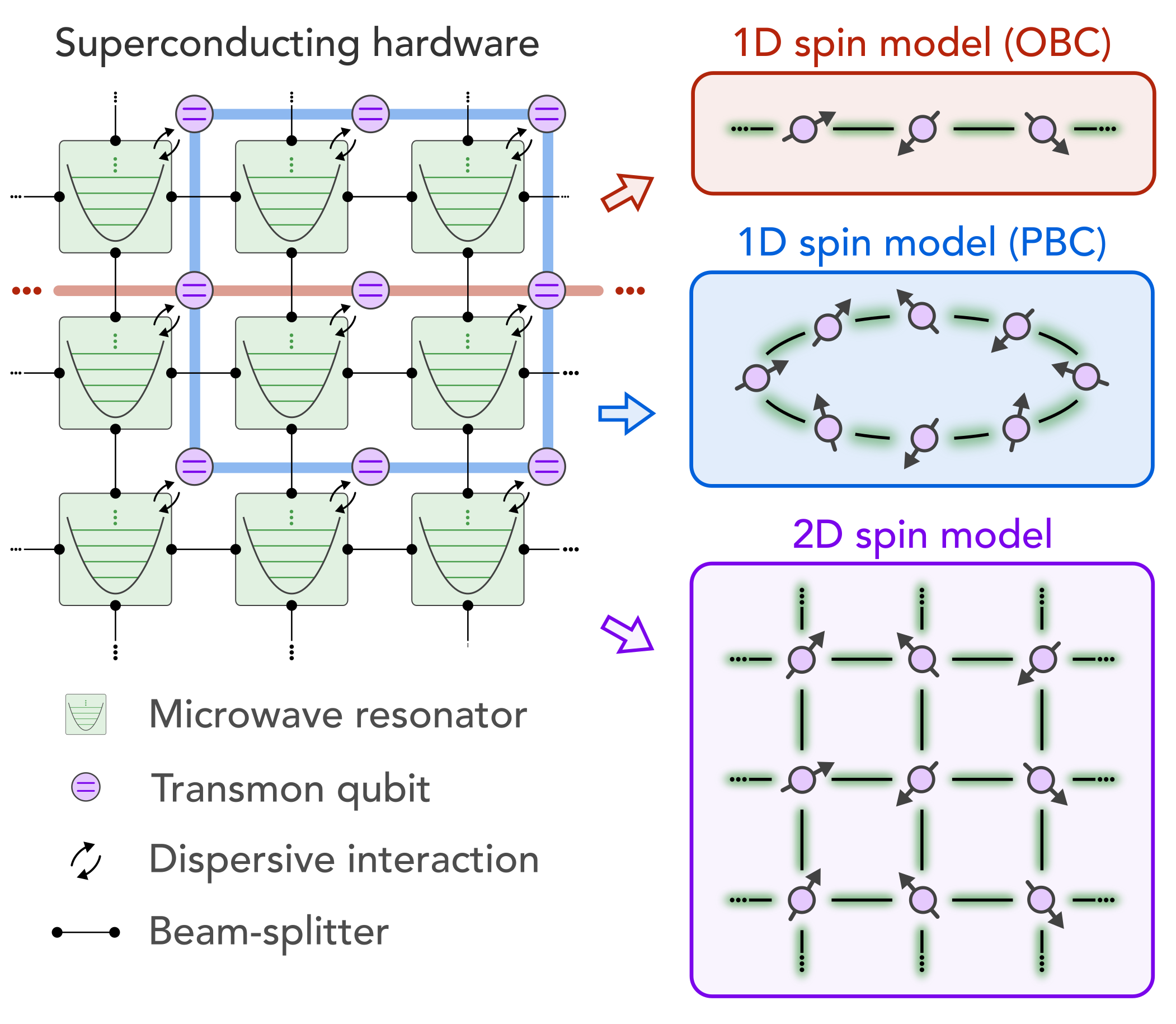}
\caption{Hardware layout of 2D superconducting quantum processor (left panel), illustrating the microwave (green) and trasmon qubit (purple) sublattices. See Figure~2 of \citenum{Liu2024} for additional trapped-ion and neutral-atom CV/DV architectures. The right column indicates embedding 1D geometries with open (OBC) and periodic (PBC) boundary conditions and a 2D generalization. Links between lattice sites are highlighted in green to emphasize that all spin-spin interactions are mediated via the oscillators.}
  \label{fig:hardware-mapping}
\end{figure}

Comparing \cref{eq:Ux0-beta} with \eqref{eq:Ux00SMG}, the general $\op{0}{\beta}_B$ block is a potentially powerful oracle, as it not only introduces a global damping factors for all $p_n$ of the mixed state, but also produces a complex shift $i\frac{\beta}{\alpha}$ for each eigenstates. This shift can be interpreted as a real-time evolution $e^{-i t \hat{H}'}$ with $t = \alpha \beta$, which commutes with the imaginary-time evolution low-energy filter. Thus, Eqs. \eqref{eq:Ux0-beta} and \eqref{eq:Ux00SMG} unify (imaginary-time) state preparation and (real-time) dynamical evolution into a single complex-time functional.

\cref{eq:photon_POVM} and \cref{eq:Ux0-beta} both generalize the VTA scattering process and also recovers our initial Gaussian spectral filter \cref{eq:filter_GSE_shift}, for example through the $m=0$ measurement. For $m\geq1$, in the eigenbasis, the wave functions spectral weight is mapped as $|c_\lambda|^2 \rightarrow \frac{(\alpha \hat{H})^{2m}}{m!}e^{-(\alpha \hat{H})^2}|c_\lambda|^2$ which provides a higher energy filter, compared with $E_{\text{s}}$. Lastly, note that $\text{Tr}_{\bf B}(|\mel{m}{U_p(\alpha)}{0}|^2) = \sum_{m=0}^{\infty} \hat{P}(m,\tau)^\dagger  \hat{P}(m,\tau) = \mathds{1}_{\bf Q}$. This means the filters expressed in \cref{eq:photon_POVM} are also Kraus operators, with each representing a distinct spectral filter path. Further, they satisfy the necessary completeness relation. One implication is that, by stochastically tracing over all photon numbers, or phase-space, we recover the logical identity $\mathds{1}_{\bf Q}$. This provides a mechanism to construct the identity operator using the oracle-level operations. Later, in \cref{sec:physical_error}, we will discuss how this enables protocols which quantify physical error rates. 

\section{Hardware Architecture and Compilation}
\label{sec:hardware}

Discovering new strategies for compilation and automating known compilation strategy via a compiler are two separate, but complementary, research goals. The compilation strategy we present below, building on the technical details of this section, is new and, therefore, addresses the former question. We emphasize that our strategy is scalable and general. On trapped ion systems, both in linear traps and in individual trapping zones such as in Quantinuum's machine, one key gate is the Molmer-Sorenson (MS) gate. At its core this is a hybrid CV-DV compilation strategy that entangles two ions (qubits) using motional CV modes, where the modes return to their original state after the gate (and therefore do not appear explicitly in the computation). From this perspective, the compilation strategies introduced below can be viewed as extensions of the MS gate. That is, we explicitly exposing the CV degrees of freedom and use them to perform a broader class of operations on DV systems which go beyond entangling gates, e.g., eigenstate filtering.

Regarding automating known compilation strategies, this has been partially addressed by one of the author's recent works~\cite{chen2025genesis}. Here, compilation is algorithmically encoded using a rule-based template matching that automates compilation of Hamiltonian simulation on hybrid systems. This work discovers new compilation strategy for state preparation, which complements the Hamiltonian simulation tasks that Ref.~\citenum{chen2025genesis} describes. In future work, we can encode the newly discovered compilation rules into Ref.~\cite{chen2025genesis} to grow it into a systematic and more optimized CV-DV compiler. 

\subsection{Layout and Connectivity}
We now compile our eigenvalue transformation oracle with respect to the hardware architecture and ISA detailed in Ref.~\cite{Liu2024}. As illustrated in \cref{fig:hardware-mapping}, the architectural layout consists of a pair of two-dimensional (2D) sub-lattices resulting in a 2D+1 spatio-temporal architecture. Superconducting transmon qubits (Q) comprise the first sub-lattice. Each qubit is directly coupled to a bosonic (B) harmonic oscillator, which is itself embedded within the second sub-lattice. The 2D qubit coordinates are indexed as $(0,0) \leq Q \coloneqq (Q_x,Q_y) \leq (N_x -1,N_y -1)$ and, following \cref{eq:tensor_product_evolution}, appear in the \textit{subscript} of hybrid CV-DV operations. Bosonic qumodes are indices go as $(0,0) \leq B \coloneqq (B_x,B_y) \leq (N_x -1,N_y -1)$ and appear in the superscript. For one-dimensional problems, we omit the $y$ index. 

\cref{tab:gates} outlines the gates used in our proposed architecture. The DV qubit gates are already well known and do appear below. The CV gates are also well studied and will, for example, involve beam-splitters at the edges connecting the vertices in \cref{fig:high-level}. The CV-DV gates are discussed in detail below. When oscillator $j$ is dispersively coupled to qubit $k$, $R^{(j)}_{Z_k}(\theta)$ and $D_{Z_k}^{(j)}(\alpha)$ are \textit{native} hardware operation. If spatially separated, the gate may be accomplished with the overhead of a SWAP network which exchanges the cavity mode from $j$ to the mode physically coupled with qubit $k$.

\subsection{CV-DV Compilation}
\label{sec:ISA}

\begin{table}[t]
\centering
\tabcolsep=0.2cm
\setlength\extrarowheight{5pt}
\begin{tabular}{|c|c|}
\hline
     \multicolumn{2}{|c|}{Qumode (CV) gates}
     \\[5pt] \hline \hline
      $R^{(j)}(\theta)$ &  $\exp \left(- i\theta \hat{a}_j^\dagger \hat{a}_j^\pdag \right) = \exp \left(- i\theta \hat{n}_j\right)$
      \\[5pt] \hline
      $F^{(j)}$ &  $R^{(j)}(\pi/2)$
      \\[5pt] \hline
      $D^{(j)}(\alpha)$ &  $\exp \left( \alpha \hat{a}_j^\dagger - \alpha^*\hat a^{\pdag}_j \right)\equiv \exp(\widetilde{\alpha}_j)$
      \\[5pt] \hline
      $\operatorname{BS}^{(j,k)}(\theta,\varphi)$ &  $\exp \left[-i\frac{\theta}{2}\left(e^{i\varphi} \hat a_j^\dagger \hat a^{\pdag}_k + e^{-i\varphi}\hat a^{\pdag}_j \hat a_k^\dagger\right)\right]$ 
      \\[5pt] \hline\hline
     \multicolumn{2}{|c|}{Qubit (DV) gates} 
      \\[5pt] \hline\hline
       $R_{\hat{m}_j}(\theta)$& $\exp \left(-i \frac{\theta}{2} \hat{m} \cdot \vec{\sigma}_j
    \right)$
      \\[5pt] \hline \hline
    \multicolumn{2}{|c|}{Hybrid (CV-DV) gates}
    \\[5pt] \hline \hline
    $R^{(j)}_{Z_k}(\theta)$ &  $\exp \left(-i\theta Z_k\otimes\hat{n}_j\right)$ 
      \\[5pt] \hline
      $P^{(j)}_{Z_k}$ &  $R^{(j)}_{Z_k}\left(\pi/2\right)$ 
      \\[5pt] \hline

      $D_{Z_k}^{(j)}(\alpha)$ & $\exp \left[Z_k\otimes \left( \alpha \hat{a}_j^\dagger - \alpha^*\hat a^{\pdag}_j\right) \right] = e^{Z_k\otimes \widetilde{\alpha}_j}$
      \\[5pt] \hline
\end{tabular}
\caption{Native CV-DV gate set corresponding to the `Phase-Space ISA' in Ref.~\cite{Liu2024}. Note that $\vec{\sigma}_j = (X_j, Y_j, Z_j)$ is the Pauli operator basis for a spin oriented along an arbitrary axis $\hat{m}_j = (m_{x}, m_{y}, m_{z})$ on the $j$th Bloch sphere.  The hybrid CV-DV gates are natively available for the case $j=k$. For $j\ne k$, the state of cavity $j$ must be swapped into cavity $k$ to perform the gate controlled by qubit $k$, and then swapped back to cavity $j$ (see \cref{eq:native_DcZ}).} \label{tab:gates}
\end{table}

Denote the CNOT gate with control on qubit $j$ and target on qubit $k$ gate as CX$_{jk}$. Two-body phase-rotations can be factorized into the product of a one-body phase rotations and a pair of CNOT gates as
\begin{align}
\label{eq:RZZ}
     R_{Z_j Z_k}(\theta) 
     \equiv e^{-i\frac{\theta}{2} Z_j Z_k}
  &= \text{CX}_{jk}^\dagger R_{Z_k}(\theta) \text{CX}_{jk}.
\end{align}

In this work the qubit-controlled (photon occupation number~\cite{Liu2024}) parity operator, defined as
\begin{align}
     P^{(j)}_{Z_k} 
  &= e^{-i \frac{\pi}{2} Z_k \otimes \hat{a}^{\dagger}_j \hat{a}^{\pdag}_j}
   = e^{-i \frac{\pi}{2} Z_k \otimes \hat{n}_j},
    \label{eq:def-CP}
\end{align}
will be analogous to the CNOT in how it compiles oscillator-controlled two-body interactions. The photon number operator is $\hat{n}_j = \hat{a}^{\dagger}_j \hat{a}^{\pdag}_j = \hat{x}_j^2 + \hat{p}_j^2 - \frac{1}{2}$, so the bosonic part of \cref{eq:def-CP} is generated by the standard quantum Harmonic oscillator interactions, including displacement. Note that $P^{(j)}_{Z_k} = R^{(j)}_{Z_k} (\pi/2)$. Under the qubit-controlled rotation, $R_{Z_k}^{(j)}(t) = {\exp}(-i t Z_k\otimes \hat{n}_j)$, the Heisenberg-picture (i.e., time-shifted) operators are:
\begin{subequations}
\begin{align}
     \mathds{1}_k \otimes \hat{a}_j(t)
  &\equiv R_{Z_k}^{(j) \dagger}(t) \mathds{1}_k \otimes \hat{a}_j R_{Z_k}^{(j)}(t) \nonumber\\
   &= e^{-i t Z_k} \otimes \hat{a}_j, \\
     \mathds{1}_k \otimes \hat{x}_j(t) 
  &= \cos(t) \mathds{1}_k \otimes \hat{x}_j + \sin(t) Z_k \otimes \hat{p}_j, \\
     \mathds{1}_k \otimes \hat{p}_j(t) 
  &= \cos(t) \mathds{1}_k \otimes \hat{p}_j - \sin(t) Z_k \otimes \hat{x}_j.
\end{align}
\end{subequations}
From now on we will omit ``$\mathds{1}_k\otimes$'' when there is no confusion. As a special case of the equations above, conjugation by a controlled-parity operator then adjoins a qubit's $Z_k=\pm1$ operator to the photon destruction operator 
\begin{subequations}
\begin{align}
  P^{(j)\dagger}_{Z_k} \hat{a}_j P^{(j)}_{Z_k} = -i Z_k \otimes \hat{a}_j,
\end{align}
while such conjugation performs Fourier transform on quadrature operators in addition to the $Z_k$ factor
\begin{align}
  P^{(j)\dagger}_{Z_k} \hat{x}_j P^{(j)}_{Z_k} &= Z_k \otimes \hat{p}_j, \\
  P^{(j)\dagger}_{Z_k} \hat{p}_j P^{(j)}_{Z_k} &= -Z_k \otimes \hat{x}_j,
\end{align}
as if the phase space is rotated by $\frac{\pi}{2}$ counterclockwise. 
\end{subequations}
Finally we can see that, in direct analogy with \cref{eq:RZZ}, the $P^{(j)}_{Z_k}$ gate maps single-qubit controlled displacements to two-qubit (Ising-interaction) controlled displacements:
\begin{align}
     \label{eq:Dn_ZZ}
     D^{(n)}_{Z_j Z_k}(\alpha)
  &= e^{-2 i \alpha  Z_j Z_k \otimes \hat{p}_n} = P^{(n)^{\dagger}}_{Z_j} D^{(n)}_{Z_k}(\alpha) P^{(n)}_{Z_j}. 
\end{align}
The $P^{(n)}_{Z_j}$ gate also enables us to compile single-qubit controlled displacements (not a native primitive in Fock space ISA) with uncontrolled displacements conjugated by the $P^{(j)}_{Z_k}$ (Fock space ISA)~\cite{Liu2024}. Local qubit rotations then complete the Heisenberg-interaction gate-set (see \cref{fig:dimerCircuits} and \cref{sec:gadget} for details). As illustrated in \cref{fig:circuits}, We may then stroboscopically apply such gates to Trotter-approximate real-time evolution. Note that such a cavity-mediated-interaction compilation strategy removes the need for two-qubit entangling gates. 

It will also be useful to exchange the qumodes within the $B$ sub-lattice. To do so, we employ the native beam-splitter gate acting on neighboring oscillators $j$ and $k$
\begin{align}
    \text{BS}^{(j,k)}(\theta, \varphi) = e^{-i \frac{\theta}{2} \left[e^{i \varphi} \hat{a}^{\dagger}_j \hat{a}^{\pdag}_k + e^{-i \varphi} \hat{a}^{\dagger}_k \hat{a}^{\pdag}_j    \right]}.
    \label{eq:def-BS}
\end{align}
This couples cavity modes and defines the bosonic (canonical commutation preserving) $\text{SWAP}^{(j,k)} \coloneqq \text{BS}^{(j, k)}(\pi, 0) F^{(j, k)}$~\footnote{This SWAP definition follows Eq.~(188) of Ref.~\cite{Liu2024}. Alternatively, following Eq.~(286) of Ref.~\cite{Liu2024} we can define $\text{SWAP}^{(j,k)} \coloneqq \text{BS}^{(j, k)}(\pi, \frac{\pi}{2}) \Pi^{(k)}$ where the photon number parity gate $\Pi^{(k)} = [F^{(k)}]^2 = (-1)^{\hat{n}_k}$, which is less symmetric with respect to index $j \leftrightarrow k$.}, where the inverse Fourier transform gate (Sec.~III.B.2 in Ref.~\cite{Liu2024}) is $F^{(j, k)} = {\exp}[i\frac{\pi}{2} (\hat{n}_j + \hat{n}_k)] = F^{(j)} F^{(k)}$. Explicitly, 
\begin{align}
  &\mathrel{\phantom{=}}
     \text{SWAP}^{(j,k)\dagger} \hat{a}_j \text{SWAP}^{(j,k)}  \notag \\
  &= F^{(j, k)\dagger} \text{BS}^{(j, k)\dagger}(\pi, 0) \hat{a}_j \text{BS}^{(j, k)}(\pi, 0) F^{(j, k)} \notag \\
  &= F^{(j, k)\dagger} (-i\hat{a}_k) F^{(j, k)} \notag \\
  &= F^{(k)\dagger} F^{(j)\dagger} (-i\hat{a}_k) F^{(j)} F^{(k)} \notag \\
  &= F^{(k)\dagger} (-i\hat{a}_k) F^{(k)} = \hat{a}_k. \label{eq:SWAP}
\end{align}

We now see that $\text{SWAP}^{(k,j)}$ enables us to convert a non-native CV-DV gate $D_{Z_k}^{(j)}(\alpha)$ for $j\neq k$ in \cref{tab:gates} into a sequence of native gates as
\begin{align}
     \text{SWAP}^{(j,k)\dagger} D_{Z_k}^{(j)}(\alpha) \text{SWAP}^{(j,k)}
  &= D_{Z_k}^{(k)}(\alpha).
  \label{eq:native_DcZ}
\end{align}
In \cref{sec:M=1} the SWAP gate is used to couple a single qumode to the entire spin lattice. Afterwards, using $M$ qumodes in \cref{sec:M=N}, we will parallelize the QITE filter.

\begin{figure}
  \centering
  \includegraphics{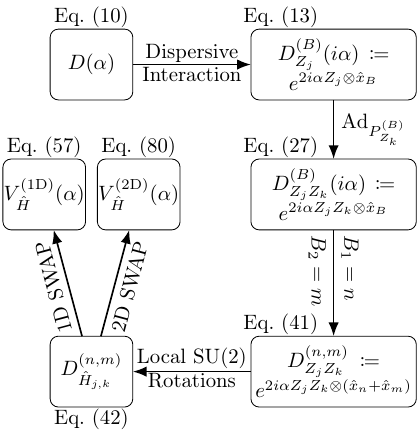}
\caption{Compilation workflow where each object represents a relevant quantum operation. The arrows transform the operations, corresponding to the labeled equations, in the Heisenberg picture.}
  \label{fig:compilation}
\end{figure}

\section{Heisenberg Model Compilation}
\label{sec:app-heisenberg-model}

\subsection{Synthesizing Time-Evolution}
\label{sec:Heisenberg_Complilation}

With \cref{sec:ISA} describing the required gate set, we proceed to constructively synthesize $\varepsilon$-approximations to the spectral filter $\hat{P}_\tau$. One possible explicit construction of \cref{eq:time-position-displacement} could be based on Lie-Trotter product formulas (LTPFs). Such LTPFs, based on convergence with respect to the Euler-Trotter identity $\lim_{r\rightarrow\infty} (e^{A/r}e^{B/r})^r = e^{A+B}$, are useful to approximate exponential functions which are difficult to analytically express due to the exponential's non-commuting generators~\cite{Childs2021,kang2023leveraging}. Our goal is to now systematically $\varepsilon$-approximate \cref{eq:filter_GSE_shift}'s ideal eigenvalue transformation oracle with a block-encoded LTPF $\hat{R}_{\hat{H}'}(\tau)$. 

\begin{thm}[Trotter-Hubbard-Stratonovich Formula] \label{thm:LTHS}
Using $U_{\hat{H}'}^{(p)}(\tau)$ as in \cref{eq:time-position-displacement}, and $V_{\hat{H}'}^{(p)}(\tau)$ as an $\varepsilon$-approximate LTPF suffices to $\varepsilon$-approximate $\hat{P}_{\hat{H}'}(\tau)$ in \cref{eq:filter_GSE_shift} with $\hat{R}_{\hat{H}'}(\tau)$ as
\begin{align}
\label{eq:LTPF_eps_bound}
    \norm{\hat{R}_{\hat{H}'}(\tau) - \hat{P}_{\hat{H}'}(\tau)} = \norm{\bra{0} V_{\hat{H}'}^{(p)}(\tau)  - U_{\hat{H}'}^{(p)}(\tau) \ket{0} } \leq \varepsilon.
\end{align}
\end{thm}
To begin the proof, consider arbitrary 2-local Hamiltonian interaction graph with partitions $\Gamma = \chi'(G)$, where $\chi'(G)$ is the chromatic index of the simple graph $G = (V, E)$ with the vertices $V$ being the lattice sites and the edges $E$ the spin-spin (magnetic exchange)  couplings. Using $\chi'$ partitions is necessary to ensure that for any $\gamma$, all weight-2 Pauli strings in $\hat{H}_\gamma$ mutually commute. Since $\Delta(G) \leq \chi'(G) \leq \Delta(G) + 1$ (Vizing's theorem), where $\Delta(G)$ is the maximum degree of $G$ (largest coordination numbers for all lattice sites), $\Gamma = \chi'(G) = \order{d}$ for physics relevant lattices in $d$ dimensions, instead of scaling with system size. 

Writing the mutually commuting Hamiltonian partitions as $\hat{H} = \sum_{\gamma=1}^{\Gamma} \hat{h}_\gamma$, we use  $U^{(\hat{p})}_{\hat{H}}(\alpha) = \lim_{r \rightarrow \infty} [\prod_\gamma D^{(j)}_{\hat{h}_\gamma}(\alpha/r)]^r \equiv \lim_{r \rightarrow \infty} V^{\hat{p}}_{\hat{H}}$ as the first-order limiting formula before taking $r$ finite. Recall the vacuum wavefunction in momentum basis is $\psi_0(p) = \ip{p}{0} = (2/\pi)^{1/4} e^{-p^2}$, we can then bound the Trotter error as  
\begin{align}
     \varepsilon 
  &= \norm{ \ev{ \prod_{\gamma=1}^\Gamma e^{-2i\alpha \hat{p} \hat{h}_\gamma} }{0}
           -\ev{ e^{-2i\alpha \hat{p} \sum_{\gamma=1}^\Gamma \hat{h}_\gamma} }{0} } \\
  &= \norm{ \int_{-\infty}^{+\infty} dp \abs{\psi_0(p)}^2 
     \qty(\prod_{\gamma=1}^\Gamma e^{-2i\alpha p \hat{h}_\gamma} - e^{-2i\alpha p \sum_{\gamma=1}^\Gamma \hat{h}_\gamma}) } \nonumber \\
  &\leq \sqrt{\frac{2}{\pi}} \int_{-\infty}^{+\infty} dp\, e^{-2p^2} \norm{
     \prod_{\gamma=1}^\Gamma e^{-2i\alpha p \hat{h}_\gamma} - e^{-2i\alpha p \sum_{\gamma=1}^\Gamma \hat{h}_\gamma} } \nonumber \\  
  &\leq \sqrt{\frac{2}{\pi}} \int_{-\infty}^{+\infty} dp\, e^{-2p^2}
     \frac{(2\alpha p)^2}{2} \norm{\sum_{\gamma=1}^{\Gamma-1} \sum_{\delta=\gamma+1}^{\Gamma} [\hat{h}_\delta, \hat{h}_\gamma]} 
    \label{eq:prop9}\\
  &= \frac{\alpha^2}{2} \norm{\sum_{\gamma=1}^{\Gamma-1} \sum_{\delta=\gamma+1}^{\Gamma} [\hat{h}_\delta, \hat{h}_\gamma]}.
\end{align}
where \cref{eq:prop9} used Proposition 9 from Ref.~\cite{Childs2021}.

We will now discuss the circuitry required to conditionally displace an oscillator mode by the energy eigenvalues of a single Hamiltonian interaction between a pair of sites. Note that this Hamiltonian, and commutators between such Hamiltonians, preserve \cref{eq:XXZ}'s symmetries. We will then provide minimal $\Gamma=2$ and $\Gamma=4$ examples in 1D and 2D respectively.  To go beyond \cref{thm:LTHS}'s analytic result, we also examine a non-color-ordered Trotter sequence which we then numerically simulate. 

\subsection{2-Local Gadget}
\label{sec:gadget}
Consider the Heisenberg model Hamiltonian
\begin{align}
     \hat{H}
  & = \sum_{\langle j,k \rangle} \hat{H}_{j,k} 
  \equiv \sum_{\langle j,k\rangle} J_x X_{j} X_{k} + J_y Y_{j} Y_{k} + J_z Z_{j} Z_{k}, \label{eq:XXZ}
\end{align}
where $\langle j,k\rangle$ indicates nearest-neighbor couplings on a lattice (i.e., a grid graph; for general graph, $\langle j,k\rangle$ denotes the set of edges). Without loss of generality, we target the XXZ spin model Hamiltonian by choosing $J_x = J_y = 1$ and $J_z = \Delta J_x = \Delta$. 

Using the native gates in Table~\ref{tab:gates} and compilation primitives in Section~\ref{sec:ISA}, we compile a displacement gate on qumode $n$, controlled by the energy of a single bond $\hat{H}_{j,k} = X_j X_k + Y_j Y_k + \Delta Z_j Z_k$, as
\begin{align}
\label{eq:Heisenberg_M=1}
     D^{(n)}_{\hat{H}_{j,k}}(\alpha)
   = D^{(n)}_{X_j X_k}(\alpha) D^{(n)}_{Y_j Y_k}(\alpha) D^{(n)}_{Z_j Z_k}(\Delta \alpha).
\end{align}
This equality leverages the $\mathfrak{su}(4)$ Cartan sub-algebra's ($\{X_j X_k, Y_j Y_k, Z_j Z_k \}$) pairwise commutativity. The first gate required to synthesize \cref{eq:Heisenberg_M=1} is the conditional displacement controlled by the joint parity $Z_j Z_k$ given in \cref{eq:Dn_ZZ}. Using transversal local qubit rotations $V=e^{i \frac{\pi}{4} Y_j} e^{i \frac{\pi}{4} Y_k}$ and $W = e^{-i \frac{\pi}{4} X_j} e^{-i \frac{\pi}{4} X_k}$, we compile displacements weighted by the Heisenberg model's off-diagonal interactions  
\begin{align}
     \label{eq:Dn_XX}
     D^{(n)}_{X_j X_k}(\alpha) 
  &= V^\dagger D^{(n)}_{Z_j Z_k} (\alpha) V,  \\
     \label{eq:Dn_YY}
     D^{(n)}_{Y_j Y_k}(\alpha)
  &= W^\dagger D^{(n)}_{Z_j Z_k} (\alpha) W.
\end{align}
Together with \cref{eq:Dn_ZZ}, these gates enable the implementation of one round of \cref{eq:Heisenberg_M=1}'s Heisenberg real-time-evolution operator in a three-stroke clock cycle which will be employed as a building block to synthesize controlled approximations to \cref{eq:XXZ}'s global energy filter.

\begin{figure}
  \centering
  \includegraphics[width=1.0\linewidth]{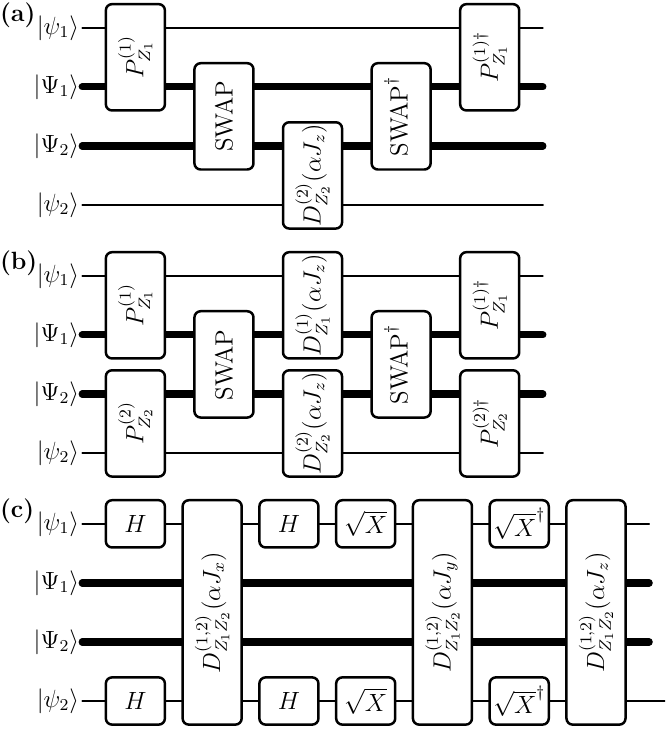}
\caption{Compiled circuits for (a) $D_{Z_1 Z_2}^{(1)}$, (b) $D_{Z_1 Z_2}^{(1,2)}$, and (c) $D_{\hat{H}_{1,2}}^{(1,2)} = D_{Z_1 Z_2}^{(1,2)} D_{Y_1 Y_2}^{(1,2)} D_{X_1 X_2}^{(1,2)}$, where $\hat{H}_{1,2}$ is defined in \cref{eq:Heisenberg_M=1}. Thicker wires denote two oscillators (need not be adjacent) in the respective quantum state $\ket{\Psi_{1}}$ and $\ket{\Psi_{2}}$; $\ket{\psi_{1}}$ and $\ket{\psi_{2}}$ denote the respective states of two qubits linked to their corresponding oscillator.}
  \label{fig:dimerCircuits}
\end{figure}

\subsection{Parallelizing Compilation in 1D}
\label{sec:1D}
Our Hamiltonian is a sum of two-local interactions 
\begin{align}
\label{eq:local_terms}
     \hat{H}
  &= \sum_{j=0}^{N'} \hat{H}_{j, j+1}.
\end{align}
For open boundary conditions (OBC), $N' = N - 2$, while for periodic boundary conditions (PBC), $N' = N - 1$ with $N'+1 \equiv 0$ (i.e., indices are modulo $N$). Now consider an additive factorization of \cref{eq:local_terms} into the two terms corresponding interactions on even ($\hat{A}$) and odd ($\hat{B}$) edges, $\hat{H}=\hat{A}+\hat{B}$. Without loss of generality we take $N$ even, such that
\begin{subequations}
\label{eq:AB_tiling}
\begin{align}
     \hat{A}
  &= \sum_{j=0}^{\frac{N}{2}-1} \hat{H}_{2j, 2j+1}
   \equiv \sum_j \hat{A}_j, \\ 
     \hat{B}
  &= \sum_{j=0}^{\frac{N}{2}-1} \hat{H}_{2j+1, 2j+2}
   \equiv \sum_j \hat{B}_j.
\end{align}
\end{subequations}
The $2$-coordinated nearest-neighbor Hamiltonian interaction can be additively factorized, in that each equation in \cref{eq:AB_tiling} is a sum over terms with \textit{disjoint} support which was not the case in \cref{eq:local_terms}. This proves that $\comm*{\hat{A}_j}{\hat{A}_k} = \comm*{\hat{B}_j}{\hat{B}_k} = 0$, a fact that will be used to parallelize a controlled operation over the global operator $\hat{A}$ or $\hat{B}$. Recalling that $\hat{H}_{j, j+1} = X_j X_{j+1} + Y_j Y_{j+1} + \Delta Z_j Z_{j+1}$ was compiled in \cref{sec:gadget}, our task is now to (i) explain how a single oscillator may play the role of the block encoding ancilla qumode and (ii) explain how this sequence can be embarrassingly parallelized using a lattice of qumodes. We now examine (i), before outlining (ii) in 1 and 2 dimensions.

\subsubsection{One-Oscillator Compilation}
\label{sec:M=1}

Before proceeding to the fully parallelized compilation, let us first compile dynamics of \cref{eq:XXZ} controlled on a single ($M=1$) qumode. Following the color ordering presented above, we displace the oscillator by $\hat{A}$ and then subsequently by $\hat{B}$. This is achieved by serially displacing by each bond in $\hat{A}$. To do so, a pair of SWAPs will route the bosonic mode around the spin chain, displacing by each term in $\hat{A}$ along the way. Repeating for each of the terms in $\hat{B}$ then realizes a first order LTPF $e^{\hat{B}} e^{\hat{A}}$ of \cref{eq:XXZ}. Using $N$ SWAPs to evolve by each colored partition, the resulting Trotter formula $V_{\hat{H}}(\alpha) = e^{-i 2\alpha \hat{p}  \hat{B}} e^{-i 2\alpha \hat{p} \hat{A}}$ is related to our target dynamics through $V_{\hat{H}}(\alpha) = U^{(p)}_{\hat{H}}(\alpha) + \mathcal{A}(\alpha)$ where the additive error is bounded in \cref{thm:LTHS}. 

\subsubsection{Two-Oscillator Compilation}

The next step in our progression is to parallelize the compilation sequences of adjacent oscillators. We will return to formally generalize this result in \cref{sec:SWAPs}. The high-level idea is to displace the second oscillator, indexed as $k$, conditioned on the commuting disjoint components of $\hat{A}$ (and subsequently $\hat{B}$). Crucially, this occurs in parallel with the displacements of the first oscillator.

Note that an Ising interaction can, since $[D^{(j)}_{Z_j Z_k}(\alpha), D^{(k)}_{Z_j Z_k}(\alpha)] =0$, control the displacements the involved oscillators in any order. Since the Ising generators commute, both adjacent oscillators can then be simultaneously displaced using the $P^{(j)}_{Z_k}$ and $P^{(k)}_{Z_j}$ gates as illustrated in \cref{fig:compilation} which results in
\begin{align}
     D^{(j,k)}_{Z_j Z_k}(\alpha)
  &= D^{(j)}_{Z_j Z_k}(\alpha) \times D^{(k)}_{Z_j Z_k}(\alpha) \notag \\
  &= {e^{-2 i \alpha  Z_j Z_k \otimes (\hat{p}_j + \hat{p}_k)}}. \label{eq:Djk_ZZ} 
\end{align}

The same is true when compiling the two-qumode displacement conditioned on a single bond's Heisenberg-interaction spectra as 
\begin{align}
     D^{(n,m)}_{\hat{H}_{jk}}(\alpha)
  &= D^{(n,m)}_{X_j X_k}(\alpha) \times  D^{(n,m)}_{Y_j Y_k}(\alpha) \times D^{(n,m)}_{Z_j Z_k} (\Delta \alpha) \notag \\
  &= e^{-2 i \alpha \left( X_j X_k + Y_j Y_k + \Delta Z_j Z_k \right) \otimes (\hat{p}_n + \hat{p}_m)} \notag \\
  &= e^{-2 i \alpha \hat{H}_{jk} \otimes \left( \hat{p}_n + \hat{p}_m \right)}.
  \label{eq:Djk_jk}
\end{align}

In addition to a color-ordered-decomposition which is helpful in simplifying the proof of \cref{thm:LTHS}, to minimize entangling gate-count pre-factors it will also be useful to consider controlled displacements made up of terms such as $D^{(j+1)}_{\hat{B}_{j}}(\alpha) D^{(j)}_{\hat{A}_{j}}(\alpha)$. Here, conditioned on different colored interactions within the same time step, adjacent oscillators are displaced. In this case, for each oscillator, $N$ SWAPS are used in order to conditionally displace each of the modes on the individual terms comprising $\hat{A}$ and $\hat{B}$ in an alternating series. That is, each oscillator sweeps around the spin chain and is displaced by each bond in serial. This is in contrast to the color-ordering above where the number of SWAPS acting on each mode is double the length of the chain. To get a picture for how the two Trotter orderings compare, see \cref{fig:visualCircuits} and \cref{fig:6siteCircuits} where panels (a) and (b) correspond to the color-ordered and $\hat{A}$-$\hat{B}$ alternating ordered Trotterizations respectively. 

Importantly, we note that the two-oscillator compilation requiring $N$ SWAPS, when implemented on the architecture in Fig.~\ref{fig:hardware-mapping}, requires the same total SWAP-depth as the single-oscillator compilation (as the oscillator must ``visit'' every spin in order to be displaced by all terms comprising $\hat{A}$ and $\hat{B}$). Thus, there is no overhead in depth for such parallelization. In \cref{sec:M=N} we will concretely specify how, with the same overall SWAP-depth, we can realize an $M$-oscillator compilation that enables a $\sqrt{M}$-reduction in each  conditional displacement length over the single-oscillator implementation, thus reducing the overall runtime of the algorithm through parallelization.

\subsection{Multi-Oscillator Compilation}
\label{sec:M=N}
In \cref{fig:high-level} the number of conditional displacements (and by extension, the protocol's total duration) scales extensively with the principal system's size. To remedy this, we can reduce the overall time by parallelizing the total displacement of a single oscillator into the product of many shorter displacements of many oscillators. 

For the hardware where each qubit is directly controlled by an associated oscillator, i.e., with $M=N$ oscillators, we fully parallelize the energy filter over the many oscillator modes. To do so we must synthesize a $\varepsilon$-approximation where \textit{all} oscillators are displaced by $\hat{H}$. Since our filter is a quadratic function of $\hat{H}' \alpha$, $\hat{P}_{\hat{H}'}(\alpha)^{\frac{1}{M}} = \hat{P}_{\hat{H}'}(\alpha/\sqrt{M})$. Thus, assigning the re-scaled time $\alpha/\sqrt{M}$ to each of the $M$ distinct spatial oscillators, one may parallelize $\tau = \alpha$ in the imaginary time evolution operator in \cref{thm:LTHS} (\nameref*{thm:LTHS}). 

Since the native gate time depends on the magnitude of the displacement, displacing individual oscillators by $\alpha / \sqrt{M}$ reduces the physical execution time of our algorithm while maintaining the same scaling of Trotter error as in the single oscillator case. Note that the first-order Trotter error per oscillator becomes $\order*{N(\alpha/\sqrt{M})^2}$, but the total Trotter error of $M$-factors scales as $M\cdot \order*{N(\alpha/\sqrt{M})^2} = \order*{N\alpha^2}$, the same as the single oscillator case. As a result, we now see how parallelizing over hardware-native oscillators reduces the absolute algorithmic runtime by a factor $\sqrt{M}$. Generalizing a single oscillator variable $\hat{p}_0$ block-encoding, to an $M$ oscillator $\{\hat p_{j}\}_{j=0}^{M-1}$ block encoding, we have:
\begin{lem}[Spacetime Parallelization] \label{thm:M-oscillator}
\begin{align}
\label{eq:1D_displaced_trotter}
  \mathrel{\phantom{=}}  \hat{P}_{\hat{H}'}(\alpha) \notag
  &= \ev{U_{\hat{H}'}^{(p_0)} (\alpha)}{0} \\
  &= \ev{\prod_{j=0}^{M-1} U_{\hat{H}'}^{(p_j)} \qty(\alpha/\sqrt{M})}{\bm{0}}  \\
  &= \bigotimes_{j=0}^{M-1} \ev{U_{\hat{H}'}^{(p_j)} \qty(\alpha/\sqrt{M})}{0_j} 
\end{align}    
\end{lem}
In the top line, one oscillator and $\sqrt{M}$ serial time resources are utilized. The second line uses $M$ oscillators as spatial resources to control spin dynamics and requires an $M$-fold tensor product measurement to generalize \cref{eq:LTPF_eps_bound} to the multi-oscillator limit. Upon measuring the vacuum configuration $\ket{{\bf 0}} = |0\rangle^{\otimes M}$, after the application of $\prod_{j=0}^{M-1} U_{\hat{H}'}^{(p_j)} \qty(\alpha/\sqrt{M})$, we have \cref{eq:filter_general_shift}. 

The last step, to complete end-to-end compilation~\cite{endtoend}, is the synthesis of a $\varepsilon$-approximate of \cref{thm:M-oscillator}. While block-encoded $\varepsilon$-approximations of unitaries offer scaling advantages, keeping near-term applications in mind, we synthesize a Trotter approximation. 
\begin{thm}[Parallelized LTHS Formula] \label{thm:parallel}
 \mycomment{
Using $U_{\hat{H}'}^{(j)}(\tau)$ as in \cref{eq:time-position-displacement}, and $V_{\hat{H}'}^{(j)}(\tau) = e^{-i 2\tau p_j  \hat{B}} e^{-i 2\tau p_j \hat{A}}$ as its LTPF, we $\varepsilon$-approximate \cref{eq:filter_GSE_shift} as 
\begin{align}
\label{eq:1DProj}
    &\mathrel{\phantom{=}}  \norm{\bra{{\bf 0}} \prod_{j=0}^{M-1} V_{\hat{H}'}^{(j)}\left(\alpha/\sqrt{M} \right) -  \prod_{j=0}^{M-1} U_{\hat{H}'}^{(j)} \qty(\alpha/\sqrt{M}) \ket{{\bf 0}} } \\
    &= \norm{\hat{R}^{\mathrm{(1D)}}_{\hat{H}'}(\tau) - \hat{P}_{\hat{H}'}(\tau)} \leq \varepsilon 
\end{align}
 }
Consider a general 2-local spin (spectrum-shifted) Hamiltonian $\hat{H} = \sum_{\gamma=1}^{\Gamma} \hat{H}_\gamma$ partitioned into $\Gamma$ parts $\{\hat{H}_\gamma \mid \hat{H}_1 = \hat{A},  \hat{H}_2 = \hat{B}, \hat{H}_3 = \hat{C}, \dots \}$ according to a proper edge coloring of the interaction graph of $\hat{H}$ so that all weight-2 Pauli strings mutually commute in each $\hat{H}_\gamma$. There exists a coloring so that $\Gamma - 1$ is less than or equal to the largest coordination number of any site in the system (Vizing's theorem), and thus the number of partitions is bounded by the system size. With controlled displacement $U_{\hat{H}}^{(p_j)}(\tau = \alpha/\sqrt{M})$ as in \cref{eq:time-position-displacement} for $M$ oscillators respectively in parallel, an exact ground state projection (GSP) operator is given by the vacuum transition probability (VTA)

\begin{align}
\label{eq:para_P}
     \hat{P}_{\hat{H}}(\alpha)
  &= \ev{\prod_{j=0}^{M-1} U_{\hat{H}}^{(p_j)} \qty(\frac{\alpha}{\sqrt{M}})}{\bf 0}
   = e^{-\alpha^2 \hat{H}^2/2}.
\end{align}
The LTHS synthesized approximation is given by
\begin{subequations}
\label{eq:para_R}
\begin{align}
     \hat{R}_{\hat{H}}(\alpha)
  &= \ev{\prod_{\gamma=1}^{\Gamma} \qty[\prod_{j=0}^{M-1} U_{\hat{H}_\gamma}^{(p_j)} \qty(\frac{\alpha}{\sqrt{M}})]}{\bf 0} \label{eq:para_R_a}\\
  &= \operatorname{\mathcal{T}_\gamma} e^{-\alpha^2 \qty(\sum_\gamma \hat{H}_\gamma)^2/2}, \label{eq:para_R_b}
\end{align}
\end{subequations}
where the $\gamma$ (color) ordering  $\mathcal{T}_\gamma$ is analogous to Feynman's time order operator.The Lie-Trotter series' commutator-scaling additive error is
\begin{align}
\label{eq:para_trotter_err}
     \varepsilon
  &= \norm{\hat{R}_{\hat{H}}(\alpha) - \hat{P}_{\hat{H}}(\alpha)}
   \leq \frac{\alpha^2}{2} \sum_{\gamma_1 < \gamma_2 } \norm{\comm{\hat{H}_{\gamma_1}}{\hat{H}_{\gamma_2}}}.
\end{align}
If $\hat{H}$ is a Heisenberg spin model, \cref{eq:para_R} can be simplified to the following form:
\begin{align}
\label{eq:para_R_Heis}
     \hat{R}_{\hat{H}}(\alpha)
  &= \sum_{\{b_{i,j}\}\in \{0,1\}^{|E|}} 
     e^{-\alpha^2 \qty[E_\text{\rm s} + \sum_{(i,j) \in E} (4b_{i,j} - 1)]^2 / 2} \notag \\
  &\mathrel{\phantom{=}} \times 
  \operatorname{\mathcal{T}_\gamma}\prod_{(i,j)\in E} \frac{(3-2b_{i,j}) + (1-2b_{i,j})\hat{H}_{i,j}}{4},
\end{align}
where the color ordering is trivial for the product of triplet (singlet) projectors $\frac{3+\hat{H}_{i,j}}{4}$ for $b_{i,j} = 0$ 
 $(\frac{1-\hat{H}_{i,j}}{4}$ for $b_{i,j} = 1)$ over the edge set $E = \{(i,j) \mid \hat{H}_{i,j} = X_i X_j + Y_i Y_j + Z_i Z_j \neq 0 \}$ with Heisenberg interactions.
\end{thm}

\begin{figure}
  \centering
  \includegraphics[width=1.0\linewidth]{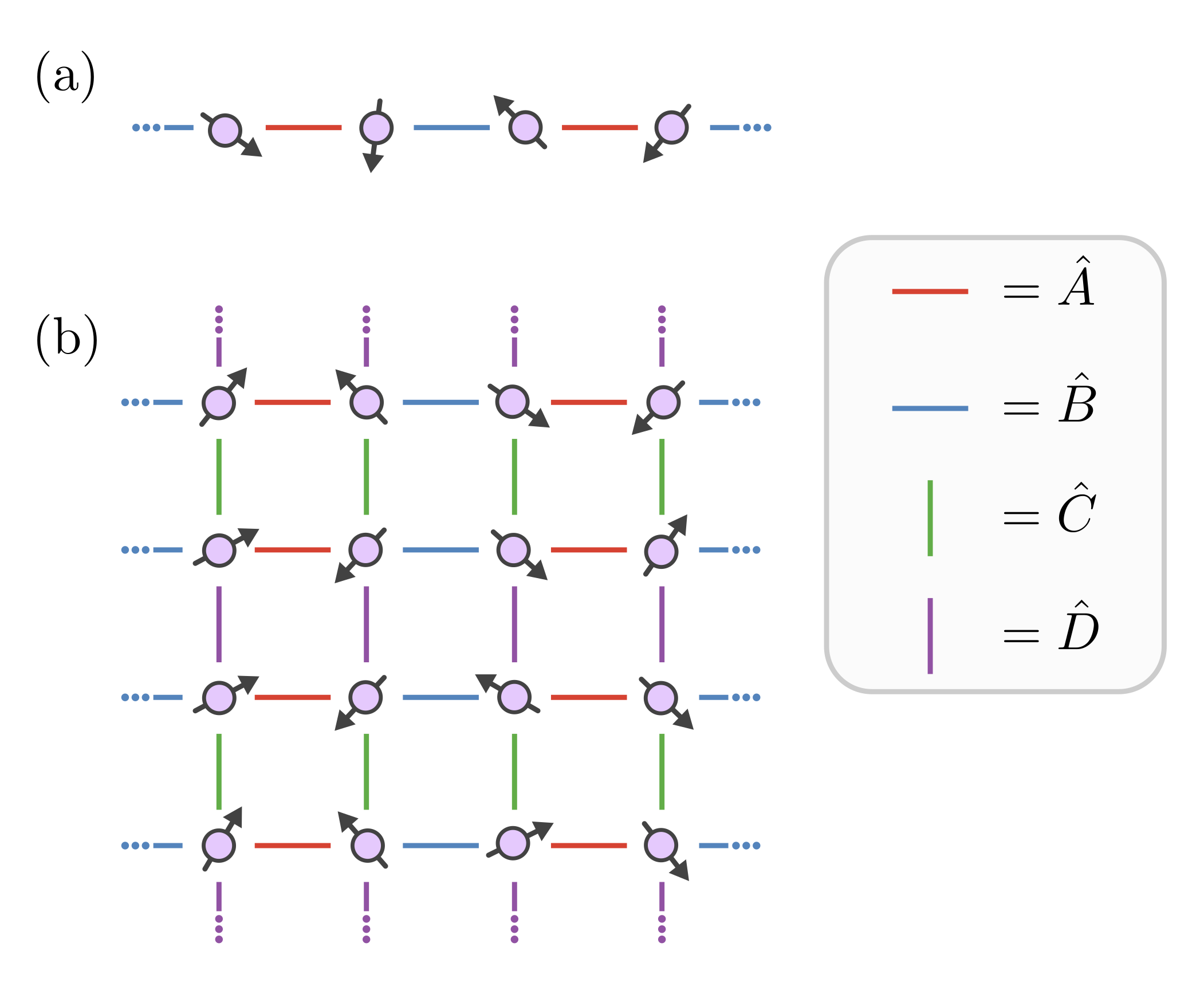}
\caption{Trotter orderings for 1D (a) and 2D (b) spin models. Each color denotes spin-spin interaction terms for which conditional oscillator displacements can be parallelized. (a) In 1D, this results in alternating sequence of displacements corresponding to red ($\hat{A}$) and blue ($\hat{B}$) links. In 2D, interaction terms are grouped by red ($\hat{A}$), blue ($\hat{B}$), green ($\hat{C}$), and purple ($\hat{D}$) links.}
  \label{fig:trotter_tilings}
\end{figure}

The proof of \cref{eq:para_R,eq:para_trotter_err,eq:para_R_Heis} is given in \cref{sec:proof-thm:parallel}. The analytical expression of $\hat{R}_{\hat{H}}(\alpha)$ in \cref{eq:para_R_Heis} for the 1D chain colored as in \cref{fig:trotter_tilings}(a) is given in \cref{sec:AnalyticalResults}.

In this scheme, the circuit depth scales extensively with the system size, similar to a single oscillator, and $\hat{R}_{\hat{H}}(\alpha)$ is compiled by parallelized gate sequences which are copies of the single oscillator circuitry. However, assuming that the gate duration for $D_{\hat{H}}^{(j)}(\tau)$ on real hardware is proportional to $\tau$~\cite{Liu2024}, the time complexity is reduced by a factor $\sqrt{M}$ from the single oscillator case. 

Importantly, with the connectivity of the architecture Fig.~\ref{fig:hardware-mapping}, realization of this reduced time-complexity in practice requires that the multi-oscillator compilation incurs no additional oscillator SWAP-depth over its single-oscillator counterpart; in the next Section, we show this to be the case. In particular, we describe a concrete implementation of a multi-oscillator SWAP network requiring total SWAP depth of $N$ -- equivalent to the single-oscillator case. Crucially, the time duration of a single oscillator SWAP is roughly an order of magnitude shorter than that of the doubly conditioned conditional displacements ($\sim 100$ ns vs.\ $\sim 2$ $\mu$s \cite{Liu2024}). Thus, while the multi-oscillator parallelization reduces only the displacement duration and not the SWAP-depth, this enables an approximate $\sqrt{M}$-reduction in total runtime in practice.

\begin{figure}
  \centering
  \includegraphics[width=1.0\linewidth]{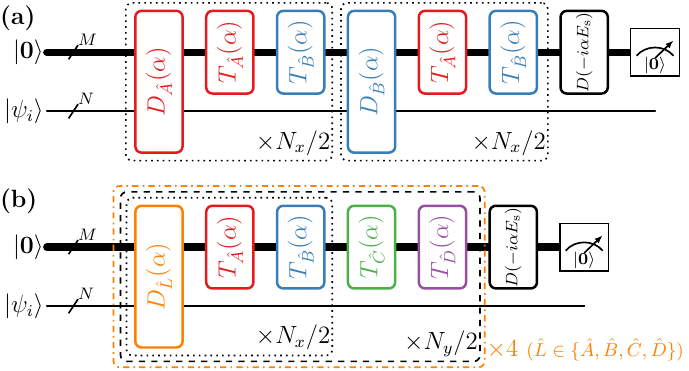}
\caption{Quantum realization of the color-ordered Trotter-approximated Hamiltonian displacement in (a) 1D \cref{eq:V^1D} and (b) 2D \cref{eq:V2D}. The initial state of the CV-DV system is $\ket{\bf 0} \otimes \ket{\psi_i}$ with $M$ qumodes in the all-vacuum state $\ket{\bf 0} = \ket{0}^{\otimes M} = \ket{0\dots 0}$ and $N$ qubits in the state $\ket{\psi_i}$. We assume $M = N$. The slash symbol denotes the bundle of quantum wires (thicker bundled wires for CV system). In (b) $\hat{L}$ is substituted with $\hat{A}$, $\hat{B}$, $\hat{C}$, or $\hat{D}$ for each of the four repetitions of the circuit part enclosed in the dash-dotted orange box.}
  \label{fig:circuits}
\end{figure}

\subsection{1D SWAP Networks}
\label{sec:SWAPs}
We now describe the role that SWAP networks (SNs) for qumodes consisting of beam splitters play in parallelized multi-oscillator compilation. Namely, the SNs spatially cycle all oscillators while synchronizing their controlled-time-evolution gates. As illustrated in \cref{fig:high-level} by the two boxes with arrows, each oscillator is displaced conditioned on a two-body interaction localized at an edge connecting two qubits on a graph, and this primitive operation is implemented by the circuit in \cref{fig:dimerCircuits}(c) for \cref{eq:AB_tiling}. 

Now, exploiting \cref{thm:M-oscillator}'s freedom and our hardware's ability to parallelize displacements of nonadjacent oscillators, we use \cref{eq:Djk_jk} and \cref{eq:AB_tiling} to define two edge-colored displacement operators
\begin{subequations}
\label{eq:color_product_dislacement}
\begin{align}
     D_{\hat{A}}(\alpha) 
     &= \prod_{j=0}^{\frac{N}{2} -1} D^{(2j,2j+1)}_{A_j} (\alpha) \\
     D_{\hat{B}}(\alpha)
     &= \prod_{j=0}^{\frac{N}{2} -1} D^{(2j+1,2j+2)}_{B_j} (\alpha).
\end{align}
\end{subequations}
This pair of operators displaces pairs of oscillators by single pieces of the $\hat{A}$ or $\hat{B}$ Hamiltonian partition. Therefore, oscillators at even (odd) cavities must be routed to even (odd) cavities if they are to be displaced by the next term of the same colored interaction at the next step. Define the transversal ($T$) bosonic SWAP operators as
\begin{subequations}
\begin{align}
  \perm_{\hat{A}} &= \bigotimes_{j=0}^{ \frac{N}{2} -1} \text{SWAP}^{(2j,2j+1)}, \label{eq:SWAP_A}\\
  \perm_{\hat{B}} &= \bigotimes_{j=0}^{ \frac{N}{2} -1}\text{SWAP}^{(2j+1,2j+2)}. \label{eq:SWAP_B}
\end{align}
\end{subequations}
Acting along the bonds illustrated in \cref{fig:trotter_tilings}, these gates permute bosonic modes on all bonds of a given color. The global effect is to exchange all modes along the edges of one color, equivalent to a total translation of one lattice constant for $A$-sublattice and $B$-sublattice, respectively, in the opposite direction. Since $\perm_{\hat{A}}^2 = \perm_{\hat{B}}^2 = \mathds{1}$, we instead define the back-to-back execution of \cref{eq:SWAP_A,eq:SWAP_B} as the composite permutation 
\begin{align}
    \perm_{\hat{A} \hat{B}} = \perm_{\hat{B}} \perm_{\hat{A}}.
\end{align}
$\perm_{\hat{A} \hat{B}}$ translates the oscillators by two lattice sites for $A$/$B$-sublattices in the opposite directions as
\begin{align}
     \perm_{\hat{A} \hat{B}}^{}(\hat{a}_j) 
  &\equiv \perm_{\hat{A} \hat{B}}^\dagger \hat{a}_j \perm_{\hat{A} \hat{B}}^{\pdag}
   =
   \begin{cases}
     \hat{a}_{j+2} & \text{if } j \text{ is even}, \\
     \hat{a}_{j-2} & \text{if } j \text{ is odd}.
   \end{cases}
\end{align}
such that $p$ iterations of $\perm_{\hat{A} \hat{B}}$ shifts by $2p$ sites.
\begin{align}
    \perm_{\hat{A} \hat{B}}^p (\hat{a}_j) =
\begin{cases}
\hat{a}_{j+2p} & \text{if } j \text{ is even}, \\
\hat{a}_{j-2p} & \text{if } j \text{ is odd}.
\end{cases}
\end{align}
Assuming even $N$, we see that $\perm_{\hat{A} \hat{B}}^{\frac{N}{2}} = \mathds{1}$, which implies that $\frac{N}{2}$ applications return each mode to its original cavity
\begin{align}
    \perm_{\hat{A} \hat{B}}^{\frac{N}{2}}(\hat{a}_j)  = \hat{a}_j.
\end{align}

Naturally, these relations lead us to assign a group structure for these permutations. We define an element of the symmetric group $\tau \in S_N$ that, after $p$ iterations of $\perm_{\hat{A} \hat{B}}$, maps oscillator indices to cavity indices as 
\begin{align}
\tau^p(j) =
    \begin{cases}
        j + 2p \pmod{N} & \text{if } j \text{ is even}, \\
        j - 2p \pmod{N} & \text{if } j \text{ is odd}. 
    \end{cases}
\end{align}

By interspersing the translation operators with \cref{eq:color_product_dislacement} we are able to use \textit{each} oscillator to control the time evolution of each color partition. Using this, the two steps comprising the 1D Trotter-factorized approximation are:
\begin{subequations}
  \label{eq:global_color_displacement}
\begin{align}
    \phantom{={}}& \left(\perm_{\hat{A} \hat{B}} D_{\hat{B}}(\alpha)\right)^{\frac{N}{2}} \notag \\
    ={}& \prod_{p=0}^{\frac{N}{2} -1 } \exp \left\{-2 i \alpha  \sum_{j}^{N} B_{j} \otimes \qty( \hat{p}_{\tau^p(j)} +\hat{p}_{\tau^p(j+1)} ) \right\}, \\
    \phantom{={}}& \left(\perm_{\hat{A} \hat{B}} D_{\hat{A}}(\alpha)\right)^{\frac{N}{2}} \notag \\
    ={}& \prod_{q=0}^{\frac{N}{2} -1 } \exp \left\{-2 i \alpha  \sum_{j}^{N} A_{j} \otimes \qty( \hat{p}_{\tau^q(j)} +  \hat{p}_{\tau^q(j+1)} ) \right\}.
\end{align}
\end{subequations}
Together, these two operators are composed to construct the Trotterized global Hamiltonian displacement. This provides an end-to-end compilation of the expressions in \cref{thm:LTHS} and \cref{thm:parallel}. In \cref{fig:circuits} we present a quantum circuit that realizes 
\begin{equation}
\label{eq:V^1D}
\left(\perm_{\hat{A} \hat{B}} D_{\hat{B}}(\alpha)\right)^{\frac{N}{2}}\times \left(\perm_{\hat{A} \hat{B}} D_{\hat{A}}(\alpha)\right)^{\frac{N}{2}} = V_{\hat{H}'} (\alpha).
\end{equation}

\subsection{Numerical Results}
\label{sec:Numerics}
\begin{figure}
  \centering
  \includegraphics{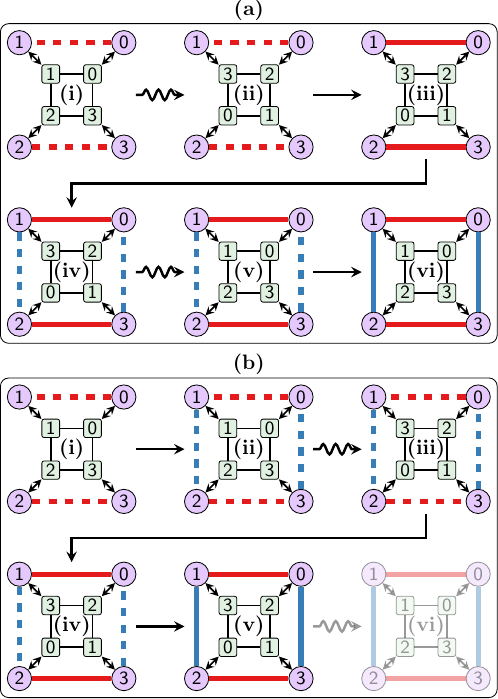}
\caption{Visual representations of the compilations for $N=4$ site Heisenberg model. Purple circular nodes are qubits and light green square nodes are qumodes. The panel (a) strictly follows the sequence of gate operations illustrated in \cref{fig:circuits}(a), while the panel (b) gives an alternative ordering of the sequence to couple $\hat{H}_\gamma$ terms of the same color to respective qumodes, resulting in a shortened circuit compared to \cref{fig:circuits}(a). Wavy-line arrows indicate that only qumode SWAP network operations are applied; straight-line arrows indicate that the circuit in \cref{fig:dimerCircuits}(c) are applied in parallel to all edges of one color [including in step (i)]. For colored edges representing $\hat{H}_\gamma$ terms, dashed-line means coupling $\hat{H}_\gamma$ to a half of the qumodes and solid line all qumodes. The final SWAP network is not necessary in panel (b).}
  \label{fig:visualCircuits}
\end{figure}

To demonstrate and verify our algorithm, we consider the case of the antiferromagnetic Heisenberg model with $N=4$ sites and give numerical simulation results below, with some additional analytical results included in \cref{sec:AnalyticalResults}. In \cref{fig:visualCircuits} the two panels present visual representations of the color-ordered (a) and alternating-ordered (b) Trotter scheme compilations. In addition, to help build further intuitive understanding of the use of SNs in each clock cycle, \cref{fig:6siteCircuits} in \cref{sec:6site} presents visual representations of the two compilations for a 6-site chain. It is easy to see that, in the 1D case with a $\Gamma = 2$ colorable chain, the number of SNs (i.e., $T_{\hat{A} \hat{B}}$) is $\Gamma (N_x/2 - 1) = N_x - 2$ for the (a)-compilation sequence and $N_x/2 - 1$ for the (b)-compilation sequence. Note that the (a)-compilation sequence strictly follows \cref{fig:circuits}(a), except that the final layers of $T_{\hat{A}}$ and $T_{\hat{B}}$ in each dashed box in \cref{fig:circuits}(a) are dropped without affecting the result since $-2 i \alpha \sum_{j}\hat{p}_j$ is permutation invariant.

\subsubsection{\texorpdfstring{$\alpha \ll 1$}{α ≪ 1} LTPF Regime}
\label{sec:trotter_error}
Using $\beta = \alpha(1 - \frac{E_{\text{s}}}{N})$, we approximately displace all oscillators by the spectrum-shifted Hamiltonian $\hat{H}'$ via the first-order product formula 
\begin{align}
    &\phantom{=}\prod_{p= 0}^{N-1} e^{-2 i \alpha \hat{B}' \otimes \hat{p}_p} \times \prod_{q=0}^{N-1}  e^{ -2 i \alpha \hat{A}' \otimes \hat{p}_q} \notag \\
    &=  \exp( -2 i \alpha \hat{B}' \otimes \sum_j \hat{p}_j) \exp( -2 i \alpha \hat{A}' \otimes \sum_j \hat{p}_j).
\end{align}
Recalling the ground-state energy shift, define
\begin{subequations}
\label{eq:AB_tiling_shift}    
\begin{align}
     \hat{A}'
  &= \left(\sum_{j=0}^{\frac{N}{2}-1} \hat{H}_{2j, 2j+1}\right) - \frac{E_{\text{s}}}{2}
   \equiv \sum_j \hat{A}_j - \frac{E_{\text{s}}}{N}, \\ 
     \hat{B}'
  &= \left( \sum_{j=0}^{\frac{N}{2}-1} \hat{H}_{2j+1, 2j+2} \right) - \frac{E_{\text{s}}}{2}
   \equiv \sum_j \hat{B}_j - \frac{E_{\text{s}}}{N},
\end{align}
\end{subequations}
such that $\hat{H}' = \hat{A}' + \hat{B}' = \hat{H} - E_{\text{s}}$. In one-dimension we partition the interactions into two sets of generators $\hat{H} = \sum_{j=1}^{j=L} H_{j,j+1} = \hat{A} + \hat{B}$. For such a system, the first-order Trotter expansion 
\begin{align}
  \label{eq:1st_order_trotter}
     e^{i \tau \hat{H}} 
  &= \qty(e^{i \tau \hat{A}/r} \cdot e^{i \tau \hat{B}/r})^r + 
     \order{\norm{ \comm{\hat{A}}{\hat{B}} }\tau^2/r}.
\end{align}
When $\norm*{[\hat{A},\hat{B}]} \tau^2/r \leq 1$ this expansion is valid, and bound to the ideal dynamics up by the above additive error. Next we apply the interaction for each set controlled on the qumode. The interactions are stroboscopically repeated to realized a bounded Lie-Trotter expansion of the target Hamiltonian's real-time evolution. 

Inserting the product formula \cref{eq:1st_order_trotter} into \cref{eq:filter_integral_transform} generates a formula for the error in approximating the target imaginary-time spectral projection given by \cref{eq:filter_general_shift}.  The Hubbard-Stratonovich transformation in Eq. \eqref{eq:filter_integral_transform} synthesizes an imaginary-time evolution as a linear superposition over real-time evolutions. This connection provides a useful tool to analyze imaginary-time propagation errors from the existing theory on Trotter errors established for real-time evolution \cite{Childs2021}. For example, for composite Hamiltonian $\hat{H}'$ as a sum of many terms, a $p$th order Trotterization of $e^{-i \hat{H}' \tau x}$ on the RHS of Eq. \eqref{eq:filter_integral_transform} for small $\tau$ effectively produces a $2p$th order imaginary-time evolution on the LHS (due to $\tau^2$) via the integral transformation with the weight $e^{-\frac{1}{2} x^2}$. 

\begin{figure}
  \centering
  \includegraphics[width=1.0\linewidth]{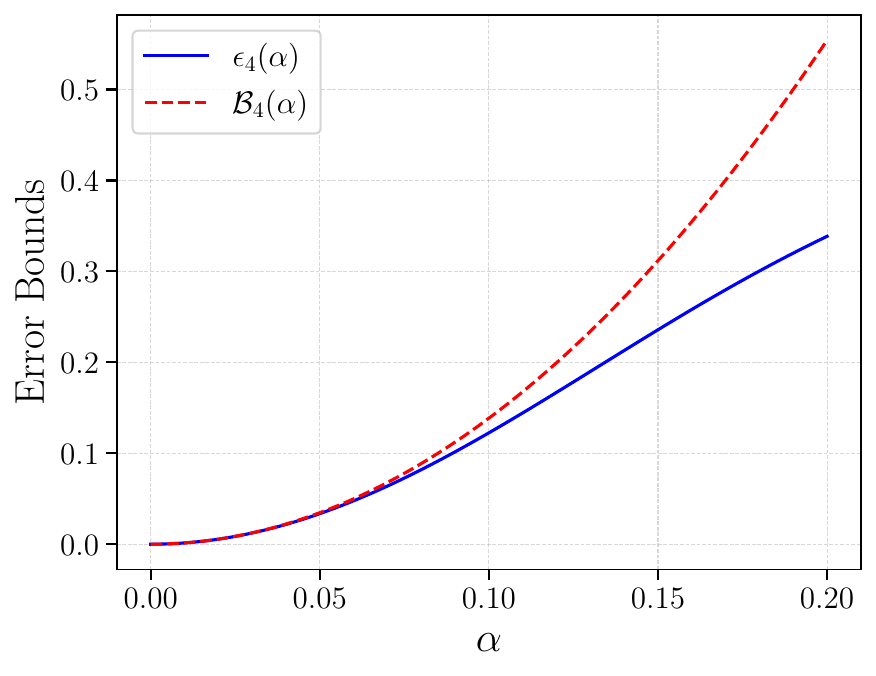}
\caption{Comparison of the additive Trotter error $\epsilon_N(\alpha) = \norm{\hat{R}_{H'}(\alpha) - \hat{P}_{H'}(\alpha)}$ and the scaling bound $\mathcal{B}_N(\alpha) = 
\frac{\alpha^2}{2} \sum_{\gamma_1 < \gamma_2} \norm{\comm{\hat{H}_{\gamma_1}}{\hat{H}_{\gamma_2}}}$ for a target system of $N = 4$ sites with $\Gamma = 2$ partitions. For these simulations we use the spectral norm $\norm{\hat{M}}_2 \equiv \sigma_{\text{max}}(\hat{M}) = \sqrt{\lambda_{\text{max}} (\hat{M}^* \hat{M})}$, i.e., the largest singular value of $\hat{M}$. 
These results agree with Eq. (\ref{eq:para_trotter_err}) in Theorem \ref{thm:parallel} since $\epsilon_N(\alpha) \leq B_N(\alpha) \ \forall \alpha \geq 0 $.}
\end{figure}

As we demonstrated in \cref{sec:app-heisenberg-model}, given qubit-oscillator hardware connectivity, Trotterization suffices to compile the real-time evolution of a Hamiltonian $\hat{H}'$ comprised of non-commuting terms. The HS transform provides a concrete way to analyze the error in our projection scheme using existing error analysis for real-time evolution. We note that the synthesis of real-time evolution with high-order Trotter errors from a linear combination of low-order product formulas via error cancellation has been explored to construct effective multi-product formulas for efficient Hamiltonian simulation \cite{vazquez2023well,Childs2012,Faehrmann2022,zhuk2023trotter}. From this perspective, the HS transform,  as a linear combination of real-time evolution, is a Gaussian multi-product formula for imaginary-time propagation.

\begin{figure}
  \centering
  \includegraphics{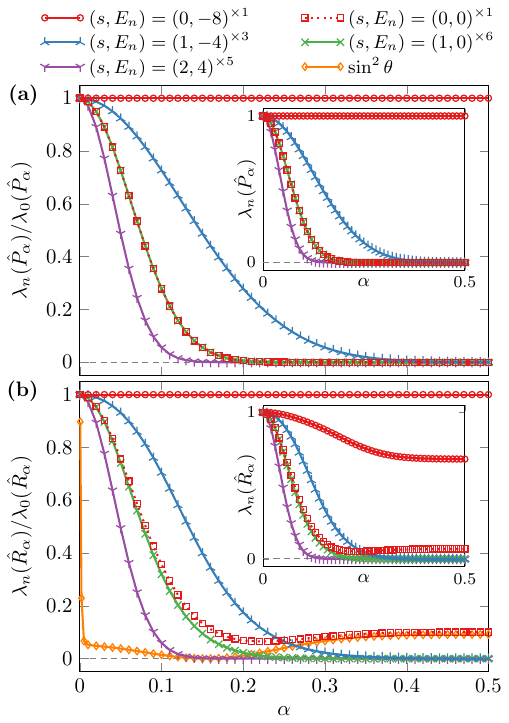}  
\caption{The normalized eigenvalues of (a) exact projector $\hat{P}_\alpha$ and (b) Trotter-approximate projector $\hat{R}_\alpha$ for $N=4$ site antiferromagnetic Heisenberg model. The insets show the unnormalized values. Here, $\lambda_n(\hat{O})$ denotes the $n$th \mbox{(right-)eigenvalue} for the \mbox{(non-)Hermitian} operator ($\hat{O}=\hat{R}_\alpha$) $\hat{O}=\hat{P}_\alpha$.
The value of normalization factor $\lambda_0(\hat{O})$ is shown by the red solid curve with circles in the insets.
Since $\hat{P}_\alpha$ and $\hat{H}$ share common eigenvectors, $\hat{P}_\alpha = \sum_n \lambda_n(\hat{P}_\alpha)\op{E_n}$ and thus eigenvalues of $\hat{P}_\alpha$ are filtering weights of each individual eigenvector projector $\op{E_n}$, where $\ket{E_0}$ denotes the ground state of Hamiltonian.
In general the ground state projector has larger \emph{relative} weight as $\alpha$ increases and $\lambda_{n\neq 0}$ decreases (except that for the red dotted curve with squares in panel (b), the maximal ratio $\lambda_0(\hat{R}_\alpha)/\lambda_n(\hat{R}_\alpha)$ is at $\alpha\sim 0.2$).
The symbols are from numerical simulation and the lines are from analytical expressions. In the legend different curves are labeled by the corresponding conserved quantum number $s$ (total spin) and $E_n$ (energy) and the degeneracy $d_g$ is marked by ``$\times d_g$'' superscript. However, in (b), the two red curves (circles and squares) of $\hat{R}_\alpha$ have slightly hybridized $E_n$ eigenvectors and the degree of mixtures is shown by the orange curve for $\sin^2\theta$ (see main text).
At $\alpha \sim 0.15$, $\sin^2\theta = 0$, so the energy is conserved and the $\alpha$ value is optimal for projection.}
  \label{fig:eigsProjOp}
\end{figure}
\subsubsection{\texorpdfstring{$\alpha \gg 1$}{α ≫ 1} Floquet-Regime}
Since the Trotterized time evolution is a stroboscopic evolution, the real time dynamics can be understood as time evolution under an effective Floquet propagator. In the limit of small displacement, the Trotter error bounds may be viewed as the Floquet propagator being localized to the ideal Hamiltonian propagator~\cite{Heyl2019}. However, in the large displacement limit a perturbative analysis is not useful. Rather, it is relevant to think about the Floquet process in terms of the underlying symmetries which are preserved and to realize that the energy eigenstates within these symmetry sectors will be hybridized in the Floquet limit. In this case, one wishes to find a sweet spot where the displacement magnitude is maximized while the hybridization between energy eigenstates is minimized. Below, in \cref{sec:opt}, we will provide a relevant example optimizing the displacement factor $\alpha$. This is illustrated in \cref{fig:eigsProjOp} for a 4-site XXZ model ($\Delta = 1$) with a single Trotter step $r=1$. We see that both the total (a) and Trotter (b) rates of spectral projection converge to a constant in the $\alpha \gg 1$ limit. Thus, the bottom panel highlights how Floquet spectrum converges to a fixed point in the large $\alpha$ limit. Crucially, the spectra share similar features, partially enforced by the unbroken $S^2$ and $S^Z_{tot}$ symmetries. One may therefore ask, which displacement magnitude is optimal?

\subsubsection{Optimization}
\label{sec:opt}

We now present numerical simulations for the energy filter. To analyze cases which are most amenable to experiments, i.e., which minimize all prefactors in the resource overhead, we analyze a filter realized by the alternating-color controlled displacement Trotterization as illustrated in panel (b) of \cref{fig:visualCircuits}. Here the number of SNs is half the spatial extent of the system, whereas the number of SNs is directly proportional to the spatial extent of the system in the color ordering. While both display the same asymptotic scaling, removing this extra pre-factor of 2 may prove useful in experimental implementations where physical errors dominate over algorithmic errors. We defer a more detailed discussion of experimental errors to \cref{sec:physical_error}.

The rate at which spectral operations, including $\hat{R}_{\alpha}$'s ground state projection, transform depends on spectral features. First, the magnitude of the projector's smallest eigenvalue $\lambda_0$, and its proximity to $E_\text{s}$, quantifies the preservation of the ground state $(\ket{\lambda_0})$ after oscillator measurement. Secondly, QITE projection depends on the proximity of eigenvalues. Denote the first-excited synthesized eigenvalue as $\lambda_1$. A large low-energy multiplicative condition number, $\kappa_{0,1} = \lambda_0/\lambda_1$, implies efficient ground state preparation. In terms of the additive spectral gap $\Delta_{0,1} = \lambda_1 - \lambda_0$, has previously been analyzed \cite{Ge2019, Lin2020, Keen2021}. This begs the question of whether there exists an optimal $\alpha$, i.e., one that optimizes state preparation. 

In our numerical simulations, of the 4-site XXZ model consisting of four qubits and the low-photon number subspace of four qumodes, we now describe convolution of the aforementioned spectral competitions with couplings between states within the projector's spin-0 sector. The detailed analytical results are given in \cref{sec:AnalyticalResults}. One can analytically show that $\hat{R}_{\alpha}$ is block-diagonal in the eigenbasis of $\hat{H}_{4}$ ~\footnote{More explicitly, $\hat{R}_{\alpha}$ is block diagonal is the simultaneous eigenbasis of $\hat{H}_4$ and its mutually commuting symmetry operators $ \{ \hat{S}_{\text{tot}}^2, \hat{S}_Z, \hat{\sigma}_{24} \}$ where $\hat{\sigma}_{24} = \frac{1}{2} \left(\mathds{1} + X_2 X_4 + Y_2 Y_4 + Z_2 Z_4 \right)$ is a swap operator between sites 2 and 4} and can be decomposed as 
\begin{equation}
    \hat{R}_{\alpha} = \hat{R}^{(0)}_{\alpha} \oplus \left( \hat{R}^{(1)}_{\alpha} \oplus \hat{R}^{(2)}_{\alpha} \right)
\end{equation}
where $\hat{R}^{(1, 2)}_{\alpha} = \hat{R}^{(1)}_{\alpha} \oplus \hat{R}^{(2)}_{\alpha} $ is a 14-dimensional diagonal matrix containing the spin-1 and spin-2 eigenstates and 
\begin{align}
    \hat{R}^{(0)}_{\alpha} &= r_{00} \op{\lambda_0} +
    r_{01} \op{\lambda_{0}}{\lambda_{1}} \\
    & + r_{10} \op{\lambda_{1}}{\lambda_{0}} + r_{11} \op{\lambda_1} \\
    & = 
    \begin{pmatrix}
            r_{00} & r_{01} \\
            r_{10} & r_{11}
    \end{pmatrix}
\end{align}
is a non-diagonal matrix containing the two spin-0 eigenstates. The off-diagonal entries in $\hat{R}^{(0)}_{\alpha}$ correspond to the state transfer terms that mix the ground state $\ket{\lambda_0}$ $(E_{\text{s}} = -8)$ and the higher energy state $\ket{\lambda_1}$ $(E_{\text{s}} = 0)$ with a coupling strength governed by $r_{01} = -r_{10}$. 

If $\abs{r_{01}}$ were minimized at some $\alpha_{\text{optimal}}$, then $\hat{R}_{\alpha}$ would be approximately, if not actually, diagonal, and hence an ideal projector. As shown in~\cref{fig:eigsProjOp}, we successfully decouple $\ket{\lambda_0}$ and $\ket{\lambda_1}$ by numerically finding that $r_{01} = r_{10} = 0$ when $\alpha \approx 0.1546$. We posit the existence of $\alpha_{\text{optimal}}$ for larger spin chains, where such an $\alpha$ can be numerically identified, as demonstrated in the 4-site model, by minimizing the vector containing all off-diagonal matrix elements coupling the ground state to higher energy states. 

\subsection{Algorithm in 2D}
\label{sec:2D}

As illustrated in the bottom right panel of \cref{fig:hardware-mapping}, in two dimensions \cref{eq:XXZ}'s nearest neighbor ($\langle i,j\rangle$) interactions still run over the horizontal $ ((i,j) , (i\pm1,j))$ lattice sites, as in the prior 1D case, and now additionally run over the vertical $ ((i,j) , (i,j\pm1))$ edges. As illustrated in \cref{fig:trotter_tilings}, the nearest neighbor interaction graph can be partitioned into four (as in 1D, equal to the spin coordination number) components:\begin{align}
 \label{eq:ABCD_tiling}
 \hat{A} &= \sum_{j=0,i=0}^{j=N -1,i=\frac{N}{2}-1} \hat{H}_{(2i  ,j   ), (2i+1,j   )}, \\
 \hat{B} &= \sum_{j=0,i=0}^{j=N -1,i=\frac{N}{2}-1} \hat{H}_{(2i+1,j   ), (2i+2,j   )}, \\
 \hat{C} &= \sum_{i=0,j=0}^{i=N -1,j=\frac{N}{2}-1} \hat{H}_{(i   ,2j  ), (i   ,2j+1)}, \\
 \hat{D} &= \sum_{i=0,j=0}^{i=N-1,j=\frac{N}{2}-1} \hat{H}_{(i   ,2j+1), (i   ,2j+1)}.
\end{align} 
Note that for a given value of $j$, $\hat{A}$, and $\hat{B}$ reduce to the one-dimensional case that we have already analyzed above. $\hat{C}$ and $\hat{D}$ likewise tile the vertical edges. 
Subsequently, the scheme to time evolve is similar to the prior 1D case. The only difference is that each spin variable is involved in four interactions as colored. Taken together, this gate sequence constitutes a 2D Trotterized time-evolution clock cycle, which is stroboscopically applied until the desired propagation time. 

The 2D model's logical displacement operators are 
\begin{align}
 D^{(n)}_{\hat{A}} &= \prod_{j=0,i=0}^{j=N-1,i=\frac{N}{2}-1} D_{\hat{H}_{(2i  ,j   ), (2i+1,j   )}}^{(n)}, \\
 D^{(n)}_{\hat{B}} &= \prod_{j=0,i=0}^{j=N-1,i=\frac{N}{2}-1} D_{\hat{H}_{(2i+1,j   ), (2i+2,j   )}}^{(n)}, \\
 D^{(n)}_{\hat{C}} &= \prod_{i=0,j=0}^{i=N-1,j=\frac{N}{2}-1} D_{\hat{H}_{(i   ,2j  ), (i   ,2j+1)}}^{(n)}, \\
 D^{(n)}_{\hat{D}} &= \prod_{i=0,j=0}^{i=N-1,j=\frac{N}{2}-1} D_{\hat{H}_{(i   ,2j+1), (i   ,2j+1)}}^{(n)}. 
\end{align}
The transversal bosonic SWAP operators, which interchange modes along bonds tiling the 2D lattice, now read
\begin{align}
 \perm_{\hat{A}} &= \bigotimes_{j=0,i=0}^{j=N-1,i=\frac{N}{2}-1} \text{SWAP}^{(2i  ,j   ), (2i+1,j   )}, \label{eq:PERM_A}\\
 \perm_{\hat{B}} &= \bigotimes_{j=0,i=0}^{j=N-1,i=\frac{N}{2}-1} \text{SWAP}^{(2i+1,j   ), (2i+2,j   )}, \label{eq:PERM_B}\\
 \perm_{\hat{C}} &= \bigotimes_{i=0,j=0}^{i=N-1,j=\frac{N}{2}-1} \text{SWAP}^{(i   ,2j  ), (i   ,2j+1)}, \label{eq:PERM_C}\\
 \perm_{\hat{D}} &= \bigotimes_{i=0,j=0}^{i=N-1,j=\frac{N}{2}-1} \text{SWAP}^{(i   ,2j+1), (i   ,2j+1)}. \label{eq:PERM_D}
\end{align}

We define the back-to-back execution of Eqs. (\ref{eq:PERM_A}, \ref{eq:PERM_B}) and Eqs. (\ref{eq:PERM_C}, \ref{eq:PERM_D}) as the permutation operators
\begin{align}
    \perm_{\hat{A} \hat{B}} &= \perm_{\hat{B}} \perm_{\hat{A}}, \\
    \perm_{\hat{C} \hat{D}} &= \perm_{\hat{D}} \perm_{\hat{C}}.
\end{align}
The Trotterized unitary displacing the oscillator controlled on the 2D spin-spectrum (see \cref{fig:circuits}) then reads,
\begin{align}
    & V^{\text{(2D)}}_{\hat{H}'} (\alpha) = \prod_{\hat{L} \in \{ \hat{A}, \hat{B}, \hat{C}, \hat{D} \}}
    \left[ \perm_{\hat{C} \hat{D}} \left(\perm_{\hat{A} \hat{B}} D_{\hat{L}}(\alpha)\right)^{\frac{N}{2}} \right]^{\frac{N}{2}} \\
    & = \prod_{p=t}^{N-1} e^{-2 i \alpha \hat{D} \otimes  \hat{p}_t} \times \prod_{s=0}^{N-1}  e^{-2 i \alpha \hat{C} \otimes  \hat{p}_s} \\
    & \times \prod_{q= 0}^{N-1} e^{-2 i \alpha \hat{B} \otimes  \hat{p}_q} \times \prod_{p=0}^{N-1}  e^{-2 i \alpha \hat{A} \otimes  \hat{p}_p} \notag
    \label{eq:V2D}
\end{align}
This is the desired expression used in the 2D versions of \cref{thm:LTHS} and \cref{thm:parallel}. A visual quantum circuit representation of this algorithm appears in the bottom panel of \cref{fig:circuits}.

\section{Physical Error Analysis}
\label{sec:physical_error}

A complete understanding of algorithmic and physical errors together in quantum computations is an open problem. In developing this new framework, we first derived an algorithm in which hardware errors were considered to be vanishingly small. While a complete analysis of the effect of physical hardware errors on the Gaussian energy projection protocol is beyond the scope of this work, here we briefly discuss the dominant error processes and their relevance to our oracles.

The important errors to consider in the experimental implementation of our Gaussian projection protocol are the (weak) amplitude damping of the oscillators at rate $\kappa$, and bit-flip and phase flip errors in the qubits \cite{Liu2024}. For short times, $\delta t$, the Kraus operators associated with amplitude damping in a given oscillator are
\begin{align}
K_0&= \sqrt{1-\kappa\,\delta t\, a^\dagger a}\approx e^{-\frac{\kappa}{2}\delta t\, \hat{a}^\dagger \hat{a}^{\pdag}}\\
    K_1&=\sqrt{\kappa\,\delta t}\,\hat{a}.
\end{align}
Here $K_1$ is the photon loss `jump operator' and $K_0$ is the `no-jump' operator describing the Bayesian update that reduces the amplitude of large photon number states when no photon escapes into the environment \cite{BinomialCodes}.

For simplicity, we will analyze the effect of the Kraus operators within the simple model of the protocol given at the beginning of Sec.~\ref{sec:Methods} in Eqs.~(\ref{eq:Intro1}-\ref{eq:Intro2}).
Interestingly, the no-jump errors associated with $K_0$ deterministically shrink the coherent amplitudes, but this simply yields a decrease in the effective value of $\beta$, so that the only effect is to lower the strength of the projection.  If a jump error occurs, the photon loss re-weights the relative amplitudes of the states. In the simplest case of a single photon loss, the (unnormalized) state is approximately given by

\begin{align}
    D_{\hat{H}'}(\beta - \widetilde{\beta})a D_{\hat{H}'}(\widetilde{\beta})|\Psi_\mathrm{in}\rangle\nonumber 
    &= \sum_n \tilde\beta_nD_{\hat{H}'}(\beta)|E_n\rangle\otimes |0\rangle \\&= \sum_n \tilde\beta_n\psi_n|E_n\rangle\otimes |\beta_n\rangle,\label{eq:photonlossstate}
\end{align}
where $\tilde\beta_n = \widetilde{\beta}(E_n-E_\text{s})$ is the value of the coherent state amplitude of the oscillator at the time of the photon jump partway through the protocol. Therefore, the resulting state is identical to $\ket{\Psi_{\textrm{out}}}$ up to the prefactors $\widetilde{\beta}_n$ that rescale the relative weight of each eigenstate $\ket{E_n}$.  This rescaling reduces the effectiveness of projection into states with energy eigenvalues close to $E_\text{s}$, but is not fatal to the protocol in the sense that eigenstates with energy eigenvalues far from $E_\text{s}$ are still strongly projected. 

We note that in writing the above, we have made the simplifying assumption that the protocol can be approximated as a continuous displacement of the qumode at a rate proportional to the spin eigenvalue.  In practice, complications can arise from photon loss during the compilation steps needed to produce the weight-2 Pauli strings in the spin Hamiltonian. In particular, photon loss during a conditional displacement or beam splitter gate will manifest similarly to the rescaling effect above; in contrast, photon loss during a controlled-parity gate will produce errors that propagate in a more complicated and generally fatal fashion.

Qubit decay and dephasing will produce additional errors, which will require modeling, beyond the scope of this work, for quantitative error emulation.  However, one can obtain a sense of the circuit depths that are achievable by considering the relevant time scales. In current hardware, beam-splitter $\operatorname{BS}^{(j,k)}(\theta,\varphi)$ (e.g., cavity SWAP) gates have durations on the order of $\sim 100$ ns \cite{chapman2022high,lu2023highfidelity,Chou2024}, controlled-parity operations, $P^{(j)}_{Z_k}$, require of order $\sim 1\, \mu$s \cite{Sun2014,Wang2020FCFs,Chou2024,deGraaf2024midcircuit}, and controlled displacements require of order $\sim 0.25-1\, \mu$s \cite{EickbuschECD}. These times are to be compared to state-of-the-art transmon lifetimes of $\sim 300-500\,\mu$s \cite{place_new_2021,wang_towards_2022}, and superconducting resonator lifetimes of $\sim 1-1000\,$ms \cite{Reagor2016,Romanenko_PhysRevApplied.13.034032,RosenblumCavity2023,ganjam_surpassing_2024,PRXQuantum.5.040307}.

Lastly, we outline a protocol to disentangle algorithmic errors from the physical errors that appear due to hardware and control imperfections. To do so, we consider a randomized benchmarking protocol at the oracular level. The main idea in randomized benchmarking of qubit unitaries is to apply `typical Haar random' circuits which are logical identity operators. The circuits should also be representative of the random gate set being benchmarked. The physical errors lead to exponentially decaying observables, with respect to the logical identity's depth, which are efficient to evaluate. Consequently, randomized benchmarking provides a simple and useful estimate of physical gate errors. RB algorithms addressing bosonic operations have recently appeared \cite{valahu2024benchmarking}. 

In addition to physical-level RB, one may, using $\text{Tr}_{\bf B}(|\mel{m}{U_x(\alpha)}{0}|^2) = \sum_{m=0}^{\infty} (\hat{P}^m_\tau)^\dagger  \hat{P}^m_\tau = \mathds{1}_{\bf Q}$, also consider a logical-level oracular randomized benchmark (ORB). The bold letters denote the composite space of all bosons or qubits. This means that our filters expressed in \cref{eq:photon_U_m} are Kraus operators, each representing a distinct spectral filter, satisfying the above completeness relation. One implication is that, by tracing over all photon numbers, we stochastically apply $\mathds{1}_{\bf Q}$. The spectral ORB protocol therefore will measure an important domain-specific function of an easy-to-prepare input state. The above trace is simply evaluated in a stochastic manner by iteratively sampling over all photon numbers measured. 

\section{Conclusion}
\label{sec:conclusion}

In this work, we explicitly constructed a spectral filter (imaginary-time propagating) that propagates a target quantum system to, for example, a ground state. We partitioned this task into the compilation and application of block-encoded quantum oracles, which take advantage of analog features such as i) the infinite-dimensional support of bosonic field variables and ii) the ease of manipulating Gaussian states. In doing so, our protocol lifts unitary DV circuits to a popular CV-DV architecture, where transmon qubits are interspersed (\cref{fig:hardware-mapping}, \cite{Liu2024}) with microwave cavity oscillators. 

To develop our framework, we first provided a theory of block encoding functions using both qubit and Gaussian CV variables. These CV-DV operators were then used to compile circuits that approximate the desired spectral filter functions, that is, acting with respect to a parent Hamiltonian. In addition to considering the ideal filter, we have also derived the contributions of algorithmic errors within the block encoding procedure, with both heterodyne detection and photon-based block encodings analyzed. Overall, our framework defines a stochastic process that can be taken as a semicoherent process \cite{Wang2024} with lower coherence requirements and can be augmented by the use of amplitude amplification as needed. Finally, we note that spectral filtering is an easier task than eigenstate preparation. More work is required to precisely relate these two tasks to one another and to also clarify how the spectral gap affects overall scaling. The complexity of preparing an input state with constant overlap with a desired output state is likewise still an open question. 

The middle part of this work was dedicated to synthesizing CV-DV circuits which $\varepsilon$-approximate the Gaussian spectral transformation oracles. We first outlined the gate set for the base interactions that the Trotter product formula employs, and how these gates are themselves synthesized in terms of physical interactions between qubits and qumodes. Afterwards, we derived the base Trotter factorizations, as well as the associated error scaling, in which imaginary time scaling is a convolution of the real-time errors. By analytically and numerically analyzing a minimal example, we optimized the displacement strength and discussed the success probabilities of the protocol. Lastly, we demonstrate that the projection is highly parallelizable for $M=N$ bosonic modes, and we generalized the spectral filtering algorithm to 2D. A future open research direction is to encode the newly discovered compilation routines as rules, similarly to Ref.~\cite{chen2025genesis}, to develop a systematic, automated, and optimized CV-DV compiler for hybrid systems. 

Our work demonstrates how complex oracular functions can be optimally encoded by a judicious selection of physical degrees of freedom and Gaussian CV-DV unitary gates and measurements. Future work may also generalize the class of integral-based, block-encoded oracles. In principle, our work can be straightforwardly generalized to simulate non-Heisenberg models. For example, one may examine spin models residing on tri-coordinated lattices or those with next-nearest-neighbor spin interactions. 

By enabling nonlinear analytic transformations on block-encoded operators, this computational model's utility as a powerful resource for efficiently compiling quantum algorithms is highlighted. Notably, the nonlinear transformations enacted by the CV qumode register form a broader class of analytic functions (for example, Gaussians inherited from the CV wave functions) beyond what a DV ancilla register can do (only polynomials) and thus constitute a potentially more powerful computation model. As a paradigmatic example of continuous eigenvalue transformation oracles, we specifically focused on spectral filtering. It will be of interest to explore the set of non-linear transformations which can be efficiently encoded with our protocol.

We note that the hybrid qubit-oscillator operations required for the implementation of our protocol are experimentally feasible with high quality and have already been demonstrated at small scales \cite{Wang2020FCFs,EickbuschECD}. This begs the question of what their role is in the grand scheme of quantum computation. Looking beyond direct implementations in the near term, we note that there are a variety of ways in which our algorithm may be generalized to be compatible with quantum error correction. In addition to qubit-only fault-tolerant implementations, where the continuous field is discretized~\cite{Keen2021}, one may consider a paradigm where, in order to minimize the overall resource cost, error-corrected logical qubits are combined with low-error oscillators. This approach is well suited to the hierarchy of error rates in the proposed hybrid CV-DV architecture, where qubit errors are the dominant source of error \cite{Liu2024}. Furthermore, the compilation techniques discussed in this work can be extended to realize controlled displacements conditioned on logical spin operators. That is, given the trade-off between the number and quality of logical qubits encoded from physical qubits, we conjecture that a semi-fault-tolerant paradigm, using high-quality-factor qumodes, can minimize overall resources by maximizing the number and fault tolerance of the encoded logical qubits. In addition, oscillator encodings may also be combined with our framework to partially suppress the noise on the oscillators and the Gaussian gates they utilize \cite{noh2020encoding}. In general, more research is required to holistically analyze and optimize the experimentally available resources. 

Finally, we provided a preliminary analysis of physical errors. In particular, we discussed how native error channels both contribute and can be characterized at various algorithmic levels. To quantify these effects, we provided physical error rates and examined their leading order contribution to our target propagator. We then examined the bosonic variants of physical level randomized benchmarking and outlined a logical-level ORB, a variant of randomized benchmarking which naturally emerges by treating all non-vacuum block-encoded operations equally and sampling over all outcomes. Reference \citenum{Varona24}, which encodes diagrammatically defined operators in terms of the VTA with similar architecture, appeared during the completion of this manuscript. 

The data and code that support the findings in this study are available at Ref.~\citenum{Bell_Github}.

\begin{acknowledgments}
We thank P. Lotshaw for helpful comments. 
L.B.\ was supported by DOE ASCR funding under the Quantum Computing Application Teams program, FWP ERKJ347. 
Y.W.\ is supported by the U.S.\ Department of Energy, Office of Science, National Quantum Information Science Research Centers, Quantum Science Center, and DOE ASCR funding under the Accelerated Research in Quantum Computing Program, FWP ERKJ445.
E.D.\ is supported by the U.S. Department of Energy, Office of Science, Advanced Scientific Research Program, Early Career Award under contract number ERKJ420.
K.S.\ was supported by the U.S. Department of Energy, Office of Science, National Quantum Information Science Research Centers, Co-design Center for Quantum Advantage under contract number DE-SC0012704.
Y.L.\ acknowledges the support by the U.S. Department of Energy, Office of Science, Advanced Scientific Computing Research, under contract number DE-SC0025384.
%
S.M.G.\ acknowledges additional support for research sponsored by the Army Research Office (ARO), and accomplished under Grant Number W911NF-23-1-0051. The views and conclusions contained in this document are those of the authors and should not be interpreted as representing the official policies, either expressed or implied, of the Army Research Office (ARO), or the U.S. Government. The U.S. Government is authorized to reproduce and distribute reprints for Government purposes notwithstanding any copyright notation herein.

\textit{External Interest Disclosure:} S.M.G.\ is a consultant for, and equity holder in, Quantum Circuits, Inc.


\textit{Author contributions:} YW, YL, ED, and SMG conceptualized initial methodologies. 
LB, KCS, YL, and SMG developed the instruction set architecture and provided hardware-specific details. 
LB, KCS, and YW contributed to visualization as well as numerical and analytical validation.  
KCS, YL, ED, and SMG analyzed physical errors. 
All authors contributed to original writing, review and editing, and formulating the compilation presented.

\end{acknowledgments}


\appendix
\section{Brief review of continuous variables}
\label{sec:bosonic_convention}
First set $\hbar = 1$ (e.g., in the time evolution operator ${\exp}(-it\hat{H}/\hbar) \to {\exp}(-it\hat{H})$). Recall the bosonic field operators for the $j^\text{th}$ mode create, acting as $\hat{a}_j^\dagger \ket{n}_j = \sqrt{n+1}\ket{n+1}_j$, or annihilate, acting as $\hat{a}_j \ket{n}_j = \sqrt{n}\ket{n-1}_j$, a photon excitation. The field operators are subject to the bosonic commutation relations $[\hat{a}_i,  \hat{a}_j^\dagger] = \delta_{i,j}$. 

Focusing on a single CV mode, and dropping the mode index, we define the hermitian quadrature operators (analogous dimensionless position and momentum operators)
\begin{subequations}
\begin{align}
    \hat{x} &= \lambda_x (\hat{a} + \hat{a}^\dagger), \\ 
    \hat{p} &= -i\lambda_p(\hat{a}-\hat{a}^\dagger),
\end{align}
\end{subequations}
where $\lambda_{x,p} \in \mathbb{R}$. The choice of $\lambda_x = \lambda_p = 1/2$ defines Wigner units and corresponds to
\begin{subequations}
 \label{eq:xp_to_a}
\begin{align}
    \hat{a} &= \hat{x} + i\hat{p}, \\
    \hat{a}^\dagger &= \hat{x} - i\hat{p}.
\end{align}
\end{subequations}
It is straightforward to verify $[\hat{x}, \hat{p}] = (i/2)[\hat{a},  \hat{a}^\dagger] = i/2$, which is as if $\hbar = 1/2$ in the canonical commutation relation $[\hat{x}, \hat{p}] = i\hbar$ where $\hat{x}$ and $\hat{p}$ are dimensional position and momentum operators, respectively.

Wigner units subsequently define the width of the momentum-space Gaussian wavefunction for the vacuum state $\ket{0}$ as $\psi_0(p) = \ip{p}{0} = \mel{p}{(\int \op{x} dx)}{0} = \int \ip{p}{x} \ip{x}{0} dx = \int \psi_p^{*}(x) \psi_0(x) dx =(2/\pi)^{1/4} e^{-p^2}$, where the position space vacuum wavefunction in Wigner unites is $\psi_0(x) = \ip{x}{0} = (2/\pi)^{1/4} e^{-x^2}$~\cite{Liu2024} and the plane-wave function $\psi_{p}(x) = \ip{x}{p} = \pi^{-1/2} e^{2ipx}$ [$\pi^{-1/2}$ factor so that orthogonality reads $\ip{p}{p'} = \int \psi_{p}(x) \psi_{p'}^{*}(x) dx = \delta(p-p')$]. 
In this way, \cref{eq:Ux00SMG} has the physical interpretation of being the expectation value of the operation $U^{(\hat{p})}_{\hat{H}'}(\alpha) = e^{-2i \alpha  \hat{H}' \otimes \hat{p}}$, with respect to the cavity electromagnetic field's vacuum state $\ket{0}$. 

\cref{eq:Ux00SMG} is derived by setting $K\rightarrow \pm \tau \hat{H}^{'}$ in the expression for the position boosted state in Eq.~36 of Ref.~\citenum{Liu2024}, multiplying by $\langle p|0\rangle$ and integrating over position coordinates.
That is the same as setting $\hat{A} = iK = \pm i\tau \hat{H}'$ in \cref{eq:HS_transform}.
As a result of Wigner units, and the factor of two coming with $i\alpha$, the ``expectation value'' (an operator in fact) of the position displacement conditioned on the principal system's Hamiltonian,
$\ev{U^{(\hat{p})}_{\hat{H}'}(\alpha)}{0} 
= \int dp \ip{0}{p} U^{(p)}_{\hat{H}'}(\alpha) \ip{p}{0}$, becomes
\begin{align}
  \label{eq:Wigner_Projector}
  &= \sqrt{\frac{2}{\pi}} \int dp e^{-2p^2} e^{-2i \alpha \hat{H}' p} \nonumber \\
  &= \sqrt{\frac{2}{\pi}} \int dp e^{-2(p^2\mathds{1} + i \alpha p \hat{H}')} \nonumber \\
  &= \sqrt{\frac{2}{\pi}} \int dp e^{-2(p\mathds{1} + i \frac{\alpha}{2} \hat{H}')^2 + 2(i\frac{\alpha}{2} \hat{H}')^2} \nonumber \\
  &= e^{2(i \frac{\alpha}{2} \hat{H}')^2} 
   = e^{-(\alpha \hat{H}')^2/2}
   = \hat{P}_{\hat{H}'}(\alpha).
\end{align}

For the interested reader, note that \Cref{eq:D_product} indicates $e^{i\theta}D(\alpha)$ forms a group (Heisenberg group) with a geometric phase $e^{i (\Re\alpha) (\Im\alpha)}$, where $(\Re\alpha) (\Im\alpha)$ is twice the area enclosed by vectors $z_1 = i\Im\alpha$, $z_2 = \Re\alpha$, and $z_3 = -\alpha$ in the complex plane. In addition, any complex displacement $D(\alpha)$ can be factorized into a momentum boost $D(i\Im\alpha)$ followed by a position displacement $D(\Re\alpha)$, together with a geometric phase \cite{Liu2024}. 

Last, we calculate the $m$-photon wavefunction in momentum basis $\psi_m(p) = \ip{p}{m}$ that is used in the \cref{ssec:oracular-int-transform}. Using $\ket{m} = \frac{1}{\sqrt{m!}} (\hat{a}^\dagger)^m\ket{0}$ and $\hat{a}^\dagger = \hat{x} - i\hat{p} = e^{\hat{p}^2} \hat{x} e^{-\hat{p}^2}$~\footnote{This can be proved by using the so-called Hadamard lemma (Campbell identity) $e^{A}Be^{-A} = B + [A,B] + \frac{1}{2!}[A,[A,B]] + \frac{1}{3!}[A,[A,[A,B]]] + \dots$ Note that a nonunitary transformation is applied to transform a Hermitian operator $\hat{x}$ to a non-Hermitian one $\hat{a}^\dagger$.} and plugging in $\hat{x} \to \frac{i}{2} \partial_p$, we have
$
\psi_m(p) = \ip{p}{m} 
= \frac{1}{\sqrt{m!}} \mel{p}{e^{\hat{p}^2} \hat{x}^m e^{-\hat{p}^2}}{0}
= \frac{1}{\sqrt{m!}} [e^{p^2}(\frac{i}{2} \partial_p)^m (e^{-p^2}\ip{p}{0})]
= \frac{(-i)^m}{\sqrt{2^m m!}} \psi_0(p) H_m(\sqrt{2} p)
$, where $\ip{p}{0} = \psi_0(p) = (2/\pi)^{1/4} e^{-p^2}$, $H_m(p) = (-1)^m e^{p^2} (\partial_p)^m e^{-p^2} = (2p - \partial_p)^m \cdot 1$ is the $m$th Hermite polynomial.

Alternatively, a simplified proof of \cref{eq:photon_POVM} goes as follows [similar to the the derivation to Eq.~(38) in Ref.~\onlinecite{Feynman1951}]. Denoting $\hat{\xi} = \tau \hat{H}'$ and using $U_{\hat{H}'}^{(\hat{p})}(\tau) = e^{-i\tau \hat{H}' 2\hat{p}} = e^{\hat{\xi} \hat{a}^\dagger - \hat{\xi} \hat{a}} = e^{-\hat{\xi}^2/2} e^{\hat{\xi}\hat{a}^\dagger} e^{-\hat{\xi}\hat{a}}$ and $\hat{a}^m e^{\hat{\xi}\hat{a}^\dagger} = (\hat{a} + \hat{\xi})^m$, we find the matrix element
\begin{align}
  &\mathrel{\phantom{=}}
    \hat{P}_{\hat{H}'}(m, \tau) = \mel{m}{U_{\hat{H}'}^{(\hat{p})}(\tau)}{0}  \label{eq:photon_U_m} \notag \\
  &= \frac{1}{\sqrt{m!}} \mel{0}{\hat{a}^m e^{\hat{\xi} \hat{a}^\dagger - \hat{\xi} \hat{a}} }{0}  \notag\\
  &= \frac{e^{-\hat{\xi}^2/2}}{\sqrt{m!}} \mel{0}{e^{\hat{\xi} \hat{a}^\dagger} (\hat{a} + \hat{\xi})^m  e^{- \hat{\xi} \hat{a}} }{0} \notag\\
  &= \frac{e^{-\hat{\xi}^2/2}}{\sqrt{m!}} \mel{0}{(\hat{a} + \hat{\xi})^m}{0}  \notag\\
  &= \frac{e^{-\hat{\xi}^2/2} \hat{\xi}^m}{\sqrt{m!}} = \frac{(\tau \hat{H}')^m}{\sqrt{m!}} e^{-\frac{\tau^2}{2} \hat{H'}^2}.
\end{align}

\section{Proof of \cref{thm:parallel}}
\label{sec:proof-thm:parallel}
We give a brief proof of \cref{eq:para_R,eq:para_trotter_err,eq:para_R_Heis} in \cref{thm:parallel}. After expressing the LTHS operator $\hat{R}_{\hat{H}}(\alpha) = \ev{\cdot}{\vb{0}}$ in the integral form $\int d\vb{p} \bra{\vb{0}} \cdot \ket{\vb{p}}\!\! \bra{\vb{p}}\ket{\vb{0}} / \pi^{M}$ (in Wigner units) and using the vacuum wavefunction in the momentum basis $\psi_0(\vb{p}) = \ip{\vb{p}}{\vb{0}} = (2\pi)^{M/4} e^{-\vb{p}\vdot \vb{p}} = (2\pi)^{M/4} \prod_{j=0}^{M-1} e^{-p_j^2}$, \cref{eq:para_R_a} becomes
\begin{align}
  &\mathrel{\phantom{=}}
     (2/\pi)^{\frac{M}{2}} \int d\vb{p}
     \prod_{\gamma=1}^{\Gamma} \prod_{j=0}^{M-1}
     e^{-2p_j^2} e^{-2i(\alpha/\sqrt{M}) \hat{H}_\gamma p_j} \\
  &= (2/\pi)^{\frac{M}{2}} \int d\vb{p}
     \operatorname{\mathcal{T}_\gamma} \prod_{j=0}^{M-1}
     e^{-2p_j^2 -2i(\alpha/\sqrt{M}) (\sum_\gamma \hat{H}_\gamma) p_j} \\
  &= \operatorname{\mathcal{T}_\gamma}
     \prod_{j=0}^{M-1} \sqrt{2/\pi}\int dp_j
     e^{-2p_j^2 -2i(\alpha/\sqrt{M}) (\sum_\gamma \hat{H}_\gamma) p_j} \\
  &= \operatorname{\mathcal{T}_\gamma} \prod_{j=0}^{M-1} e^{-\alpha^2 (\sum_\gamma \hat{H}_\gamma)^2/(2M)} \\
  &= \operatorname{\mathcal{T}_\gamma} e^{-\alpha^2 (\sum_\gamma \hat{H}_\gamma)^2/2}.
\end{align}
This gives \cref{eq:para_R_b}.

To prove \cref{eq:para_trotter_err} we again apply the HS transformation with a variable $k$ to $\hat{P}_{\hat{H}}$ and $\hat{R}_{\hat{H}}$ respectively:
\begin{subequations}
\begin{align}
     \hat{P}_{\hat{H}}(\alpha)
  &= \sqrt{\frac{2}{\pi}} \int dk\, e^{-2k^2} e^{-2i\alpha \hat{H} k}, \\
     \hat{R}_{\hat{H}}(\alpha)
  &= \operatorname{\mathcal{T}_\gamma} \sqrt{\frac{2}{\pi}} \int dk\, e^{-2k^2} e^{-2i\alpha (\sum_\gamma \hat{H}_\gamma) k} \notag \\
  &= \sqrt{\frac{2}{\pi}} \int dk\, e^{-2k^2} \prod_{\gamma=1}^{\Gamma} e^{-2i\alpha \hat{H}_\gamma k}. \label{eq:para_R_int}
\end{align}
\end{subequations}
As shown by the integral form \cref{eq:para_R_int}, the LTHS synthesized $\hat{R}_{\hat{H}}(\alpha)$ operator by displacing $M$ oscillators with $\alpha/\sqrt{M}$ in parallel is identical to the LTHS synthesized operator by displacing a single oscillator with $\alpha$, given our choice of parallel gate sequences (a different choice may result numerically similar LTHS approximation for $\hat{R}_{\hat{H}}$ but more complicated analytical expression).

Using triangle inequality for integrals, we have
\begin{align}
  &\mathrel{\phantom{=}}  \norm{\hat{R}_{\hat{H}}(\alpha) - \hat{P}_{\hat{H}}(\alpha)} \notag\\
  &\leq \sqrt{\frac{2}{\pi}} \int dk\, e^{-2k^2} \norm{\prod_{\gamma=1}^{\Gamma} e^{-2i\alpha \hat{H}_\gamma k} - e^{-2i\alpha (\sum_\gamma \hat{H}_\gamma) k}} \notag\\
  &\leq \sqrt{\frac{2}{\pi}} \int dk\, e^{-2k^2} 
   2\alpha^2 k^2 \sum_{\gamma_1 = 1}^{\Gamma} \norm{\sum_{\gamma_2 = \gamma_1 + 1}^{\Gamma} \!\!\comm{H_{\gamma_2}}{H_{\gamma_1}}} \label{eq:prop9_Childs} \\
  &\leq  \frac{\alpha^2}{2} \sum_{\gamma_1 < \gamma_2} \norm{\comm{H_{\gamma_1}}{H_{\gamma_2}}}. \notag
\end{align}

In \cref{eq:prop9_Childs} we used the Proposition 9 from Ref.~\onlinecite{Childs2021}.

Finally, to obtain \cref{eq:para_R_Heis} for Heisenberg interactions, we begin with the integral form \cref{eq:para_R_int}. First, by definition, each $\hat{H}_\gamma = \sum_{\delta=1}^{\Delta_\gamma}\hat{H}_{\gamma,\delta}$ consists of mutually commuting Heisenberg coupling bonds $\hat{H}_{\gamma,\delta} = X_{j_1} X_{j_2} + Y_{j_1} Y_{j_2}  + Z_{j_1} Z_{j_2} $, where the qubit indices $(j_1, j_2)$ are uniquely determined by the indices $(\gamma, \delta)$ and $\delta \in \{1,\dots, \Delta_\gamma \}$ labels the edges of the same color indexed by $\gamma$. $\hat{H}_{\gamma,\delta}$ has the eigenvalue $+1$ ($-3$) on the local triplet (singlet) subspace $\operatorname{span}\{ \ket{00}, \ket{11}, \ket{01} + \ket{10}\}$ ($\operatorname{span}\{\ket{01} - \ket{10}\}$), resulting in the triplet (singlet) projector $\pi^0_{\gamma,\delta} = (3 + \hat{H}_{\gamma,\delta})/4$ and $\pi^1_{\gamma,\delta} = (1 - \hat{H}_{\gamma,\delta})/4$. Since the projectors are idempotent, $(\pi^0_{\gamma,\delta})^2 = \pi^0_{\gamma,\delta}$ and $(\pi^1_{\gamma,\delta})^2 = \pi^1_{\gamma,\delta}$, and orthogonal, $\pi^0_{\gamma,\delta} \pi^1_{\gamma,\delta} = 0$, the exponential factor $e^{-2i\alpha k \hat{H}_\gamma} = \prod_{\delta=1}^{\Delta_\gamma} e^{-2i\alpha k \hat{H}_{\gamma,\delta}} = \prod_{\delta=1}^{\Delta_\gamma} e^{-2i\alpha k (\pi^0_{\gamma,\delta} - 3\pi^1_{\gamma,\delta})} = \prod_{\delta=1}^{\Delta_\gamma} (e^{-2i\alpha k}\pi^0_{\gamma,\delta} + e^{6i\alpha k}\pi^1_{\gamma,\delta})$, which can then be expanded as homogeneous polynomial of degree $\Delta_\gamma$. Thus, with an explicit constant spectrum shift $\sum_\gamma \hat{H}_\gamma \to (\sum_\gamma \hat{H}_\gamma) - E_\text{s}$, \cref{eq:para_R_int} becomes
\begin{align}
  &\mathrel{\phantom{=}}
    \sqrt{\frac{2}{\pi}} \int dk\, e^{-2k^2 + 2i\alpha E_\text{s}k} \prod_{\gamma=1}^{\Gamma}
     \prod_{\delta=1}^{\Delta_\gamma}(e^{-2i\alpha k}\pi^0_{\gamma,\delta} + e^{6i\alpha k}\pi^1_{\gamma,\delta}) \notag \\
  &= \sum_{\{b_{\gamma, \delta}\} \in \{0,1\}^{|E(G)|}}
     w(\alpha; \{b_{\gamma, \delta}\})
     \prod_{\gamma=1}^{\Gamma} \prod_{\delta=1}^{\Delta_\gamma} \pi^{b_{\gamma, \delta}}_{\gamma,\delta}, \label{eq:R_Heis_proof_a}
\end{align}
where the sum runs through all $|E(G)|$ bit-variables $b_{\gamma, \delta}$ [one for each edge of Heisenberg interaction graph $G = (V, E)$] and the Gaussian weight factors
\begin{align}
  &\mathrel{\phantom{=}}
     w(\alpha; \{b_{\gamma, \delta}\}) \notag\\
  &= \sqrt{\frac{2}{\pi}} \int dk\,
     e^{-2k^2 + 2i\alpha E_\text{s}k +
     \sum_{\gamma, \delta} [6i\alpha k b_{\gamma, \delta} - 2i\alpha k (1-b_{\gamma, \delta})]} \notag \\
  &= e^{-\alpha^2 \qty[E_\text{s} + \sum_{\gamma, \delta} (4b_{\gamma, \delta} - 1)]^2 / 2}.  \label{eq:R_Heis_proof_b}
\end{align}

Considering the indices $(\gamma, \delta) \leftrightarrow (j_1, j_2)$ labeling the same set of edges with Heisenberg interactions, and the explicitly color-ordered product $\prod_\gamma$ can be indicated with color ordering operator $\operatorname{\mathcal{T}_\gamma}$, the \cref{eq:R_Heis_proof_a,eq:R_Heis_proof_b} prove the \cref{eq:para_R_Heis}.

\section{Analytical Results for 1D Chain}
\label{sec:AnalyticalResults}
For 1D chain colored as in \cref{fig:trotter_tilings}(a), we write out the specific analytical forms for $\hat{R}_{\hat{H}}(\alpha)$ given by \cref{eq:R_Heis_proof_a,eq:R_Heis_proof_b} [equivalently, \cref{eq:para_R_Heis}].

For an even $N$-site chain with periodic boundary condition, label the $N$ sites $0,1,\dots,N-1$ and color the edges $\{(2j, 2j+1)\}_{j=0}^{N/2-1}$ red and the edges $\{(2j+1, 2j+2)\}_{j=0}^{N/2-1}$ blue (index $N$ also labels the site $0$), as shown in \cref{fig:trotter_tilings}(a). Using \cref{eq:R_Heis_proof_a,eq:R_Heis_proof_b}, we find
\begin{align}
     \hat{R}_{\hat{H}}(\alpha)
  &= \sum_{ \vb{b} \in \{0,1\}^{N}}
     e^{-\alpha^2 \qty(E_\text{s} - N + 4\sum_{j=0}^{N-1} b_{j})^2 / 2}  \notag \\
  &\mathrel{\phantom{=}} \times
     \qty(\prod_{j=0}^{\frac{N}{2}-1} \pi^{b_{j+\frac{N}{2}}}_{2j+1,2j+2})
     \qty(\prod_{j=0}^{\frac{N}{2}-1} \pi^{b_{j}}_{2j,2j+1}).
  \label{eq:AR_1D_Nsites}
\end{align}
Here, the bit-variables define a vector $\vb{b} = (b_0,b_1,\dots,b_{N-1}) \in \{0,1\}^{N}$ and, equivalently, a bit string $b = \overline{b_{N-1}\cdots b_1 b_0} \in \{0,\dots, 2^N - 1\}$.

For $N=4$, 
\cref{eq:AR_1D_Nsites} becomes
\begin{align}
     \hat{R}_{\hat{H}}(\alpha)
  &= \sum_{ \vb{b} \in \{0,1\}^{4}}
     e^{-\alpha^2 \qty(E_\text{s} - 4 + 4\sum_j b_{j})^2/2}
     \pi^{b_{3}}_{3,0} \pi^{b_{2}}_{1,2}
     \pi^{b_{1}}_{2,3} \pi^{b_{0}}_{0,1}. \label{eq:R_4site_MPO}
\end{align}

Since we use the 4-site example to illustrate our algorithm, it is desirable to have more detailed analytical result in energy eigenbasis $\{\ket{E_n}\}_{n=0}^{16}$ so that the eigenstate projection action by the LTHS operator $\hat{R}_{\hat{H}'}(\alpha)$, where $\hat{H}' =  \hat{H} - E_\text{s}$, can be understood more easily with respect to the parameters $\alpha$ and $E_\text{s}$. For all degeneracies, it is sufficient to use simultaneous eigenstates for the following conserved symmetry (i.e., commuting with $\hat{H}$) operators $(\vb{S}^2, \vb{S}_z, \textsc{swap}_{1,3}$), where $\vb{S}_z = \sum_j Z_j/2$, $\vb{S}^2 = \vb{S}_x^2 + \vb{S}_y^2 + \vb{S}_z^2$, and $\textsc{swap}_{1,3} = (\hat{H}_{1,3} + \mathds{1})/2$.

\begin{table}[tb]
 \centering
  \setlength{\tabcolsep}{10pt}
  \renewcommand{\arraystretch}{1.2}
  \newcolumntype{R}{>{$}r<{$}} 
  \newcolumntype{H}{>{\color{red}$\bf}r<{$}} 
  \newcolumntype{!}{>{\global\let\currentrowstyle\relax}}
  \newcolumntype{^}{>{\currentrowstyle}}
  \newcommand{\rowstyle}[1]{\gdef\currentrowstyle{#1}#1\ignorespaces}  
\begin{tabular}{!R^R^R^R^R}
\hline \hline
 n  &E_n    &s      &s_z    &\sigma_{1,3}  \\ \hline
\rowstyle{\color{red} \bf}
 0  &-8     &0      & 0     & 1     \\
 1  &-4     &1      &-1     & 1     \\
 2  &-4     &1      & 0     & 1     \\
 3  &-4     &1      & 1     & 1     \\
\rowstyle{\color{red} \bf}
 4  & 0     &0      & 0     &-1     \\
 5  & 0     &1      &-1     &-1     \\
 6  & 0     &1      &-1     & 1     \\
 7  & 0     &1      & 0     &-1     \\
 8  & 0     &1      & 0     & 1     \\
 9  & 0     &1      & 1     &-1     \\
10  & 0     &1      & 1     & 1     \\
11  & 4     &2      &-2     & 1     \\
12  & 4     &2      &-1     & 1     \\
13  & 4     &2      & 0     & 1     \\
14  & 4     &2      & 1     & 1     \\
15  & 4     &2      & 2     & 1     \\
\hline \hline
\end{tabular}
\caption{Eigenbasis for 4-site Heisenberg chain, labeled by energies and conserved quantum numbers $(s, s_z, \sigma_{1,3})$ corresponding to total $\vb{S}$, total $\vb{S}_z$, and parity under $\textsc{swap}_{1,3}$ (symmetry with respect to the mirror axis connecting site 0 and 2).}
 \label{tab:H4_eigs}
\end{table}

In this energy eigenbasis listed in \cref{tab:H4_eigs}, \cref{eq:R_4site_MPO} becomes
\begin{align}
     \hat{R}_{\hat{H}'}(\alpha)
  ={}& \sum_{i,j \in \{0,4\}} g_{i,j}(\alpha; E_\text{s})\op{E_i}{E_j} \notag \\
     &+ g_\text{t}(\alpha; E_\text{s})\sum_{k=1}^{3} \op{E_k} \notag \\
     &+ g_\text{s}(\alpha; E_\text{s})\sum_{k=5}^{10} \op{E_k} \notag \\
     &+ g_\text{q}(\alpha; E_\text{s})\sum_{k=11}^{15} \op{E_k}.
     \label{eq:R_4site_eigen}
\end{align}
where $g_\text{t}$, $g_\text{s}$, $g_\text{q}$ are weights for the triple-, sextuple-, and quintuple-rank projectors, and the expressions for these weights are:
\begin{subequations}
\begin{align}
   g_{00}
&= \frac{-e_{0} + 6 e_{2} + 3 e_{4}}{8},\\
   g_{44}
&= \frac{3 e_{0} + 6 e_{2} - e_{4}}{8},\\
   g_{04}
&= \frac{\sqrt{3} \left(e_{0} - 2 e_{2} + e_{4}\right)}{8},\\
   g_{40}
&= - g_{04},\\
   g_\text{t}
&= e_2,\
   g_\text{s}
 = \frac{e_{1}}{2},\
   g_\text{q}
 = e_{0},
\end{align}
where
\begin{align}
    e_k &= e^{-\alpha^2 \qty(E_\text{s} - 4 + 4k)^2/2},\ k\in\{0,1,2,3\}.
\end{align}
\end{subequations}

Note that $g_{04} < 0$, $\forall \alpha \neq 0$%
~\footnote{Jensen's inequality can be used to show $g_{04} < 0$, that is, $(e_0 + e_4)/2 < e_2$ as follows: $\forall \alpha \neq 0$, $(e_0 + e_4)/2 = [e^{-\alpha^2 (E_\text{s} - 4)^2/2} + e^{-\alpha^2 (E_\text{s} + 12)^2/2}]/2 < e^{-\alpha^2 [(E_\text{s} - 4)^2 + (E_\text{s} + 12)^2]/4} \leq e^{-\alpha^2 [(E_\text{s} + 4)^2]/2} = e_2$}.%
So for any finite $\alpha$, the excited state $\ket{E_4}$ will be mixed in the projection of ground state $\ket{E_0}$ regardless of the choice of $E_\text{s}$ or $\alpha$. From the symmetry point of view, the highlighted eigenstates $\ket{E_0}$ and $\ket{E_4}$ in \cref{tab:H4_eigs} have the same spin symmetry $s=s_z=0$ despite large energy difference, so $\ket{E_4}$ is usually not fully filtered out by $\hat{R}_{\hat{H}'}(\alpha)$ operator that preserves the spin symmetry. One solution is to apply a projector $\frac{1}{2}(\textsc{swap}_{1,3} + \mathds{1}) = \frac{1}{4}(\hat{H}_{1,3} + 3)$ to select the ground state with even symmetry $\sigma_{1,3} = +1$.

However, for infinitesimal $\alpha$, it is easy to verify that
\begin{align}
  &\mathrel{\phantom{=}}  \hat{R}_{\hat{H}'}(\alpha) \notag\\
  &= \sum_{k=0}^{15}  \qty[1 -\alpha^2 \qty(E_k - E_\text{s})^2/2] \op{E_k} + \order{\alpha^4} \\
  &= \sum_{k=0}^{15}  e^{-\alpha^2 \qty(E_k - E_\text{s})^2/2} \op{E_k} + \order{\alpha^4} \\
  &= \hat{P}_{\hat{H}'}(\alpha) + \order{\alpha^4},
\end{align}
where $\hat{P}_{\hat{H}'}(\alpha)$ is the exact ground state projector for a shift $E_\text{s} = E_0$. Thus, the LTHS is always a valid ground (eigen-) state filter with controllable error for small $\alpha$.

\section{Visual Compilations for 6-site Heisenberg Chain}
\label{sec:6site}

\begin{figure}[!ht]
  \centering
  \includegraphics[scale=0.9]{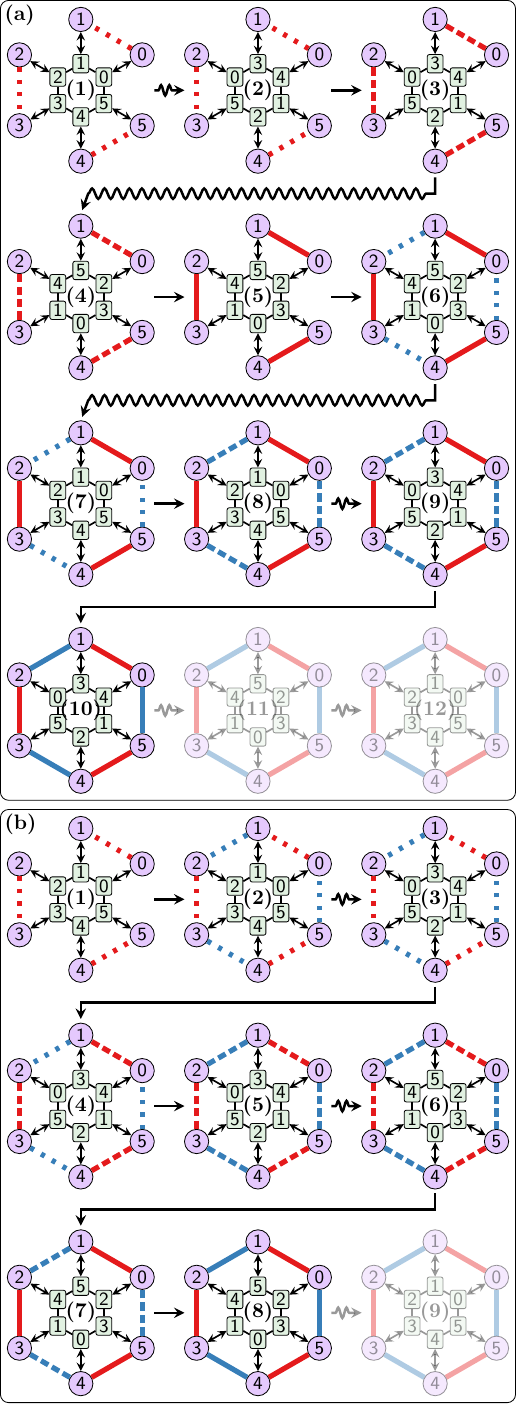}
\caption{Visual representations of the compilations for $N = 6$
site Heisenberg model. See \cref{fig:visualCircuits} caption for notations.}
  \label{fig:6siteCircuits}
\end{figure}

\clearpage


\begin{thebibliography}{82}%
\makeatletter
\providecommand \@ifxundefined [1]{%
 \@ifx{#1\undefined}
}%
\providecommand \@ifnum [1]{%
 \ifnum #1\expandafter \@firstoftwo
 \else \expandafter \@secondoftwo
 \fi
}%
\providecommand \@ifx [1]{%
 \ifx #1\expandafter \@firstoftwo
 \else \expandafter \@secondoftwo
 \fi
}%
\providecommand \natexlab [1]{#1}%
\providecommand \enquote  [1]{``#1''}%
\providecommand \bibnamefont  [1]{#1}%
\providecommand \bibfnamefont [1]{#1}%
\providecommand \citenamefont [1]{#1}%
\providecommand \href@noop [0]{\@secondoftwo}%
\providecommand \href [0]{\begingroup \@sanitize@url \@href}%
\providecommand \@href[1]{\@@startlink{#1}\@@href}%
\providecommand \@@href[1]{\endgroup#1\@@endlink}%
\providecommand \@sanitize@url [0]{\catcode `\\12\catcode `\$12\catcode `\&12\catcode `\#12\catcode `\^12\catcode `\_12\catcode `\%12\relax}%
\providecommand \@@startlink[1]{}%
\providecommand \@@endlink[0]{}%
\providecommand \url  [0]{\begingroup\@sanitize@url \@url }%
\providecommand \@url [1]{\endgroup\@href {#1}{\urlprefix }}%
\providecommand \urlprefix  [0]{URL }%
\providecommand \Eprint [0]{\href }%
\providecommand \doibase [0]{https://doi.org/}%
\providecommand \selectlanguage [0]{\@gobble}%
\providecommand \bibinfo  [0]{\@secondoftwo}%
\providecommand \bibfield  [0]{\@secondoftwo}%
\providecommand \translation [1]{[#1]}%
\providecommand \BibitemOpen [0]{}%
\providecommand \bibitemStop [0]{}%
\providecommand \bibitemNoStop [0]{.\EOS\space}%
\providecommand \EOS [0]{\spacefactor3000\relax}%
\providecommand \BibitemShut  [1]{\csname bibitem#1\endcsname}%
\let\auto@bib@innerbib\@empty
\bibitem [{\citenamefont {Gily\'{e}n}\ \emph {et~al.}(2019)\citenamefont {Gily\'{e}n}, \citenamefont {Su}, \citenamefont {Low},\ and\ \citenamefont {Wiebe}}]{Gilyen2019}%
  \BibitemOpen
  \bibfield  {author} {\bibinfo {author} {\bibfnamefont {A.}~\bibnamefont {Gily\'{e}n}}, \bibinfo {author} {\bibfnamefont {Y.}~\bibnamefont {Su}}, \bibinfo {author} {\bibfnamefont {G.~H.}\ \bibnamefont {Low}},\ and\ \bibinfo {author} {\bibfnamefont {N.}~\bibnamefont {Wiebe}},\ }\bibfield  {title} {\bibinfo {title} {Quantum singular value transformation and beyond: exponential improvements for quantum matrix arithmetics},\ }in\ \href {https://doi.org/10.1145/3313276.3316366} {\emph {\bibinfo {booktitle} {Proceedings of the 51st Annual {ACM} {SIGACT} Symposium on Theory of Computing}}},\ \bibinfo {series and number} {STOC 2019}\ (\bibinfo  {publisher} {Association for Computing Machinery},\ \bibinfo {address} {New York, NY, USA},\ \bibinfo {year} {2019})\ pp.\ \bibinfo {pages} {193--204},\ \Eprint {https://arxiv.org/abs/1806.01838} {arXiv:1806.01838 [quant-ph]} \BibitemShut {NoStop}%
\bibitem [{\citenamefont {Martyn}\ \emph {et~al.}(2021)\citenamefont {Martyn}, \citenamefont {Rossi}, \citenamefont {Tan},\ and\ \citenamefont {Chuang}}]{Martyn2021}%
  \BibitemOpen
  \bibfield  {author} {\bibinfo {author} {\bibfnamefont {J.~M.}\ \bibnamefont {Martyn}}, \bibinfo {author} {\bibfnamefont {Z.~M.}\ \bibnamefont {Rossi}}, \bibinfo {author} {\bibfnamefont {A.~K.}\ \bibnamefont {Tan}},\ and\ \bibinfo {author} {\bibfnamefont {I.~L.}\ \bibnamefont {Chuang}},\ }\bibfield  {title} {\bibinfo {title} {Grand unification of quantum algorithms},\ }\href {https://doi.org/10.1103/PRXQuantum.2.040203} {\bibfield  {journal} {\bibinfo  {journal} {PRX Quantum}\ }\textbf {\bibinfo {volume} {2}},\ \bibinfo {pages} {040203} (\bibinfo {year} {2021})}\BibitemShut {NoStop}%
\bibitem [{\citenamefont {Irmejs}\ \emph {et~al.}(2024)\citenamefont {Irmejs}, \citenamefont {Bañuls},\ and\ \citenamefont {Cirac}}]{Irmejs2024}%
  \BibitemOpen
  \bibfield  {author} {\bibinfo {author} {\bibfnamefont {R.}~\bibnamefont {Irmejs}}, \bibinfo {author} {\bibfnamefont {M.~C.}\ \bibnamefont {Bañuls}},\ and\ \bibinfo {author} {\bibfnamefont {J.~I.}\ \bibnamefont {Cirac}},\ }\bibfield  {title} {\bibinfo {title} {Efficient quantum algorithm for filtering product states},\ }\href {https://doi.org/10.22331/q-2024-06-27-1389} {\bibfield  {journal} {\bibinfo  {journal} {Quantum}\ }\textbf {\bibinfo {volume} {8}},\ \bibinfo {pages} {1389} (\bibinfo {year} {2024})}\BibitemShut {NoStop}%
\bibitem [{\citenamefont {Litteken}\ \emph {et~al.}(2023)\citenamefont {Litteken}, \citenamefont {Seifert}, \citenamefont {Chadwick}, \citenamefont {Nottingham}, \citenamefont {Roy}, \citenamefont {Li}, \citenamefont {Schuster}, \citenamefont {Chong},\ and\ \citenamefont {Baker}}]{Litteken2023}%
  \BibitemOpen
  \bibfield  {author} {\bibinfo {author} {\bibfnamefont {A.}~\bibnamefont {Litteken}}, \bibinfo {author} {\bibfnamefont {L.~M.}\ \bibnamefont {Seifert}}, \bibinfo {author} {\bibfnamefont {J.~D.}\ \bibnamefont {Chadwick}}, \bibinfo {author} {\bibfnamefont {N.}~\bibnamefont {Nottingham}}, \bibinfo {author} {\bibfnamefont {T.}~\bibnamefont {Roy}}, \bibinfo {author} {\bibfnamefont {Z.}~\bibnamefont {Li}}, \bibinfo {author} {\bibfnamefont {D.}~\bibnamefont {Schuster}}, \bibinfo {author} {\bibfnamefont {F.~T.}\ \bibnamefont {Chong}},\ and\ \bibinfo {author} {\bibfnamefont {J.~M.}\ \bibnamefont {Baker}},\ }\bibfield  {title} {\bibinfo {title} {Dancing the quantum waltz: Compiling three-qubit gates on four level architectures},\ }in\ \href {https://doi.org/10.1145/3579371.3589106} {\emph {\bibinfo {booktitle} {Proceedings of the 50th Annual International Symposium on Computer Architecture}}},\ \bibinfo {series and number} {ISCA '23}\ (\bibinfo  {publisher} {Association for Computing Machinery},\ \bibinfo {address}
  {New York, NY, USA},\ \bibinfo {year} {2023})\BibitemShut {NoStop}%
\bibitem [{\citenamefont {Menicucci}\ \emph {et~al.}(2008)\citenamefont {Menicucci}, \citenamefont {Flammia},\ and\ \citenamefont {Pfister}}]{Menicucci2008}%
  \BibitemOpen
  \bibfield  {author} {\bibinfo {author} {\bibfnamefont {N.~C.}\ \bibnamefont {Menicucci}}, \bibinfo {author} {\bibfnamefont {S.~T.}\ \bibnamefont {Flammia}},\ and\ \bibinfo {author} {\bibfnamefont {O.}~\bibnamefont {Pfister}},\ }\bibfield  {title} {\bibinfo {title} {One-way quantum computing in the optical frequency comb},\ }\href {https://doi.org/10.1103/PhysRevLett.101.130501} {\bibfield  {journal} {\bibinfo  {journal} {Phys. Rev. Lett.}\ }\textbf {\bibinfo {volume} {101}},\ \bibinfo {pages} {130501} (\bibinfo {year} {2008})}\BibitemShut {NoStop}%
\bibitem [{\citenamefont {Liu}\ \emph {et~al.}(2025)\citenamefont {Liu}, \citenamefont {Singh}, \citenamefont {Smith}, \citenamefont {Crane}, \citenamefont {Martyn}, \citenamefont {Eickbusch}, \citenamefont {Schuckert}, \citenamefont {Li}, \citenamefont {Sinanan-Singh}, \citenamefont {Soley}, \citenamefont {Tsunoda}, \citenamefont {Chuang}, \citenamefont {Wiebe},\ and\ \citenamefont {Girvin}}]{Liu2024}%
  \BibitemOpen
  \bibfield  {author} {\bibinfo {author} {\bibfnamefont {Y.}~\bibnamefont {Liu}}, \bibinfo {author} {\bibfnamefont {S.}~\bibnamefont {Singh}}, \bibinfo {author} {\bibfnamefont {K.~C.}\ \bibnamefont {Smith}}, \bibinfo {author} {\bibfnamefont {E.}~\bibnamefont {Crane}}, \bibinfo {author} {\bibfnamefont {J.~M.}\ \bibnamefont {Martyn}}, \bibinfo {author} {\bibfnamefont {A.}~\bibnamefont {Eickbusch}}, \bibinfo {author} {\bibfnamefont {A.}~\bibnamefont {Schuckert}}, \bibinfo {author} {\bibfnamefont {R.~D.}\ \bibnamefont {Li}}, \bibinfo {author} {\bibfnamefont {J.}~\bibnamefont {Sinanan-Singh}}, \bibinfo {author} {\bibfnamefont {M.~B.}\ \bibnamefont {Soley}}, \bibinfo {author} {\bibfnamefont {T.}~\bibnamefont {Tsunoda}}, \bibinfo {author} {\bibfnamefont {I.~L.}\ \bibnamefont {Chuang}}, \bibinfo {author} {\bibfnamefont {N.}~\bibnamefont {Wiebe}},\ and\ \bibinfo {author} {\bibfnamefont {S.~M.}\ \bibnamefont {Girvin}},\ }\bibfield  {title} {\bibinfo {title} {Hybrid oscillator-qubit quantum processors: Instruction set
  architectures, abstract machine models, and applications},\ }\href {https://doi.org/10.1103/4rf7-9tfx} {\bibfield  {journal} {\bibinfo  {journal} {PRX Quantum (in press)}\ ,\ } (\bibinfo {year} {2025})},\ \Eprint {https://arxiv.org/abs/2407.10381} {arXiv:2407.10381 [quant-ph]} \BibitemShut {NoStop}%
\bibitem [{\citenamefont {S\o{}rensen}\ and\ \citenamefont {M\o{}lmer}(1999)}]{sorensen1999quantum}%
  \BibitemOpen
  \bibfield  {author} {\bibinfo {author} {\bibfnamefont {A.}~\bibnamefont {S\o{}rensen}}\ and\ \bibinfo {author} {\bibfnamefont {K.}~\bibnamefont {M\o{}lmer}},\ }\bibfield  {title} {\bibinfo {title} {Quantum computation with ions in thermal motion},\ }\href {https://doi.org/10.1103/PhysRevLett.82.1971} {\bibfield  {journal} {\bibinfo  {journal} {Phys. Rev. Lett.}\ }\textbf {\bibinfo {volume} {82}},\ \bibinfo {pages} {1971} (\bibinfo {year} {1999})}\BibitemShut {NoStop}%
\bibitem [{\citenamefont {Martinez}\ \emph {et~al.}(2016)\citenamefont {Martinez}, \citenamefont {Muschik}, \citenamefont {Schindler}, \citenamefont {Nigg}, \citenamefont {Erhard}, \citenamefont {Heyl}, \citenamefont {Hauke}, \citenamefont {Dalmonte}, \citenamefont {Monz}, \citenamefont {Zoller},\ and\ \citenamefont {Blatt}}]{Martinez2016}%
  \BibitemOpen
  \bibfield  {author} {\bibinfo {author} {\bibfnamefont {E.~A.}\ \bibnamefont {Martinez}}, \bibinfo {author} {\bibfnamefont {C.~A.}\ \bibnamefont {Muschik}}, \bibinfo {author} {\bibfnamefont {P.}~\bibnamefont {Schindler}}, \bibinfo {author} {\bibfnamefont {D.}~\bibnamefont {Nigg}}, \bibinfo {author} {\bibfnamefont {A.}~\bibnamefont {Erhard}}, \bibinfo {author} {\bibfnamefont {M.}~\bibnamefont {Heyl}}, \bibinfo {author} {\bibfnamefont {P.}~\bibnamefont {Hauke}}, \bibinfo {author} {\bibfnamefont {M.}~\bibnamefont {Dalmonte}}, \bibinfo {author} {\bibfnamefont {T.}~\bibnamefont {Monz}}, \bibinfo {author} {\bibfnamefont {P.}~\bibnamefont {Zoller}},\ and\ \bibinfo {author} {\bibfnamefont {R.}~\bibnamefont {Blatt}},\ }\bibfield  {title} {\bibinfo {title} {Real-time dynamics of lattice gauge theories with a few-qubit quantum computer},\ }\href {https://doi.org/10.1038/nature18318} {\bibfield  {journal} {\bibinfo  {journal} {Nature}\ }\textbf {\bibinfo {volume} {534}},\ \bibinfo {pages} {516–519} (\bibinfo {year}
  {2016})}\BibitemShut {NoStop}%
\bibitem [{\citenamefont {Katz}\ \emph {et~al.}(2023)\citenamefont {Katz}, \citenamefont {Feng}, \citenamefont {Risinger}, \citenamefont {Monroe},\ and\ \citenamefont {Cetina}}]{katz2023demonstration}%
  \BibitemOpen
  \bibfield  {author} {\bibinfo {author} {\bibfnamefont {O.}~\bibnamefont {Katz}}, \bibinfo {author} {\bibfnamefont {L.}~\bibnamefont {Feng}}, \bibinfo {author} {\bibfnamefont {A.}~\bibnamefont {Risinger}}, \bibinfo {author} {\bibfnamefont {C.}~\bibnamefont {Monroe}},\ and\ \bibinfo {author} {\bibfnamefont {M.}~\bibnamefont {Cetina}},\ }\bibfield  {title} {\bibinfo {title} {Demonstration of three-and four-body interactions between trapped-ion spins},\ }\href@noop {} {\bibfield  {journal} {\bibinfo  {journal} {Nature Physics}\ }\textbf {\bibinfo {volume} {19}},\ \bibinfo {pages} {1452} (\bibinfo {year} {2023})}\BibitemShut {NoStop}%
\bibitem [{\citenamefont {Davoudi}\ \emph {et~al.}(2021)\citenamefont {Davoudi}, \citenamefont {Linke},\ and\ \citenamefont {Pagano}}]{Davoudi2021}%
  \BibitemOpen
  \bibfield  {author} {\bibinfo {author} {\bibfnamefont {Z.}~\bibnamefont {Davoudi}}, \bibinfo {author} {\bibfnamefont {N.~M.}\ \bibnamefont {Linke}},\ and\ \bibinfo {author} {\bibfnamefont {G.}~\bibnamefont {Pagano}},\ }\bibfield  {title} {\bibinfo {title} {Toward simulating quantum field theories with controlled phonon-ion dynamics: A hybrid analog-digital approach},\ }\bibfield  {journal} {\bibinfo  {journal} {Physical Review Research}\ }\textbf {\bibinfo {volume} {3}},\ \href {https://doi.org/10.1103/physrevresearch.3.043072} {10.1103/physrevresearch.3.043072} (\bibinfo {year} {2021})\BibitemShut {NoStop}%
\bibitem [{\citenamefont {Crane}\ \emph {et~al.}(2024)\citenamefont {Crane}, \citenamefont {Smith}, \citenamefont {Tomesh}, \citenamefont {Eickbusch}, \citenamefont {Martyn}, \citenamefont {Kühn}, \citenamefont {Funcke}, \citenamefont {DeMarco}, \citenamefont {Chuang}, \citenamefont {Wiebe}, \citenamefont {Schuckert},\ and\ \citenamefont {Girvin}}]{Crane2024}%
  \BibitemOpen
  \bibfield  {author} {\bibinfo {author} {\bibfnamefont {E.}~\bibnamefont {Crane}}, \bibinfo {author} {\bibfnamefont {K.~C.}\ \bibnamefont {Smith}}, \bibinfo {author} {\bibfnamefont {T.}~\bibnamefont {Tomesh}}, \bibinfo {author} {\bibfnamefont {A.}~\bibnamefont {Eickbusch}}, \bibinfo {author} {\bibfnamefont {J.~M.}\ \bibnamefont {Martyn}}, \bibinfo {author} {\bibfnamefont {S.}~\bibnamefont {Kühn}}, \bibinfo {author} {\bibfnamefont {L.}~\bibnamefont {Funcke}}, \bibinfo {author} {\bibfnamefont {M.~A.}\ \bibnamefont {DeMarco}}, \bibinfo {author} {\bibfnamefont {I.~L.}\ \bibnamefont {Chuang}}, \bibinfo {author} {\bibfnamefont {N.}~\bibnamefont {Wiebe}}, \bibinfo {author} {\bibfnamefont {A.}~\bibnamefont {Schuckert}},\ and\ \bibinfo {author} {\bibfnamefont {S.~M.}\ \bibnamefont {Girvin}},\ }\href@noop {} {\bibinfo {title} {Hybrid oscillator-qubit quantum processors: Simulating fermions, bosons, and gauge fields}} (\bibinfo {year} {2024}),\ \Eprint {https://arxiv.org/abs/2409.03747} {arXiv:2409.03747 [quant-ph]}
  \BibitemShut {NoStop}%
\bibitem [{\citenamefont {Liu}\ \emph {et~al.}(2016)\citenamefont {Liu}, \citenamefont {Thompson}, \citenamefont {Weedbrook}, \citenamefont {Lloyd}, \citenamefont {Vedral}, \citenamefont {Gu},\ and\ \citenamefont {Modi}}]{Liu2016e}%
  \BibitemOpen
  \bibfield  {author} {\bibinfo {author} {\bibfnamefont {N.}~\bibnamefont {Liu}}, \bibinfo {author} {\bibfnamefont {J.}~\bibnamefont {Thompson}}, \bibinfo {author} {\bibfnamefont {C.}~\bibnamefont {Weedbrook}}, \bibinfo {author} {\bibfnamefont {S.}~\bibnamefont {Lloyd}}, \bibinfo {author} {\bibfnamefont {V.}~\bibnamefont {Vedral}}, \bibinfo {author} {\bibfnamefont {M.}~\bibnamefont {Gu}},\ and\ \bibinfo {author} {\bibfnamefont {K.}~\bibnamefont {Modi}},\ }\bibfield  {title} {\bibinfo {title} {Power of one qumode for quantum computation},\ }\href {https://doi.org/10.1103/PhysRevA.93.052304} {\bibfield  {journal} {\bibinfo  {journal} {Phys. Rev. A}\ }\textbf {\bibinfo {volume} {93}},\ \bibinfo {pages} {052304} (\bibinfo {year} {2016})}\BibitemShut {NoStop}%
\bibitem [{\citenamefont {Zhang}\ \emph {et~al.}(2021)\citenamefont {Zhang}, \citenamefont {Zhang}, \citenamefont {Xue}, \citenamefont {Zhu},\ and\ \citenamefont {Wang}}]{zhang2021cvthermoQS}%
  \BibitemOpen
  \bibfield  {author} {\bibinfo {author} {\bibfnamefont {D.-B.}\ \bibnamefont {Zhang}}, \bibinfo {author} {\bibfnamefont {G.-Q.}\ \bibnamefont {Zhang}}, \bibinfo {author} {\bibfnamefont {Z.-Y.}\ \bibnamefont {Xue}}, \bibinfo {author} {\bibfnamefont {S.-L.}\ \bibnamefont {Zhu}},\ and\ \bibinfo {author} {\bibfnamefont {Z.~D.}\ \bibnamefont {Wang}},\ }\bibfield  {title} {\bibinfo {title} {Continuous-variable assisted thermal quantum simulation},\ }\href {https://doi.org/10.1103/PhysRevLett.127.020502} {\bibfield  {journal} {\bibinfo  {journal} {Phys. Rev. Lett.}\ }\textbf {\bibinfo {volume} {127}},\ \bibinfo {pages} {020502} (\bibinfo {year} {2021})}\BibitemShut {NoStop}%
\bibitem [{\citenamefont {Teoh}\ \emph {et~al.}(2023)\citenamefont {Teoh}, \citenamefont {Winkel}, \citenamefont {Babla}, \citenamefont {Chapman}, \citenamefont {Claes}, \citenamefont {de~Graaf}, \citenamefont {Garmon}, \citenamefont {Kalfus}, \citenamefont {Lu}, \citenamefont {Maiti}, \citenamefont {Sahay}, \citenamefont {Thakur}, \citenamefont {Tsunoda}, \citenamefont {Xue}, \citenamefont {Frunzio}, \citenamefont {Girvin}, \citenamefont {Puri},\ and\ \citenamefont {Schoelkopf}}]{Teoh2023}%
  \BibitemOpen
  \bibfield  {author} {\bibinfo {author} {\bibfnamefont {J.~D.}\ \bibnamefont {Teoh}}, \bibinfo {author} {\bibfnamefont {P.}~\bibnamefont {Winkel}}, \bibinfo {author} {\bibfnamefont {H.~K.}\ \bibnamefont {Babla}}, \bibinfo {author} {\bibfnamefont {B.~J.}\ \bibnamefont {Chapman}}, \bibinfo {author} {\bibfnamefont {J.}~\bibnamefont {Claes}}, \bibinfo {author} {\bibfnamefont {S.~J.}\ \bibnamefont {de~Graaf}}, \bibinfo {author} {\bibfnamefont {J.~W.~O.}\ \bibnamefont {Garmon}}, \bibinfo {author} {\bibfnamefont {W.~D.}\ \bibnamefont {Kalfus}}, \bibinfo {author} {\bibfnamefont {Y.}~\bibnamefont {Lu}}, \bibinfo {author} {\bibfnamefont {A.}~\bibnamefont {Maiti}}, \bibinfo {author} {\bibfnamefont {K.}~\bibnamefont {Sahay}}, \bibinfo {author} {\bibfnamefont {N.}~\bibnamefont {Thakur}}, \bibinfo {author} {\bibfnamefont {T.}~\bibnamefont {Tsunoda}}, \bibinfo {author} {\bibfnamefont {S.~H.}\ \bibnamefont {Xue}}, \bibinfo {author} {\bibfnamefont {L.}~\bibnamefont {Frunzio}}, \bibinfo {author} {\bibfnamefont {S.~M.}\
  \bibnamefont {Girvin}}, \bibinfo {author} {\bibfnamefont {S.}~\bibnamefont {Puri}},\ and\ \bibinfo {author} {\bibfnamefont {R.~J.}\ \bibnamefont {Schoelkopf}},\ }\bibfield  {title} {\bibinfo {title} {Dual-rail encoding with superconducting cavities},\ }\href {https://doi.org/10.1073/pnas.2221736120} {\bibfield  {journal} {\bibinfo  {journal} {Proc. Natl. Acad. Sci.}\ }\textbf {\bibinfo {volume} {120}},\ \bibinfo {pages} {e2221736120} (\bibinfo {year} {2023})}\BibitemShut {NoStop}%
\bibitem [{\citenamefont {Chou}\ \emph {et~al.}(2024)\citenamefont {Chou}, \citenamefont {Shemma}, \citenamefont {McCarrick}, \citenamefont {Chien}, \citenamefont {Teoh}, \citenamefont {Winkel}, \citenamefont {Anderson}, \citenamefont {Chen}, \citenamefont {Curtis}, \citenamefont {de~Graaf}, \citenamefont {Garmon}, \citenamefont {Gudlewski}, \citenamefont {Kalfus}, \citenamefont {Keen}, \citenamefont {Khedkar}, \citenamefont {Lei}, \citenamefont {Liu}, \citenamefont {Lu}, \citenamefont {Lu}, \citenamefont {Maiti}, \citenamefont {Mastalli-Kelly}, \citenamefont {Mehta}, \citenamefont {Mundhada}, \citenamefont {Narla}, \citenamefont {Noh}, \citenamefont {Tsunoda}, \citenamefont {Xue}, \citenamefont {Yuan}, \citenamefont {Frunzio}, \citenamefont {Aumentado}, \citenamefont {Puri}, \citenamefont {Girvin}, \citenamefont {Moseley},\ and\ \citenamefont {Schoelkopf}}]{Chou2024}%
  \BibitemOpen
  \bibfield  {author} {\bibinfo {author} {\bibfnamefont {K.~S.}\ \bibnamefont {Chou}}, \bibinfo {author} {\bibfnamefont {T.}~\bibnamefont {Shemma}}, \bibinfo {author} {\bibfnamefont {H.}~\bibnamefont {McCarrick}}, \bibinfo {author} {\bibfnamefont {T.-C.}\ \bibnamefont {Chien}}, \bibinfo {author} {\bibfnamefont {J.~D.}\ \bibnamefont {Teoh}}, \bibinfo {author} {\bibfnamefont {P.}~\bibnamefont {Winkel}}, \bibinfo {author} {\bibfnamefont {A.}~\bibnamefont {Anderson}}, \bibinfo {author} {\bibfnamefont {J.}~\bibnamefont {Chen}}, \bibinfo {author} {\bibfnamefont {J.~C.}\ \bibnamefont {Curtis}}, \bibinfo {author} {\bibfnamefont {S.~J.}\ \bibnamefont {de~Graaf}}, \bibinfo {author} {\bibfnamefont {J.~W.~O.}\ \bibnamefont {Garmon}}, \bibinfo {author} {\bibfnamefont {B.}~\bibnamefont {Gudlewski}}, \bibinfo {author} {\bibfnamefont {W.~D.}\ \bibnamefont {Kalfus}}, \bibinfo {author} {\bibfnamefont {T.}~\bibnamefont {Keen}}, \bibinfo {author} {\bibfnamefont {N.}~\bibnamefont {Khedkar}}, \bibinfo {author} {\bibfnamefont
  {C.~U.}\ \bibnamefont {Lei}}, \bibinfo {author} {\bibfnamefont {G.}~\bibnamefont {Liu}}, \bibinfo {author} {\bibfnamefont {P.}~\bibnamefont {Lu}}, \bibinfo {author} {\bibfnamefont {Y.}~\bibnamefont {Lu}}, \bibinfo {author} {\bibfnamefont {A.}~\bibnamefont {Maiti}}, \bibinfo {author} {\bibfnamefont {L.}~\bibnamefont {Mastalli-Kelly}}, \bibinfo {author} {\bibfnamefont {N.}~\bibnamefont {Mehta}}, \bibinfo {author} {\bibfnamefont {S.~O.}\ \bibnamefont {Mundhada}}, \bibinfo {author} {\bibfnamefont {A.}~\bibnamefont {Narla}}, \bibinfo {author} {\bibfnamefont {T.}~\bibnamefont {Noh}}, \bibinfo {author} {\bibfnamefont {T.}~\bibnamefont {Tsunoda}}, \bibinfo {author} {\bibfnamefont {S.~H.}\ \bibnamefont {Xue}}, \bibinfo {author} {\bibfnamefont {J.~O.}\ \bibnamefont {Yuan}}, \bibinfo {author} {\bibfnamefont {L.}~\bibnamefont {Frunzio}}, \bibinfo {author} {\bibfnamefont {J.}~\bibnamefont {Aumentado}}, \bibinfo {author} {\bibfnamefont {S.}~\bibnamefont {Puri}}, \bibinfo {author} {\bibfnamefont {S.~M.}\ \bibnamefont
  {Girvin}}, \bibinfo {author} {\bibfnamefont {S.~H.}\ \bibnamefont {Moseley}},\ and\ \bibinfo {author} {\bibfnamefont {R.~J.}\ \bibnamefont {Schoelkopf}},\ }\bibfield  {title} {\bibinfo {title} {A superconducting dual-rail cavity qubit with erasure-detected logical measurements},\ }\href {https://doi.org/10.1038/s41567-024-02539-4} {\bibfield  {journal} {\bibinfo  {journal} {Nat. Phys.}\ }\textbf {\bibinfo {volume} {20}},\ \bibinfo {pages} {1454} (\bibinfo {year} {2024})}\BibitemShut {NoStop}%
\bibitem [{\citenamefont {Eickbusch}\ \emph {et~al.}(2022)\citenamefont {Eickbusch}, \citenamefont {Sivak}, \citenamefont {Ding}, \citenamefont {Elder}, \citenamefont {Jha}, \citenamefont {Venkatraman}, \citenamefont {Royer}, \citenamefont {Girvin}, \citenamefont {Schoelkopf},\ and\ \citenamefont {Devoret}}]{EickbuschECD}%
  \BibitemOpen
  \bibfield  {author} {\bibinfo {author} {\bibfnamefont {A.}~\bibnamefont {Eickbusch}}, \bibinfo {author} {\bibfnamefont {V.}~\bibnamefont {Sivak}}, \bibinfo {author} {\bibfnamefont {A.~Z.}\ \bibnamefont {Ding}}, \bibinfo {author} {\bibfnamefont {S.~S.}\ \bibnamefont {Elder}}, \bibinfo {author} {\bibfnamefont {S.~R.}\ \bibnamefont {Jha}}, \bibinfo {author} {\bibfnamefont {J.}~\bibnamefont {Venkatraman}}, \bibinfo {author} {\bibfnamefont {B.}~\bibnamefont {Royer}}, \bibinfo {author} {\bibfnamefont {S.~M.}\ \bibnamefont {Girvin}}, \bibinfo {author} {\bibfnamefont {R.~J.}\ \bibnamefont {Schoelkopf}},\ and\ \bibinfo {author} {\bibfnamefont {M.~H.}\ \bibnamefont {Devoret}},\ }\bibfield  {title} {\bibinfo {title} {Fast universal control of an oscillator with weak dispersive coupling to a qubit},\ }\bibfield  {journal} {\bibinfo  {journal} {Nature Physics}\ }\href {https://doi.org/10.1038/s41567-022-01776-9} {10.1038/s41567-022-01776-9} (\bibinfo {year} {2022})\BibitemShut {NoStop}%
\bibitem [{\citenamefont {Wang}\ \emph {et~al.}(2020)\citenamefont {Wang}, \citenamefont {Curtis}, \citenamefont {Lester}, \citenamefont {Zhang}, \citenamefont {Gao}, \citenamefont {Freeze}, \citenamefont {Batista}, \citenamefont {Vaccaro}, \citenamefont {Chuang}, \citenamefont {Frunzio}, \citenamefont {Jiang}, \citenamefont {Girvin},\ and\ \citenamefont {Schoelkopf}}]{Wang2020FCFs}%
  \BibitemOpen
  \bibfield  {author} {\bibinfo {author} {\bibfnamefont {C.~S.}\ \bibnamefont {Wang}}, \bibinfo {author} {\bibfnamefont {J.~C.}\ \bibnamefont {Curtis}}, \bibinfo {author} {\bibfnamefont {B.~J.}\ \bibnamefont {Lester}}, \bibinfo {author} {\bibfnamefont {Y.}~\bibnamefont {Zhang}}, \bibinfo {author} {\bibfnamefont {Y.~Y.}\ \bibnamefont {Gao}}, \bibinfo {author} {\bibfnamefont {J.}~\bibnamefont {Freeze}}, \bibinfo {author} {\bibfnamefont {V.~S.}\ \bibnamefont {Batista}}, \bibinfo {author} {\bibfnamefont {P.~H.}\ \bibnamefont {Vaccaro}}, \bibinfo {author} {\bibfnamefont {I.~L.}\ \bibnamefont {Chuang}}, \bibinfo {author} {\bibfnamefont {L.}~\bibnamefont {Frunzio}}, \bibinfo {author} {\bibfnamefont {L.}~\bibnamefont {Jiang}}, \bibinfo {author} {\bibfnamefont {S.~M.}\ \bibnamefont {Girvin}},\ and\ \bibinfo {author} {\bibfnamefont {R.~J.}\ \bibnamefont {Schoelkopf}},\ }\bibfield  {title} {\bibinfo {title} {Efficient multiphoton sampling of molecular vibronic spectra on a superconducting bosonic processor},\ }\href
  {https://doi.org/10.1103/PhysRevX.10.021060} {\bibfield  {journal} {\bibinfo  {journal} {Phys. Rev. X}\ }\textbf {\bibinfo {volume} {10}},\ \bibinfo {pages} {021060} (\bibinfo {year} {2020})}\BibitemShut {NoStop}%
\bibitem [{\citenamefont {Wang}\ \emph {et~al.}(2022{\natexlab{a}})\citenamefont {Wang}, \citenamefont {Frattini}, \citenamefont {Chapman}, \citenamefont {Puri}, \citenamefont {Girvin}, \citenamefont {Devoret},\ and\ \citenamefont {Schoelkopf}}]{WangConicalIntersection}%
  \BibitemOpen
  \bibfield  {author} {\bibinfo {author} {\bibfnamefont {C.~S.}\ \bibnamefont {Wang}}, \bibinfo {author} {\bibfnamefont {N.~E.}\ \bibnamefont {Frattini}}, \bibinfo {author} {\bibfnamefont {B.~J.}\ \bibnamefont {Chapman}}, \bibinfo {author} {\bibfnamefont {S.}~\bibnamefont {Puri}}, \bibinfo {author} {\bibfnamefont {S.~M.}\ \bibnamefont {Girvin}}, \bibinfo {author} {\bibfnamefont {M.~H.}\ \bibnamefont {Devoret}},\ and\ \bibinfo {author} {\bibfnamefont {R.~J.}\ \bibnamefont {Schoelkopf}},\ }\bibfield  {title} {\bibinfo {title} {Observation of wave-packet branching through an engineered conical intersection},\ }\href@noop {} {\bibfield  {journal} {\bibinfo  {journal} {arXiv:2202.02364}\ } (\bibinfo {year} {2022}{\natexlab{a}})}\BibitemShut {NoStop}%
\bibitem [{\citenamefont {Araz}\ \emph {et~al.}(2025)\citenamefont {Araz}, \citenamefont {Grau}, \citenamefont {Montgomery},\ and\ \citenamefont {Ringer}}]{Araz2025}%
  \BibitemOpen
  \bibfield  {author} {\bibinfo {author} {\bibfnamefont {J.~Y.}\ \bibnamefont {Araz}}, \bibinfo {author} {\bibfnamefont {M.}~\bibnamefont {Grau}}, \bibinfo {author} {\bibfnamefont {J.}~\bibnamefont {Montgomery}},\ and\ \bibinfo {author} {\bibfnamefont {F.}~\bibnamefont {Ringer}},\ }\bibfield  {title} {\bibinfo {title} {Hybrid quantum simulations with qubits and qumodes on trapped-ion platforms},\ }\href {https://doi.org/10.1103/kbv4-jj51} {\bibfield  {journal} {\bibinfo  {journal} {Phys. Rev. A}\ }\textbf {\bibinfo {volume} {112}},\ \bibinfo {pages} {012620} (\bibinfo {year} {2025})}\BibitemShut {NoStop}%
\bibitem [{\citenamefont {Stavenger}\ \emph {et~al.}(2022)\citenamefont {Stavenger}, \citenamefont {Crane}, \citenamefont {Smith}, \citenamefont {Kang}, \citenamefont {Girvin},\ and\ \citenamefont {Wiebe}}]{BosonicQiskit}%
  \BibitemOpen
  \bibfield  {author} {\bibinfo {author} {\bibfnamefont {T.~J.}\ \bibnamefont {Stavenger}}, \bibinfo {author} {\bibfnamefont {E.}~\bibnamefont {Crane}}, \bibinfo {author} {\bibfnamefont {K.~C.}\ \bibnamefont {Smith}}, \bibinfo {author} {\bibfnamefont {C.~T.}\ \bibnamefont {Kang}}, \bibinfo {author} {\bibfnamefont {S.~M.}\ \bibnamefont {Girvin}},\ and\ \bibinfo {author} {\bibfnamefont {N.}~\bibnamefont {Wiebe}},\ }\bibfield  {title} {\bibinfo {title} {{C2QA} -- {B}osonic {Q}iskit},\ }in\ \href {https://doi.org/10.1109/HPEC55821.2022.9926318} {\emph {\bibinfo {booktitle} {2022 IEEE High Performance Extreme Computing Conference (HPEC)}}}\ (\bibinfo {year} {2022})\ pp.\ \bibinfo {pages} {1--8}\BibitemShut {NoStop}%
\bibitem [{Note1()}]{Note1}%
  \BibitemOpen
  \bibinfo {note} {In addition to being widely used in describing many magnetic materials, this model and its variants also find broad applications in many-body localizations, spin dynamics, and even quantitative finance~\cite {Ciceri2025}, and is thus often used as a benchmark for algorithmic advantage from quantum computing~\cite {Childs2018}.}\BibitemShut {Stop}%
\bibitem [{\citenamefont {Verstraete}\ and\ \citenamefont {Cirac}(2005)}]{Verstraete_2005}%
  \BibitemOpen
  \bibfield  {author} {\bibinfo {author} {\bibfnamefont {F.}~\bibnamefont {Verstraete}}\ and\ \bibinfo {author} {\bibfnamefont {J.~I.}\ \bibnamefont {Cirac}},\ }\bibfield  {title} {\bibinfo {title} {Mapping local hamiltonians of fermions to local hamiltonians of spins},\ }\href {https://doi.org/10.1088/1742-5468/2005/09/P09012} {\bibfield  {journal} {\bibinfo  {journal} {Journal of Statistical Mechanics: Theory and Experiment}\ }\textbf {\bibinfo {volume} {2005}},\ \bibinfo {pages} {P09012} (\bibinfo {year} {2005})}\BibitemShut {NoStop}%
\bibitem [{\citenamefont {Whitfield}\ \emph {et~al.}(2016)\citenamefont {Whitfield}, \citenamefont {Havl\'{\i}\ifmmode~\check{c}\else \v{c}\fi{}ek},\ and\ \citenamefont {Troyer}}]{Whitfield2016}%
  \BibitemOpen
  \bibfield  {author} {\bibinfo {author} {\bibfnamefont {J.~D.}\ \bibnamefont {Whitfield}}, \bibinfo {author} {\bibfnamefont {V.~c.~v.}\ \bibnamefont {Havl\'{\i}\ifmmode~\check{c}\else \v{c}\fi{}ek}},\ and\ \bibinfo {author} {\bibfnamefont {M.}~\bibnamefont {Troyer}},\ }\bibfield  {title} {\bibinfo {title} {Local spin operators for fermion simulations},\ }\href {https://doi.org/10.1103/PhysRevA.94.030301} {\bibfield  {journal} {\bibinfo  {journal} {Phys. Rev. A}\ }\textbf {\bibinfo {volume} {94}},\ \bibinfo {pages} {030301} (\bibinfo {year} {2016})}\BibitemShut {NoStop}%
\bibitem [{\citenamefont {Jiang}\ \emph {et~al.}(2019)\citenamefont {Jiang}, \citenamefont {McClean}, \citenamefont {Babbush},\ and\ \citenamefont {Neven}}]{Jiang2019}%
  \BibitemOpen
  \bibfield  {author} {\bibinfo {author} {\bibfnamefont {Z.}~\bibnamefont {Jiang}}, \bibinfo {author} {\bibfnamefont {J.}~\bibnamefont {McClean}}, \bibinfo {author} {\bibfnamefont {R.}~\bibnamefont {Babbush}},\ and\ \bibinfo {author} {\bibfnamefont {H.}~\bibnamefont {Neven}},\ }\bibfield  {title} {\bibinfo {title} {Majorana loop stabilizer codes for error mitigation in fermionic quantum simulations},\ }\href {https://doi.org/10.1103/PhysRevApplied.12.064041} {\bibfield  {journal} {\bibinfo  {journal} {Phys. Rev. Appl.}\ }\textbf {\bibinfo {volume} {12}},\ \bibinfo {pages} {064041} (\bibinfo {year} {2019})}\BibitemShut {NoStop}%
\bibitem [{\citenamefont {Derby}\ \emph {et~al.}(2021)\citenamefont {Derby}, \citenamefont {Klassen}, \citenamefont {Bausch},\ and\ \citenamefont {Cubitt}}]{Derby2021}%
  \BibitemOpen
  \bibfield  {author} {\bibinfo {author} {\bibfnamefont {C.}~\bibnamefont {Derby}}, \bibinfo {author} {\bibfnamefont {J.}~\bibnamefont {Klassen}}, \bibinfo {author} {\bibfnamefont {J.}~\bibnamefont {Bausch}},\ and\ \bibinfo {author} {\bibfnamefont {T.}~\bibnamefont {Cubitt}},\ }\bibfield  {title} {\bibinfo {title} {Compact fermion to qubit mappings},\ }\href {https://doi.org/10.1103/PhysRevB.104.035118} {\bibfield  {journal} {\bibinfo  {journal} {Phys. Rev. B}\ }\textbf {\bibinfo {volume} {104}},\ \bibinfo {pages} {035118} (\bibinfo {year} {2021})}\BibitemShut {NoStop}%
\bibitem [{\citenamefont {Heyl}\ \emph {et~al.}(2019)\citenamefont {Heyl}, \citenamefont {Hauke},\ and\ \citenamefont {Zoller}}]{Heyl2019}%
  \BibitemOpen
  \bibfield  {author} {\bibinfo {author} {\bibfnamefont {M.}~\bibnamefont {Heyl}}, \bibinfo {author} {\bibfnamefont {P.}~\bibnamefont {Hauke}},\ and\ \bibinfo {author} {\bibfnamefont {P.}~\bibnamefont {Zoller}},\ }\bibfield  {title} {\bibinfo {title} {Quantum localization bounds trotter errors in digital quantum simulation},\ }\href {https://doi.org/10.1126/sciadv.aau8342} {\bibfield  {journal} {\bibinfo  {journal} {Science Advances}\ }\textbf {\bibinfo {volume} {5}},\ \bibinfo {pages} {eaau8342} (\bibinfo {year} {2019})},\ \Eprint {https://arxiv.org/abs/https://www.science.org/doi/pdf/10.1126/sciadv.aau8342} {https://www.science.org/doi/pdf/10.1126/sciadv.aau8342} \BibitemShut {NoStop}%
\bibitem [{Note2()}]{Note2}%
  \BibitemOpen
  \bibinfo {note} {The method can be applied to degenerate ground states with some small adaptation}\BibitemShut {NoStop}%
\bibitem [{\citenamefont {Lin}\ and\ \citenamefont {Tong}(2020)}]{Lin2020}%
  \BibitemOpen
  \bibfield  {author} {\bibinfo {author} {\bibfnamefont {L.}~\bibnamefont {Lin}}\ and\ \bibinfo {author} {\bibfnamefont {Y.}~\bibnamefont {Tong}},\ }\bibfield  {title} {\bibinfo {title} {Near-optimal ground state preparation},\ }\href {https://doi.org/10.22331/q-2020-12-14-372} {\bibfield  {journal} {\bibinfo  {journal} {{Quantum}}\ }\textbf {\bibinfo {volume} {4}},\ \bibinfo {pages} {372} (\bibinfo {year} {2020})}\BibitemShut {NoStop}%
\bibitem [{\citenamefont {Camps}\ and\ \citenamefont {Van~Beeumen}(2022)}]{camps2022fable}%
  \BibitemOpen
  \bibfield  {author} {\bibinfo {author} {\bibfnamefont {D.}~\bibnamefont {Camps}}\ and\ \bibinfo {author} {\bibfnamefont {R.}~\bibnamefont {Van~Beeumen}},\ }\bibfield  {title} {\bibinfo {title} {{FABLE}: Fast approximate quantum circuits for block-encodings},\ }in\ \href {https://doi.org/10.1109/QCE53715.2022.00029} {\emph {\bibinfo {booktitle} {2022 {IEEE} International Conference on Quantum Computing and Engineering ({QCE})}}}\ (\bibinfo {year} {2022})\ pp.\ \bibinfo {pages} {104--113}\BibitemShut {NoStop}%
\bibitem [{\citenamefont {Kuklinski}\ and\ \citenamefont {Rempfer}(2024)}]{kuklinski2024s}%
  \BibitemOpen
  \bibfield  {author} {\bibinfo {author} {\bibfnamefont {P.}~\bibnamefont {Kuklinski}}\ and\ \bibinfo {author} {\bibfnamefont {B.}~\bibnamefont {Rempfer}},\ }\href@noop {} {\bibinfo {title} {{S-FABLE} and {LS-FABLE}: Fast approximate block-encoding algorithms for unstructured sparse matrices}} (\bibinfo {year} {2024}),\ \Eprint {https://arxiv.org/abs/2401.04234} {arXiv:2401.04234 [quant-ph]} \BibitemShut {NoStop}%
\bibitem [{\citenamefont {Camps}\ \emph {et~al.}(2024)\citenamefont {Camps}, \citenamefont {Lin}, \citenamefont {Van~Beeumen},\ and\ \citenamefont {Yang}}]{camps2024explicit}%
  \BibitemOpen
  \bibfield  {author} {\bibinfo {author} {\bibfnamefont {D.}~\bibnamefont {Camps}}, \bibinfo {author} {\bibfnamefont {L.}~\bibnamefont {Lin}}, \bibinfo {author} {\bibfnamefont {R.}~\bibnamefont {Van~Beeumen}},\ and\ \bibinfo {author} {\bibfnamefont {C.}~\bibnamefont {Yang}},\ }\bibfield  {title} {\bibinfo {title} {Explicit quantum circuits for block encodings of certain sparse matrices},\ }\href {https://doi.org/10.1137/22M1484298} {\bibfield  {journal} {\bibinfo  {journal} {SIAM J. Matrix Anal. Appl.}\ }\textbf {\bibinfo {volume} {45}},\ \bibinfo {pages} {801} (\bibinfo {year} {2024})}\BibitemShut {NoStop}%
\bibitem [{\citenamefont {Sinanan-Singh}\ \emph {et~al.}(2023)\citenamefont {Sinanan-Singh}, \citenamefont {Mintzer}, \citenamefont {Chuang},\ and\ \citenamefont {Liu}}]{sinanan2023single}%
  \BibitemOpen
  \bibfield  {author} {\bibinfo {author} {\bibfnamefont {J.}~\bibnamefont {Sinanan-Singh}}, \bibinfo {author} {\bibfnamefont {G.~L.}\ \bibnamefont {Mintzer}}, \bibinfo {author} {\bibfnamefont {I.~L.}\ \bibnamefont {Chuang}},\ and\ \bibinfo {author} {\bibfnamefont {Y.}~\bibnamefont {Liu}},\ }\href@noop {} {\bibinfo {title} {Single-shot quantum signal processing interferometry}} (\bibinfo {year} {2023}),\ \Eprint {https://arxiv.org/abs/2311.13703} {arXiv:2311.13703 [quant-ph]} \BibitemShut {NoStop}%
\bibitem [{\citenamefont {Rossi}\ \emph {et~al.}(2023)\citenamefont {Rossi}, \citenamefont {Bastidas}, \citenamefont {Munro},\ and\ \citenamefont {Chuang}}]{Rossi23}%
  \BibitemOpen
  \bibfield  {author} {\bibinfo {author} {\bibfnamefont {Z.~M.}\ \bibnamefont {Rossi}}, \bibinfo {author} {\bibfnamefont {V.~M.}\ \bibnamefont {Bastidas}}, \bibinfo {author} {\bibfnamefont {W.~J.}\ \bibnamefont {Munro}},\ and\ \bibinfo {author} {\bibfnamefont {I.~L.}\ \bibnamefont {Chuang}},\ }\href {https://doi.org/10.48550/ARXIV.2304.14383} {\bibinfo {title} {Quantum signal processing with continuous variables}} (\bibinfo {year} {2023})\BibitemShut {NoStop}%
\bibitem [{\citenamefont {Childs}\ and\ \citenamefont {Wiebe}(2012)}]{Childs2012}%
  \BibitemOpen
  \bibfield  {author} {\bibinfo {author} {\bibfnamefont {A.~M.}\ \bibnamefont {Childs}}\ and\ \bibinfo {author} {\bibfnamefont {N.}~\bibnamefont {Wiebe}},\ }\bibfield  {title} {\bibinfo {title} {{Hamiltonian} simulation using linear combinations of unitary operations},\ }\href {https://doi.org/10.26421/QIC12.11-12} {\bibfield  {journal} {\bibinfo  {journal} {Quantum Info. Comput.}\ }\textbf {\bibinfo {volume} {12}},\ \bibinfo {pages} {901} (\bibinfo {year} {2012})}\BibitemShut {NoStop}%
\bibitem [{\citenamefont {Dalzell}\ \emph {et~al.}(2025)\citenamefont {Dalzell}, \citenamefont {McArdle}, \citenamefont {Berta}, \citenamefont {Bienias}, \citenamefont {Chen}, \citenamefont {Gily{\'{e}}n}, \citenamefont {Hann}, \citenamefont {Kastoryano}, \citenamefont {Khabiboulline}, \citenamefont {Kubica}, \citenamefont {Salton}, \citenamefont {Wang},\ and\ \citenamefont {Brand{\~{a}}o}}]{Dalzell2025}%
  \BibitemOpen
  \bibfield  {author} {\bibinfo {author} {\bibfnamefont {A.~M.}\ \bibnamefont {Dalzell}}, \bibinfo {author} {\bibfnamefont {S.}~\bibnamefont {McArdle}}, \bibinfo {author} {\bibfnamefont {M.}~\bibnamefont {Berta}}, \bibinfo {author} {\bibfnamefont {P.}~\bibnamefont {Bienias}}, \bibinfo {author} {\bibfnamefont {C.-F.}\ \bibnamefont {Chen}}, \bibinfo {author} {\bibfnamefont {A.}~\bibnamefont {Gily{\'{e}}n}}, \bibinfo {author} {\bibfnamefont {C.~T.}\ \bibnamefont {Hann}}, \bibinfo {author} {\bibfnamefont {M.~J.}\ \bibnamefont {Kastoryano}}, \bibinfo {author} {\bibfnamefont {E.~T.}\ \bibnamefont {Khabiboulline}}, \bibinfo {author} {\bibfnamefont {A.}~\bibnamefont {Kubica}}, \bibinfo {author} {\bibfnamefont {G.}~\bibnamefont {Salton}}, \bibinfo {author} {\bibfnamefont {S.}~\bibnamefont {Wang}},\ and\ \bibinfo {author} {\bibfnamefont {F.~G. S.~L.}\ \bibnamefont {Brand{\~{a}}o}},\ }\href {https://doi.org/10.1017/9781009639651} {\emph {\bibinfo {title} {Quantum Algorithms: A Survey of Applications and End-to-end
  Complexities}}}\ (\bibinfo  {publisher} {Cambridge University Press},\ \bibinfo {address} {Cambridge},\ \bibinfo {year} {2025})\ \Eprint {https://arxiv.org/abs/2310.03011} {arXiv:2310.03011 [quant-ph]} \BibitemShut {NoStop}%
\bibitem [{\citenamefont {Keen}\ \emph {et~al.}(2021)\citenamefont {Keen}, \citenamefont {Dumitrescu},\ and\ \citenamefont {Wang}}]{Keen2021}%
  \BibitemOpen
  \bibfield  {author} {\bibinfo {author} {\bibfnamefont {T.}~\bibnamefont {Keen}}, \bibinfo {author} {\bibfnamefont {E.}~\bibnamefont {Dumitrescu}},\ and\ \bibinfo {author} {\bibfnamefont {Y.}~\bibnamefont {Wang}},\ }\href@noop {} {\bibinfo {title} {Quantum algorithms for ground-state preparation and green's function calculation}} (\bibinfo {year} {2021}),\ \Eprint {https://arxiv.org/abs/2112.05731} {arXiv:2112.05731 [quant-ph]} \BibitemShut {NoStop}%
\bibitem [{\citenamefont {Ge}\ \emph {et~al.}(2019)\citenamefont {Ge}, \citenamefont {Tura},\ and\ \citenamefont {Cirac}}]{Ge2019}%
  \BibitemOpen
  \bibfield  {author} {\bibinfo {author} {\bibfnamefont {Y.}~\bibnamefont {Ge}}, \bibinfo {author} {\bibfnamefont {J.}~\bibnamefont {Tura}},\ and\ \bibinfo {author} {\bibfnamefont {J.~I.}\ \bibnamefont {Cirac}},\ }\bibfield  {title} {\bibinfo {title} {Faster ground state preparation and high-precision ground energy estimation with fewer qubits},\ }\href {https://doi.org/10.1063/1.5027484} {\bibfield  {journal} {\bibinfo  {journal} {J. Math. Phys.}\ }\textbf {\bibinfo {volume} {60}},\ \bibinfo {pages} {022202} (\bibinfo {year} {2019})}\BibitemShut {NoStop}%
\bibitem [{\citenamefont {Motta}\ \emph {et~al.}(2020)\citenamefont {Motta}, \citenamefont {Sun}, \citenamefont {Tan}, \citenamefont {O{\textquoteright}Rourke}, \citenamefont {Ye}, \citenamefont {Minnich}, \citenamefont {Brand{\~{a}}o},\ and\ \citenamefont {Chan}}]{Motta2020}%
  \BibitemOpen
  \bibfield  {author} {\bibinfo {author} {\bibfnamefont {M.}~\bibnamefont {Motta}}, \bibinfo {author} {\bibfnamefont {C.}~\bibnamefont {Sun}}, \bibinfo {author} {\bibfnamefont {A.~T.~K.}\ \bibnamefont {Tan}}, \bibinfo {author} {\bibfnamefont {M.~J.}\ \bibnamefont {O{\textquoteright}Rourke}}, \bibinfo {author} {\bibfnamefont {E.}~\bibnamefont {Ye}}, \bibinfo {author} {\bibfnamefont {A.~J.}\ \bibnamefont {Minnich}}, \bibinfo {author} {\bibfnamefont {F.~G. S.~L.}\ \bibnamefont {Brand{\~{a}}o}},\ and\ \bibinfo {author} {\bibfnamefont {G.~K.-L.}\ \bibnamefont {Chan}},\ }\bibfield  {title} {\bibinfo {title} {Determining eigenstates and thermal states on a quantum computer using quantum imaginary time evolution},\ }\href {https://doi.org/10.1038/s41567-019-0704-4} {\bibfield  {journal} {\bibinfo  {journal} {Nat. Phys.}\ }\textbf {\bibinfo {volume} {16}},\ \bibinfo {pages} {205} (\bibinfo {year} {2020})}\BibitemShut {NoStop}%
\bibitem [{\citenamefont {Getelina}\ \emph {et~al.}(2023)\citenamefont {Getelina}, \citenamefont {Gomes}, \citenamefont {Iadecola}, \citenamefont {Orth},\ and\ \citenamefont {Yao}}]{Getelina2023}%
  \BibitemOpen
  \bibfield  {author} {\bibinfo {author} {\bibfnamefont {J.~C.}\ \bibnamefont {Getelina}}, \bibinfo {author} {\bibfnamefont {N.}~\bibnamefont {Gomes}}, \bibinfo {author} {\bibfnamefont {T.}~\bibnamefont {Iadecola}}, \bibinfo {author} {\bibfnamefont {P.~P.}\ \bibnamefont {Orth}},\ and\ \bibinfo {author} {\bibfnamefont {Y.-X.}\ \bibnamefont {Yao}},\ }\bibfield  {title} {\bibinfo {title} {Adaptive variational quantum minimally entangled typical thermal states for finite temperature simulations},\ }\bibfield  {journal} {\bibinfo  {journal} {SciPost Physics}\ }\textbf {\bibinfo {volume} {15}},\ \href {https://doi.org/10.21468/scipostphys.15.3.102} {10.21468/scipostphys.15.3.102} (\bibinfo {year} {2023})\BibitemShut {NoStop}%
\bibitem [{Note3()}]{Note3}%
  \BibitemOpen
  \bibinfo {note} {The CCR indicates that unbounded CV operators cannot reside in a finite dimensional Hilbert space~\cite {[][{, Example 1.}] Gieres2000, [][{, p95.}] Weyl1931} like qubit systems. Suppose such a trace were well defined. Then $\Tr ([\protect \hat {x}, \protect \hat {p}]) = \Tr (\protect \hat {x}\protect \hat {p}) - \Tr (\protect \hat {p}\protect \hat {x}) = 0$ by the cyclic property, which contradicts the CCR.}\BibitemShut {Stop}%
\bibitem [{Note4()}]{Note4}%
  \BibitemOpen
  \bibinfo {note} {The eigenvectors corresponding to these eigenvalues live in a rigged (meaning equipped) Hilbert space so the spectrum need not be quantized (or countable).}\BibitemShut {Stop}%
\bibitem [{Note5()}]{Note5}%
  \BibitemOpen
  \bibinfo {note} {Ref.~\protect \rev@citealp {Chen2020} being an inspirational \protect \textit {exception} to this trend}\BibitemShut {NoStop}%
\bibitem [{\citenamefont {Feynman}(1951)}]{Feynman1951}%
  \BibitemOpen
  \bibfield  {author} {\bibinfo {author} {\bibfnamefont {R.~P.}\ \bibnamefont {Feynman}},\ }\bibfield  {title} {\bibinfo {title} {An operator calculus having applications in quantum electrodynamics},\ }\href {https://doi.org/10.1103/PhysRev.84.108} {\bibfield  {journal} {\bibinfo  {journal} {Phys. Rev.}\ }\textbf {\bibinfo {volume} {84}},\ \bibinfo {pages} {108} (\bibinfo {year} {1951})}\BibitemShut {NoStop}%
\bibitem [{\citenamefont {Bespalova}\ and\ \citenamefont {Kyriienko}(2021)}]{Bespalova2021}%
  \BibitemOpen
  \bibfield  {author} {\bibinfo {author} {\bibfnamefont {T.~A.}\ \bibnamefont {Bespalova}}\ and\ \bibinfo {author} {\bibfnamefont {O.}~\bibnamefont {Kyriienko}},\ }\bibfield  {title} {\bibinfo {title} {Hamiltonian operator approximation for energy measurement and ground-state preparation},\ }\bibfield  {journal} {\bibinfo  {journal} {PRX Quantum}\ }\textbf {\bibinfo {volume} {2}},\ \href {https://doi.org/10.1103/prxquantum.2.030318} {10.1103/prxquantum.2.030318} (\bibinfo {year} {2021})\BibitemShut {NoStop}%
\bibitem [{Note6()}]{Note6}%
  \BibitemOpen
  \bibinfo {note} {Strictly speaking, the Rodrigues formula in \protect \cref {eq:photon_POVM} defines the complex It\^{o}-Hermite polynomial basis $H_{m,n}(z,z^*)$.}\BibitemShut {Stop}%
\bibitem [{\citenamefont {Low}\ and\ \citenamefont {Chuang}(2019)}]{Low2019}%
  \BibitemOpen
  \bibfield  {author} {\bibinfo {author} {\bibfnamefont {G.~H.}\ \bibnamefont {Low}}\ and\ \bibinfo {author} {\bibfnamefont {I.~L.}\ \bibnamefont {Chuang}},\ }\bibfield  {title} {\bibinfo {title} {Hamiltonian simulation by qubitization},\ }\href {https://doi.org/10.22331/q-2019-07-12-163} {\bibfield  {journal} {\bibinfo  {journal} {Quantum}\ }\textbf {\bibinfo {volume} {3}},\ \bibinfo {pages} {163} (\bibinfo {year} {2019})}\BibitemShut {NoStop}%
\bibitem [{\citenamefont {Chabaud}\ \emph {et~al.}(2019)\citenamefont {Chabaud}, \citenamefont {Douce}, \citenamefont {Grosshans}, \citenamefont {Kashefi},\ and\ \citenamefont {Markham}}]{Chabaud2019}%
  \BibitemOpen
  \bibfield  {author} {\bibinfo {author} {\bibfnamefont {U.}~\bibnamefont {Chabaud}}, \bibinfo {author} {\bibfnamefont {T.}~\bibnamefont {Douce}}, \bibinfo {author} {\bibfnamefont {F.}~\bibnamefont {Grosshans}}, \bibinfo {author} {\bibfnamefont {E.}~\bibnamefont {Kashefi}},\ and\ \bibinfo {author} {\bibfnamefont {D.}~\bibnamefont {Markham}},\ }\href@noop {} {\bibinfo {title} {Building trust for continuous variable quantum states}} (\bibinfo {year} {2019}),\ \Eprint {https://arxiv.org/abs/1905.12700} {arXiv:1905.12700 [quant-ph]} \BibitemShut {NoStop}%
\bibitem [{\citenamefont {Jackson}(2022)}]{Jackson2022}%
  \BibitemOpen
  \bibfield  {author} {\bibinfo {author} {\bibfnamefont {C.~S.}\ \bibnamefont {Jackson}},\ }\href@noop {} {\bibinfo {title} {The photodetector, the heterodyne instrument, and the principle of instrument autonomy}} (\bibinfo {year} {2022}),\ \Eprint {https://arxiv.org/abs/2210.11100} {arXiv:2210.11100 [quant-ph]} \BibitemShut {NoStop}%
\bibitem [{\citenamefont {Jackson}\ and\ \citenamefont {Caves}(2023)}]{Jackson2023}%
  \BibitemOpen
  \bibfield  {author} {\bibinfo {author} {\bibfnamefont {C.~S.}\ \bibnamefont {Jackson}}\ and\ \bibinfo {author} {\bibfnamefont {C.~M.}\ \bibnamefont {Caves}},\ }\bibfield  {title} {\bibinfo {title} {How to perform the coherent measurement of a curved phase space by continuous isotropic measurement. {I}. {S}pin and the {K}raus-operator geometry of {$\mathrm{SL}(2,\mathbb{C})$}},\ }\href {https://doi.org/10.22331/q-2023-08-16-1085} {\bibfield  {journal} {\bibinfo  {journal} {Quantum}\ }\textbf {\bibinfo {volume} {7}},\ \bibinfo {pages} {1085} (\bibinfo {year} {2023})}\BibitemShut {NoStop}%
\bibitem [{\citenamefont {Chen}\ \emph {et~al.}(2025)\citenamefont {Chen}, \citenamefont {Li}, \citenamefont {Guo}, \citenamefont {Chen}, \citenamefont {Li}, \citenamefont {Bierman}, \citenamefont {Huang}, \citenamefont {Zhou}, \citenamefont {Liu},\ and\ \citenamefont {Zhang}}]{chen2025genesis}%
  \BibitemOpen
  \bibfield  {author} {\bibinfo {author} {\bibfnamefont {Z.}~\bibnamefont {Chen}}, \bibinfo {author} {\bibfnamefont {J.}~\bibnamefont {Li}}, \bibinfo {author} {\bibfnamefont {M.}~\bibnamefont {Guo}}, \bibinfo {author} {\bibfnamefont {H.}~\bibnamefont {Chen}}, \bibinfo {author} {\bibfnamefont {Z.}~\bibnamefont {Li}}, \bibinfo {author} {\bibfnamefont {J.}~\bibnamefont {Bierman}}, \bibinfo {author} {\bibfnamefont {Y.}~\bibnamefont {Huang}}, \bibinfo {author} {\bibfnamefont {H.}~\bibnamefont {Zhou}}, \bibinfo {author} {\bibfnamefont {Y.}~\bibnamefont {Liu}},\ and\ \bibinfo {author} {\bibfnamefont {E.~Z.}\ \bibnamefont {Zhang}},\ }\bibfield  {title} {\bibinfo {title} {Genesis: A compiler for hamiltonian simulation on hybrid cv-dv quantum computers},\ }in\ \href@noop {} {\emph {\bibinfo {booktitle} {Proceedings of the 52nd Annual International Symposium on Computer Architecture}}}\ (\bibinfo {year} {2025})\ pp.\ \bibinfo {pages} {1583--1597}\BibitemShut {NoStop}%
\bibitem [{Note7()}]{Note7}%
  \BibitemOpen
  \bibinfo {note} {This SWAP definition follows Eq.~(188) of Ref.~\cite {Liu2024}. Alternatively, following Eq.~(286) of Ref.~\cite {Liu2024} we can define $\protect \text {SWAP}^{(j,k)} \mathrel {\mathop :}\mathrel {\mkern -1.2mu}=\protect \text {BS}^{(j, k)}(\pi , \protect \frac {\pi }{2}) \Pi ^{(k)}$ where the photon number parity gate $\Pi ^{(k)} = [F^{(k)}]^2 = (-1)^{\protect \hat {n}_k}$, which is less symmetric with respect to index $j \leftrightarrow k$.}\BibitemShut {Stop}%
\bibitem [{\citenamefont {Childs}\ \emph {et~al.}(2021)\citenamefont {Childs}, \citenamefont {Su}, \citenamefont {Tran}, \citenamefont {Wiebe},\ and\ \citenamefont {Zhu}}]{Childs2021}%
  \BibitemOpen
  \bibfield  {author} {\bibinfo {author} {\bibfnamefont {A.~M.}\ \bibnamefont {Childs}}, \bibinfo {author} {\bibfnamefont {Y.}~\bibnamefont {Su}}, \bibinfo {author} {\bibfnamefont {M.~C.}\ \bibnamefont {Tran}}, \bibinfo {author} {\bibfnamefont {N.}~\bibnamefont {Wiebe}},\ and\ \bibinfo {author} {\bibfnamefont {S.}~\bibnamefont {Zhu}},\ }\bibfield  {title} {\bibinfo {title} {Theory of {T}rotter error with commutator scaling},\ }\href {https://doi.org/10.1103/PhysRevX.11.011020} {\bibfield  {journal} {\bibinfo  {journal} {Phys. Rev. X}\ }\textbf {\bibinfo {volume} {11}},\ \bibinfo {pages} {011020} (\bibinfo {year} {2021})}\BibitemShut {NoStop}%
\bibitem [{\citenamefont {Kang}\ \emph {et~al.}(2023)\citenamefont {Kang}, \citenamefont {Soley}, \citenamefont {Crane}, \citenamefont {Girvin},\ and\ \citenamefont {Wiebe}}]{kang2023leveraging}%
  \BibitemOpen
  \bibfield  {author} {\bibinfo {author} {\bibfnamefont {C.}~\bibnamefont {Kang}}, \bibinfo {author} {\bibfnamefont {M.~B.}\ \bibnamefont {Soley}}, \bibinfo {author} {\bibfnamefont {E.}~\bibnamefont {Crane}}, \bibinfo {author} {\bibfnamefont {S.}~\bibnamefont {Girvin}},\ and\ \bibinfo {author} {\bibfnamefont {N.}~\bibnamefont {Wiebe}},\ }\bibfield  {title} {\bibinfo {title} {Leveraging hamiltonian simulation techniques to compile operations on bosonic devices},\ }\href@noop {} {\bibfield  {journal} {\bibinfo  {journal} {arXiv preprint arXiv:2303.15542}\ } (\bibinfo {year} {2023})}\BibitemShut {NoStop}%
\bibitem [{\citenamefont {Ding}\ \emph {et~al.}(2025)\citenamefont {Ding}, \citenamefont {Zhan}, \citenamefont {Preskill},\ and\ \citenamefont {Lin}}]{endtoend}%
  \BibitemOpen
  \bibfield  {author} {\bibinfo {author} {\bibfnamefont {Z.}~\bibnamefont {Ding}}, \bibinfo {author} {\bibfnamefont {Y.}~\bibnamefont {Zhan}}, \bibinfo {author} {\bibfnamefont {J.}~\bibnamefont {Preskill}},\ and\ \bibinfo {author} {\bibfnamefont {L.}~\bibnamefont {Lin}},\ }\href@noop {} {\bibinfo {title} {End-to-end efficient quantum thermal and ground state preparation made simple}} (\bibinfo {year} {2025}),\ \Eprint {https://arxiv.org/abs/2508.05703} {arXiv:2508.05703 [quant-ph]} \BibitemShut {NoStop}%
\bibitem [{\citenamefont {Carrera~Vazquez}\ \emph {et~al.}(2023)\citenamefont {Carrera~Vazquez}, \citenamefont {Egger}, \citenamefont {Ochsner},\ and\ \citenamefont {Woerner}}]{vazquez2023well}%
  \BibitemOpen
  \bibfield  {author} {\bibinfo {author} {\bibfnamefont {A.}~\bibnamefont {Carrera~Vazquez}}, \bibinfo {author} {\bibfnamefont {D.~J.}\ \bibnamefont {Egger}}, \bibinfo {author} {\bibfnamefont {D.}~\bibnamefont {Ochsner}},\ and\ \bibinfo {author} {\bibfnamefont {S.}~\bibnamefont {Woerner}},\ }\bibfield  {title} {\bibinfo {title} {Well-conditioned multi-product formulas for hardware-friendly {H}amiltonian simulation},\ }\href {https://doi.org/10.22331/q-2023-07-25-1067} {\bibfield  {journal} {\bibinfo  {journal} {{Quantum}}\ }\textbf {\bibinfo {volume} {7}},\ \bibinfo {pages} {1067} (\bibinfo {year} {2023})}\BibitemShut {NoStop}%
\bibitem [{\citenamefont {Faehrmann}\ \emph {et~al.}(2022)\citenamefont {Faehrmann}, \citenamefont {Steudtner}, \citenamefont {Kueng}, \citenamefont {Kieferova},\ and\ \citenamefont {Eisert}}]{Faehrmann2022}%
  \BibitemOpen
  \bibfield  {author} {\bibinfo {author} {\bibfnamefont {P.~K.}\ \bibnamefont {Faehrmann}}, \bibinfo {author} {\bibfnamefont {M.}~\bibnamefont {Steudtner}}, \bibinfo {author} {\bibfnamefont {R.}~\bibnamefont {Kueng}}, \bibinfo {author} {\bibfnamefont {M.}~\bibnamefont {Kieferova}},\ and\ \bibinfo {author} {\bibfnamefont {J.}~\bibnamefont {Eisert}},\ }\bibfield  {title} {\bibinfo {title} {Randomizing multi-product formulas for {H}amiltonian simulation},\ }\href {https://doi.org/10.22331/q-2022-09-19-806} {\bibfield  {journal} {\bibinfo  {journal} {{Quantum}}\ }\textbf {\bibinfo {volume} {6}},\ \bibinfo {pages} {806} (\bibinfo {year} {2022})}\BibitemShut {NoStop}%
\bibitem [{\citenamefont {Zhuk}\ \emph {et~al.}(2023)\citenamefont {Zhuk}, \citenamefont {Robertson},\ and\ \citenamefont {Bravyi}}]{zhuk2023trotter}%
  \BibitemOpen
  \bibfield  {author} {\bibinfo {author} {\bibfnamefont {S.}~\bibnamefont {Zhuk}}, \bibinfo {author} {\bibfnamefont {N.}~\bibnamefont {Robertson}},\ and\ \bibinfo {author} {\bibfnamefont {S.}~\bibnamefont {Bravyi}},\ }\href@noop {} {\bibinfo {title} {{T}rotter error bounds and dynamic multi-product formulas for {H}amiltonian simulation}} (\bibinfo {year} {2023}),\ \Eprint {https://arxiv.org/abs/2306.12569} {arXiv:2306.12569 [quant-ph]} \BibitemShut {NoStop}%
\bibitem [{Note8()}]{Note8}%
  \BibitemOpen
  \bibinfo {note} {More explicitly, $\protect \hat {R}_{\alpha }$ is block diagonal is the simultaneous eigenbasis of $\protect \hat {H}_4$ and its mutually commuting symmetry operators $ \{ \protect \hat {S}_{\protect \text {tot}}^2, \protect \hat {S}_Z, \protect \hat {\sigma }_{24} \}$ where $\protect \hat {\sigma }_{24} = \protect \frac {1}{2} \left (\protect \mathds {1} + X_2 X_4 + Y_2 Y_4 + Z_2 Z_4 \right )$ is a swap operator between sites 2 and 4}\BibitemShut {NoStop}%
\bibitem [{\citenamefont {Michael}\ \emph {et~al.}(2016)\citenamefont {Michael}, \citenamefont {Silveri}, \citenamefont {Brierley}, \citenamefont {Albert}, \citenamefont {Salmilehto}, \citenamefont {Jiang},\ and\ \citenamefont {Girvin}}]{BinomialCodes}%
  \BibitemOpen
  \bibfield  {author} {\bibinfo {author} {\bibfnamefont {M.~H.}\ \bibnamefont {Michael}}, \bibinfo {author} {\bibfnamefont {M.}~\bibnamefont {Silveri}}, \bibinfo {author} {\bibfnamefont {R.~T.}\ \bibnamefont {Brierley}}, \bibinfo {author} {\bibfnamefont {V.~V.}\ \bibnamefont {Albert}}, \bibinfo {author} {\bibfnamefont {J.}~\bibnamefont {Salmilehto}}, \bibinfo {author} {\bibfnamefont {L.}~\bibnamefont {Jiang}},\ and\ \bibinfo {author} {\bibfnamefont {S.~M.}\ \bibnamefont {Girvin}},\ }\bibfield  {title} {\bibinfo {title} {New class of quantum error-correcting codes for a bosonic mode},\ }\href {https://doi.org/10.1103/PhysRevX.6.031006} {\bibfield  {journal} {\bibinfo  {journal} {Phys. Rev. X}\ }\textbf {\bibinfo {volume} {6}},\ \bibinfo {pages} {031006} (\bibinfo {year} {2016})}\BibitemShut {NoStop}%
\bibitem [{\citenamefont {Chapman}\ \emph {et~al.}(2023)\citenamefont {Chapman}, \citenamefont {de~Graaf}, \citenamefont {Xue}, \citenamefont {Zhang}, \citenamefont {Teoh}, \citenamefont {Curtis}, \citenamefont {Tsunoda}, \citenamefont {Eickbusch}, \citenamefont {Read}, \citenamefont {Koottandavida}, \citenamefont {Mundhada}, \citenamefont {Frunzio}, \citenamefont {Devoret}, \citenamefont {Girvin},\ and\ \citenamefont {Schoelkopf}}]{chapman2022high}%
  \BibitemOpen
  \bibfield  {author} {\bibinfo {author} {\bibfnamefont {B.~J.}\ \bibnamefont {Chapman}}, \bibinfo {author} {\bibfnamefont {S.~J.}\ \bibnamefont {de~Graaf}}, \bibinfo {author} {\bibfnamefont {S.~H.}\ \bibnamefont {Xue}}, \bibinfo {author} {\bibfnamefont {Y.}~\bibnamefont {Zhang}}, \bibinfo {author} {\bibfnamefont {J.}~\bibnamefont {Teoh}}, \bibinfo {author} {\bibfnamefont {J.~C.}\ \bibnamefont {Curtis}}, \bibinfo {author} {\bibfnamefont {T.}~\bibnamefont {Tsunoda}}, \bibinfo {author} {\bibfnamefont {A.}~\bibnamefont {Eickbusch}}, \bibinfo {author} {\bibfnamefont {A.~P.}\ \bibnamefont {Read}}, \bibinfo {author} {\bibfnamefont {A.}~\bibnamefont {Koottandavida}}, \bibinfo {author} {\bibfnamefont {S.~O.}\ \bibnamefont {Mundhada}}, \bibinfo {author} {\bibfnamefont {L.}~\bibnamefont {Frunzio}}, \bibinfo {author} {\bibfnamefont {M.~H.}\ \bibnamefont {Devoret}}, \bibinfo {author} {\bibfnamefont {S.~M.}\ \bibnamefont {Girvin}},\ and\ \bibinfo {author} {\bibfnamefont {R.~J.}\ \bibnamefont {Schoelkopf}},\ }\bibfield
  {title} {\bibinfo {title} {High-on-off-ratio beam-splitter interaction for gates on bosonically encoded qubits},\ }\href {https://doi.org/10.1103/PRXQuantum.4.020355} {\bibfield  {journal} {\bibinfo  {journal} {PRX Quantum}\ }\textbf {\bibinfo {volume} {4}},\ \bibinfo {pages} {020355} (\bibinfo {year} {2023})}\BibitemShut {NoStop}%
\bibitem [{\citenamefont {Lu}\ \emph {et~al.}(2023)\citenamefont {Lu}, \citenamefont {Maiti}, \citenamefont {Garmon}, \citenamefont {Ganjam}, \citenamefont {Zhang}, \citenamefont {Claes}, \citenamefont {Frunzio}, \citenamefont {Girvin},\ and\ \citenamefont {Schoelkopf}}]{lu2023highfidelity}%
  \BibitemOpen
  \bibfield  {author} {\bibinfo {author} {\bibfnamefont {Y.}~\bibnamefont {Lu}}, \bibinfo {author} {\bibfnamefont {A.}~\bibnamefont {Maiti}}, \bibinfo {author} {\bibfnamefont {J.~W.~O.}\ \bibnamefont {Garmon}}, \bibinfo {author} {\bibfnamefont {S.}~\bibnamefont {Ganjam}}, \bibinfo {author} {\bibfnamefont {Y.}~\bibnamefont {Zhang}}, \bibinfo {author} {\bibfnamefont {J.}~\bibnamefont {Claes}}, \bibinfo {author} {\bibfnamefont {L.}~\bibnamefont {Frunzio}}, \bibinfo {author} {\bibfnamefont {S.~M.}\ \bibnamefont {Girvin}},\ and\ \bibinfo {author} {\bibfnamefont {R.~J.}\ \bibnamefont {Schoelkopf}},\ }\bibfield  {title} {\bibinfo {title} {High-fidelity parametric beamsplitting with a parity-protected converter},\ }\href {https://doi.org/10.1038/s41467-023-41104-0} {\bibfield  {journal} {\bibinfo  {journal} {Nature Communications}\ }\textbf {\bibinfo {volume} {14}},\ \bibinfo {pages} {5767} (\bibinfo {year} {2023})}\BibitemShut {NoStop}%
\bibitem [{\citenamefont {Sun}\ \emph {et~al.}(2014)\citenamefont {Sun}, \citenamefont {Petrenko}, \citenamefont {Leghtas}, \citenamefont {Vlastakis}, \citenamefont {Kirchmair}, \citenamefont {Sliwa}, \citenamefont {Narla}, \citenamefont {Hatridge}, \citenamefont {Shankar}, \citenamefont {Blumoff}, \citenamefont {Frunzio}, \citenamefont {Mirrahimi}, \citenamefont {Devoret},\ and\ \citenamefont {Schoelkopf}}]{Sun2014}%
  \BibitemOpen
  \bibfield  {author} {\bibinfo {author} {\bibfnamefont {L.}~\bibnamefont {Sun}}, \bibinfo {author} {\bibfnamefont {A.}~\bibnamefont {Petrenko}}, \bibinfo {author} {\bibfnamefont {Z.}~\bibnamefont {Leghtas}}, \bibinfo {author} {\bibfnamefont {B.}~\bibnamefont {Vlastakis}}, \bibinfo {author} {\bibfnamefont {G.}~\bibnamefont {Kirchmair}}, \bibinfo {author} {\bibfnamefont {K.~M.}\ \bibnamefont {Sliwa}}, \bibinfo {author} {\bibfnamefont {A.}~\bibnamefont {Narla}}, \bibinfo {author} {\bibfnamefont {M.}~\bibnamefont {Hatridge}}, \bibinfo {author} {\bibfnamefont {S.}~\bibnamefont {Shankar}}, \bibinfo {author} {\bibfnamefont {J.}~\bibnamefont {Blumoff}}, \bibinfo {author} {\bibfnamefont {L.}~\bibnamefont {Frunzio}}, \bibinfo {author} {\bibfnamefont {M.}~\bibnamefont {Mirrahimi}}, \bibinfo {author} {\bibfnamefont {M.~H.}\ \bibnamefont {Devoret}},\ and\ \bibinfo {author} {\bibfnamefont {R.~J.}\ \bibnamefont {Schoelkopf}},\ }\bibfield  {title} {\bibinfo {title} {Tracking photon jumps with repeated quantum non-demolition
  parity measurements},\ }\href {https://doi.org/doi:10.1038/nature13436} {\bibfield  {journal} {\bibinfo  {journal} {Nature}\ }\textbf {\bibinfo {volume} {511}},\ \bibinfo {pages} {444} (\bibinfo {year} {2014})}\BibitemShut {NoStop}%
\bibitem [{\citenamefont {de~Graaf}\ \emph {et~al.}(2024)\citenamefont {de~Graaf}, \citenamefont {Xue}, \citenamefont {Chapman}, \citenamefont {Teoh}, \citenamefont {Tsunoda}, \citenamefont {Winkel}, \citenamefont {Garmon}, \citenamefont {Chang}, \citenamefont {Frunzio}, \citenamefont {Puri} \emph {et~al.}}]{deGraaf2024midcircuit}%
  \BibitemOpen
  \bibfield  {author} {\bibinfo {author} {\bibfnamefont {S.~J.}\ \bibnamefont {de~Graaf}}, \bibinfo {author} {\bibfnamefont {S.~H.}\ \bibnamefont {Xue}}, \bibinfo {author} {\bibfnamefont {B.~J.}\ \bibnamefont {Chapman}}, \bibinfo {author} {\bibfnamefont {J.~D.}\ \bibnamefont {Teoh}}, \bibinfo {author} {\bibfnamefont {T.}~\bibnamefont {Tsunoda}}, \bibinfo {author} {\bibfnamefont {P.}~\bibnamefont {Winkel}}, \bibinfo {author} {\bibfnamefont {J.~W.}\ \bibnamefont {Garmon}}, \bibinfo {author} {\bibfnamefont {K.~M.}\ \bibnamefont {Chang}}, \bibinfo {author} {\bibfnamefont {L.}~\bibnamefont {Frunzio}}, \bibinfo {author} {\bibfnamefont {S.}~\bibnamefont {Puri}}, \emph {et~al.},\ }\bibfield  {title} {\bibinfo {title} {A mid-circuit erasure check on a dual-rail cavity qubit using the joint-photon number-splitting regime of circuit qed},\ }\href {https://doi.org/10.48550/arXiv.2406.14621} {\bibfield  {journal} {\bibinfo  {journal} {arXiv:2406.14621}\ } (\bibinfo {year} {2024})}\BibitemShut {NoStop}%
\bibitem [{\citenamefont {Place}\ \emph {et~al.}(2021)\citenamefont {Place}, \citenamefont {Rodgers}, \citenamefont {Mundada}, \citenamefont {Smitham}, \citenamefont {Fitzpatrick}, \citenamefont {Leng}, \citenamefont {Premkumar}, \citenamefont {Bryon}, \citenamefont {Vrajitoarea}, \citenamefont {Sussman}, \citenamefont {Cheng}, \citenamefont {Madhavan}, \citenamefont {Babla}, \citenamefont {Le}, \citenamefont {Gang}, \citenamefont {Jäck}, \citenamefont {Gyenis}, \citenamefont {Yao}, \citenamefont {Cava}, \citenamefont {de~Leon},\ and\ \citenamefont {Houck}}]{place_new_2021}%
  \BibitemOpen
  \bibfield  {author} {\bibinfo {author} {\bibfnamefont {A.~P.~M.}\ \bibnamefont {Place}}, \bibinfo {author} {\bibfnamefont {L.~V.~H.}\ \bibnamefont {Rodgers}}, \bibinfo {author} {\bibfnamefont {P.}~\bibnamefont {Mundada}}, \bibinfo {author} {\bibfnamefont {B.~M.}\ \bibnamefont {Smitham}}, \bibinfo {author} {\bibfnamefont {M.}~\bibnamefont {Fitzpatrick}}, \bibinfo {author} {\bibfnamefont {Z.}~\bibnamefont {Leng}}, \bibinfo {author} {\bibfnamefont {A.}~\bibnamefont {Premkumar}}, \bibinfo {author} {\bibfnamefont {J.}~\bibnamefont {Bryon}}, \bibinfo {author} {\bibfnamefont {A.}~\bibnamefont {Vrajitoarea}}, \bibinfo {author} {\bibfnamefont {S.}~\bibnamefont {Sussman}}, \bibinfo {author} {\bibfnamefont {G.}~\bibnamefont {Cheng}}, \bibinfo {author} {\bibfnamefont {T.}~\bibnamefont {Madhavan}}, \bibinfo {author} {\bibfnamefont {H.~K.}\ \bibnamefont {Babla}}, \bibinfo {author} {\bibfnamefont {X.~H.}\ \bibnamefont {Le}}, \bibinfo {author} {\bibfnamefont {Y.}~\bibnamefont {Gang}}, \bibinfo {author} {\bibfnamefont
  {B.}~\bibnamefont {Jäck}}, \bibinfo {author} {\bibfnamefont {A.}~\bibnamefont {Gyenis}}, \bibinfo {author} {\bibfnamefont {N.}~\bibnamefont {Yao}}, \bibinfo {author} {\bibfnamefont {R.~J.}\ \bibnamefont {Cava}}, \bibinfo {author} {\bibfnamefont {N.~P.}\ \bibnamefont {de~Leon}},\ and\ \bibinfo {author} {\bibfnamefont {A.~A.}\ \bibnamefont {Houck}},\ }\bibfield  {title} {\bibinfo {title} {New material platform for superconducting transmon qubits with coherence times exceeding 0.3 milliseconds},\ }\href {https://doi.org/10.1038/s41467-021-22030-5} {\bibfield  {journal} {\bibinfo  {journal} {Nature Communications}\ }\textbf {\bibinfo {volume} {12}},\ \bibinfo {pages} {1779} (\bibinfo {year} {2021})}\BibitemShut {NoStop}%
\bibitem [{\citenamefont {Wang}\ \emph {et~al.}(2022{\natexlab{b}})\citenamefont {Wang}, \citenamefont {Li}, \citenamefont {Xu}, \citenamefont {Li}, \citenamefont {Wang}, \citenamefont {Yang}, \citenamefont {Mi}, \citenamefont {Liang}, \citenamefont {Su}, \citenamefont {Yang}, \citenamefont {Wang}, \citenamefont {Wang}, \citenamefont {Li}, \citenamefont {Chen}, \citenamefont {Li}, \citenamefont {Linghu}, \citenamefont {Han}, \citenamefont {Zhang}, \citenamefont {Feng}, \citenamefont {Song}, \citenamefont {Ma}, \citenamefont {Zhang}, \citenamefont {Wang}, \citenamefont {Zhao}, \citenamefont {Liu}, \citenamefont {Xue}, \citenamefont {Jin},\ and\ \citenamefont {Yu}}]{wang_towards_2022}%
  \BibitemOpen
  \bibfield  {author} {\bibinfo {author} {\bibfnamefont {C.}~\bibnamefont {Wang}}, \bibinfo {author} {\bibfnamefont {X.}~\bibnamefont {Li}}, \bibinfo {author} {\bibfnamefont {H.}~\bibnamefont {Xu}}, \bibinfo {author} {\bibfnamefont {Z.}~\bibnamefont {Li}}, \bibinfo {author} {\bibfnamefont {J.}~\bibnamefont {Wang}}, \bibinfo {author} {\bibfnamefont {Z.}~\bibnamefont {Yang}}, \bibinfo {author} {\bibfnamefont {Z.}~\bibnamefont {Mi}}, \bibinfo {author} {\bibfnamefont {X.}~\bibnamefont {Liang}}, \bibinfo {author} {\bibfnamefont {T.}~\bibnamefont {Su}}, \bibinfo {author} {\bibfnamefont {C.}~\bibnamefont {Yang}}, \bibinfo {author} {\bibfnamefont {G.}~\bibnamefont {Wang}}, \bibinfo {author} {\bibfnamefont {W.}~\bibnamefont {Wang}}, \bibinfo {author} {\bibfnamefont {Y.}~\bibnamefont {Li}}, \bibinfo {author} {\bibfnamefont {M.}~\bibnamefont {Chen}}, \bibinfo {author} {\bibfnamefont {C.}~\bibnamefont {Li}}, \bibinfo {author} {\bibfnamefont {K.}~\bibnamefont {Linghu}}, \bibinfo {author} {\bibfnamefont {J.}~\bibnamefont
  {Han}}, \bibinfo {author} {\bibfnamefont {Y.}~\bibnamefont {Zhang}}, \bibinfo {author} {\bibfnamefont {Y.}~\bibnamefont {Feng}}, \bibinfo {author} {\bibfnamefont {Y.}~\bibnamefont {Song}}, \bibinfo {author} {\bibfnamefont {T.}~\bibnamefont {Ma}}, \bibinfo {author} {\bibfnamefont {J.}~\bibnamefont {Zhang}}, \bibinfo {author} {\bibfnamefont {R.}~\bibnamefont {Wang}}, \bibinfo {author} {\bibfnamefont {P.}~\bibnamefont {Zhao}}, \bibinfo {author} {\bibfnamefont {W.}~\bibnamefont {Liu}}, \bibinfo {author} {\bibfnamefont {G.}~\bibnamefont {Xue}}, \bibinfo {author} {\bibfnamefont {Y.}~\bibnamefont {Jin}},\ and\ \bibinfo {author} {\bibfnamefont {H.}~\bibnamefont {Yu}},\ }\bibfield  {title} {\bibinfo {title} {Towards practical quantum computers: transmon qubit with a lifetime approaching 0.5 milliseconds},\ }\href {https://doi.org/10.1038/s41534-021-00510-2} {\bibfield  {journal} {\bibinfo  {journal} {npj Quantum Information}\ }\textbf {\bibinfo {volume} {8}},\ \bibinfo {pages} {3} (\bibinfo {year}
  {2022}{\natexlab{b}})}\BibitemShut {NoStop}%
\bibitem [{\citenamefont {Reagor}\ \emph {et~al.}(2016)\citenamefont {Reagor}, \citenamefont {Pfaff}, \citenamefont {Axline}, \citenamefont {Heeres}, \citenamefont {Ofek}, \citenamefont {Sliwa}, \citenamefont {Holland}, \citenamefont {Wang}, \citenamefont {Blumoff}, \citenamefont {Chou}, \citenamefont {Hatridge}, \citenamefont {Frunzio}, \citenamefont {Devoret}, \citenamefont {Jiang},\ and\ \citenamefont {Schoelkopf}}]{Reagor2016}%
  \BibitemOpen
  \bibfield  {author} {\bibinfo {author} {\bibfnamefont {M.}~\bibnamefont {Reagor}}, \bibinfo {author} {\bibfnamefont {W.}~\bibnamefont {Pfaff}}, \bibinfo {author} {\bibfnamefont {C.}~\bibnamefont {Axline}}, \bibinfo {author} {\bibfnamefont {R.~W.}\ \bibnamefont {Heeres}}, \bibinfo {author} {\bibfnamefont {N.}~\bibnamefont {Ofek}}, \bibinfo {author} {\bibfnamefont {K.}~\bibnamefont {Sliwa}}, \bibinfo {author} {\bibfnamefont {E.}~\bibnamefont {Holland}}, \bibinfo {author} {\bibfnamefont {C.}~\bibnamefont {Wang}}, \bibinfo {author} {\bibfnamefont {J.}~\bibnamefont {Blumoff}}, \bibinfo {author} {\bibfnamefont {K.}~\bibnamefont {Chou}}, \bibinfo {author} {\bibfnamefont {M.~J.}\ \bibnamefont {Hatridge}}, \bibinfo {author} {\bibfnamefont {L.}~\bibnamefont {Frunzio}}, \bibinfo {author} {\bibfnamefont {M.~H.}\ \bibnamefont {Devoret}}, \bibinfo {author} {\bibfnamefont {L.}~\bibnamefont {Jiang}},\ and\ \bibinfo {author} {\bibfnamefont {R.~J.}\ \bibnamefont {Schoelkopf}},\ }\bibfield  {title} {\bibinfo {title} {Quantum
  memory with millisecond coherence in circuit {QED}},\ }\href {https://doi.org/10.1103/physrevb.94.014506} {\bibfield  {journal} {\bibinfo  {journal} {Physical Review B}\ }\textbf {\bibinfo {volume} {94}},\ \bibinfo {pages} {014506} (\bibinfo {year} {2016})}\BibitemShut {NoStop}%
\bibitem [{\citenamefont {Romanenko}\ \emph {et~al.}(2020)\citenamefont {Romanenko}, \citenamefont {Pilipenko}, \citenamefont {Zorzetti}, \citenamefont {Frolov}, \citenamefont {Awida}, \citenamefont {Belomestnykh}, \citenamefont {Posen},\ and\ \citenamefont {Grassellino}}]{Romanenko_PhysRevApplied.13.034032}%
  \BibitemOpen
  \bibfield  {author} {\bibinfo {author} {\bibfnamefont {A.}~\bibnamefont {Romanenko}}, \bibinfo {author} {\bibfnamefont {R.}~\bibnamefont {Pilipenko}}, \bibinfo {author} {\bibfnamefont {S.}~\bibnamefont {Zorzetti}}, \bibinfo {author} {\bibfnamefont {D.}~\bibnamefont {Frolov}}, \bibinfo {author} {\bibfnamefont {M.}~\bibnamefont {Awida}}, \bibinfo {author} {\bibfnamefont {S.}~\bibnamefont {Belomestnykh}}, \bibinfo {author} {\bibfnamefont {S.}~\bibnamefont {Posen}},\ and\ \bibinfo {author} {\bibfnamefont {A.}~\bibnamefont {Grassellino}},\ }\bibfield  {title} {\bibinfo {title} {Three-dimensional superconducting resonators at $t<20$ {mK} with photon lifetimes up to $\ensuremath{\tau}=2$ s},\ }\href {https://doi.org/10.1103/PhysRevApplied.13.034032} {\bibfield  {journal} {\bibinfo  {journal} {Physical Review Applied}\ }\textbf {\bibinfo {volume} {13}},\ \bibinfo {pages} {034032} (\bibinfo {year} {2020})}\BibitemShut {NoStop}%
\bibitem [{\citenamefont {Milul}\ \emph {et~al.}(2023)\citenamefont {Milul}, \citenamefont {Guttel}, \citenamefont {Goldblatt}, \citenamefont {Hazanov}, \citenamefont {Joshi}, \citenamefont {Chausovsky}, \citenamefont {Kahn}, \citenamefont {Çiftyürek}, \citenamefont {Lafont},\ and\ \citenamefont {Rosenblum}}]{RosenblumCavity2023}%
  \BibitemOpen
  \bibfield  {author} {\bibinfo {author} {\bibfnamefont {O.}~\bibnamefont {Milul}}, \bibinfo {author} {\bibfnamefont {B.}~\bibnamefont {Guttel}}, \bibinfo {author} {\bibfnamefont {U.}~\bibnamefont {Goldblatt}}, \bibinfo {author} {\bibfnamefont {S.}~\bibnamefont {Hazanov}}, \bibinfo {author} {\bibfnamefont {L.~M.}\ \bibnamefont {Joshi}}, \bibinfo {author} {\bibfnamefont {D.}~\bibnamefont {Chausovsky}}, \bibinfo {author} {\bibfnamefont {N.}~\bibnamefont {Kahn}}, \bibinfo {author} {\bibfnamefont {E.}~\bibnamefont {Çiftyürek}}, \bibinfo {author} {\bibfnamefont {F.}~\bibnamefont {Lafont}},\ and\ \bibinfo {author} {\bibfnamefont {S.}~\bibnamefont {Rosenblum}},\ }\bibfield  {title} {\bibinfo {title} {A superconducting quantum memory with tens of milliseconds coherence time},\ }\href {https://doi.org/10.1103/PRXQuantum.4.030336} {\bibfield  {journal} {\bibinfo  {journal} {PRX Quantum}\ }\textbf {\bibinfo {volume} {4}},\ \bibinfo {pages} {030336} (\bibinfo {year} {2023})}\BibitemShut {NoStop}%
\bibitem [{\citenamefont {Ganjam}\ \emph {et~al.}(2024)\citenamefont {Ganjam}, \citenamefont {Wang}, \citenamefont {Lu}, \citenamefont {Banerjee}, \citenamefont {Lei}, \citenamefont {Krayzman}, \citenamefont {Kisslinger}, \citenamefont {Zhou}, \citenamefont {Li}, \citenamefont {Jia}, \citenamefont {Liu}, \citenamefont {Frunzio},\ and\ \citenamefont {Schoelkopf}}]{ganjam_surpassing_2024}%
  \BibitemOpen
  \bibfield  {author} {\bibinfo {author} {\bibfnamefont {S.}~\bibnamefont {Ganjam}}, \bibinfo {author} {\bibfnamefont {Y.}~\bibnamefont {Wang}}, \bibinfo {author} {\bibfnamefont {Y.}~\bibnamefont {Lu}}, \bibinfo {author} {\bibfnamefont {A.}~\bibnamefont {Banerjee}}, \bibinfo {author} {\bibfnamefont {C.~U.}\ \bibnamefont {Lei}}, \bibinfo {author} {\bibfnamefont {L.}~\bibnamefont {Krayzman}}, \bibinfo {author} {\bibfnamefont {K.}~\bibnamefont {Kisslinger}}, \bibinfo {author} {\bibfnamefont {C.}~\bibnamefont {Zhou}}, \bibinfo {author} {\bibfnamefont {R.}~\bibnamefont {Li}}, \bibinfo {author} {\bibfnamefont {Y.}~\bibnamefont {Jia}}, \bibinfo {author} {\bibfnamefont {M.}~\bibnamefont {Liu}}, \bibinfo {author} {\bibfnamefont {L.}~\bibnamefont {Frunzio}},\ and\ \bibinfo {author} {\bibfnamefont {R.~J.}\ \bibnamefont {Schoelkopf}},\ }\bibfield  {title} {\bibinfo {title} {Surpassing millisecond coherence in on chip superconducting quantum memories by optimizing materials and circuit design},\ }\href
  {https://doi.org/10.1038/s41467-024-47857-6} {\bibfield  {journal} {\bibinfo  {journal} {Nature Communications}\ }\textbf {\bibinfo {volume} {15}},\ \bibinfo {pages} {3687} (\bibinfo {year} {2024})}\BibitemShut {NoStop}%
\bibitem [{\citenamefont {Pietik\"ainen}\ \emph {et~al.}(2024)\citenamefont {Pietik\"ainen}, \citenamefont {\ifmmode~\check{C}\else \v{C}\fi{}ernot\'{\i}k}, \citenamefont {Eickbusch}, \citenamefont {Maiti}, \citenamefont {Garmon}, \citenamefont {Filip},\ and\ \citenamefont {Girvin}}]{PRXQuantum.5.040307}%
  \BibitemOpen
  \bibfield  {author} {\bibinfo {author} {\bibfnamefont {I.}~\bibnamefont {Pietik\"ainen}}, \bibinfo {author} {\bibfnamefont {O.~c.~v.}\ \bibnamefont {\ifmmode~\check{C}\else \v{C}\fi{}ernot\'{\i}k}}, \bibinfo {author} {\bibfnamefont {A.}~\bibnamefont {Eickbusch}}, \bibinfo {author} {\bibfnamefont {A.}~\bibnamefont {Maiti}}, \bibinfo {author} {\bibfnamefont {J.~W.}\ \bibnamefont {Garmon}}, \bibinfo {author} {\bibfnamefont {R.}~\bibnamefont {Filip}},\ and\ \bibinfo {author} {\bibfnamefont {S.~M.}\ \bibnamefont {Girvin}},\ }\bibfield  {title} {\bibinfo {title} {Strategies and trade-offs for controllability and memory time of ultra-high-quality microwave cavities in circuit quantum electrodynamics},\ }\href {https://doi.org/10.1103/PRXQuantum.5.040307} {\bibfield  {journal} {\bibinfo  {journal} {PRX Quantum}\ }\textbf {\bibinfo {volume} {5}},\ \bibinfo {pages} {040307} (\bibinfo {year} {2024})}\BibitemShut {NoStop}%
\bibitem [{\citenamefont {Valahu}\ \emph {et~al.}(2024)\citenamefont {Valahu}, \citenamefont {Navickas}, \citenamefont {Biercuk},\ and\ \citenamefont {Tan}}]{valahu2024benchmarking}%
  \BibitemOpen
  \bibfield  {author} {\bibinfo {author} {\bibfnamefont {C.~H.}\ \bibnamefont {Valahu}}, \bibinfo {author} {\bibfnamefont {T.}~\bibnamefont {Navickas}}, \bibinfo {author} {\bibfnamefont {M.~J.}\ \bibnamefont {Biercuk}},\ and\ \bibinfo {author} {\bibfnamefont {T.~R.}\ \bibnamefont {Tan}},\ }\bibfield  {title} {\bibinfo {title} {Benchmarking bosonic modes for quantum information with randomized displacements},\ }\href {https://arxiv.org/abs/2405.15237} {\bibfield  {journal} {\bibinfo  {journal} {arXiv preprint arXiv:2405.15237}\ } (\bibinfo {year} {2024})}\BibitemShut {NoStop}%
\bibitem [{\citenamefont {Wang}\ \emph {et~al.}(2024)\citenamefont {Wang}, \citenamefont {Chehade},\ and\ \citenamefont {Dumitrescu}}]{Wang2024}%
  \BibitemOpen
  \bibfield  {author} {\bibinfo {author} {\bibfnamefont {Y.}~\bibnamefont {Wang}}, \bibinfo {author} {\bibfnamefont {S.}~\bibnamefont {Chehade}},\ and\ \bibinfo {author} {\bibfnamefont {E.}~\bibnamefont {Dumitrescu}},\ }\bibfield  {title} {\bibinfo {title} {Semicoherent symmetric quantum processes: Theory and applications},\ }\bibfield  {journal} {\bibinfo  {journal} {AVS Quantum Science}\ }\textbf {\bibinfo {volume} {6}},\ \href {https://doi.org/10.1116/5.0215919} {10.1116/5.0215919} (\bibinfo {year} {2024})\BibitemShut {NoStop}%
\bibitem [{\citenamefont {Noh}\ \emph {et~al.}(2020)\citenamefont {Noh}, \citenamefont {Girvin},\ and\ \citenamefont {Jiang}}]{noh2020encoding}%
  \BibitemOpen
  \bibfield  {author} {\bibinfo {author} {\bibfnamefont {K.}~\bibnamefont {Noh}}, \bibinfo {author} {\bibfnamefont {S.~M.}\ \bibnamefont {Girvin}},\ and\ \bibinfo {author} {\bibfnamefont {L.}~\bibnamefont {Jiang}},\ }\bibfield  {title} {\bibinfo {title} {Encoding an oscillator into many oscillators},\ }\href {https://doi.org/10.1103/PhysRevLett.125.080503} {\bibfield  {journal} {\bibinfo  {journal} {Phys. Rev. Lett.}\ }\textbf {\bibinfo {volume} {125}},\ \bibinfo {pages} {080503} (\bibinfo {year} {2020})}\BibitemShut {NoStop}%
\bibitem [{\citenamefont {Varona}\ \emph {et~al.}(2024)\citenamefont {Varona}, \citenamefont {Saner}, \citenamefont {az\u avan}, \citenamefont {Araneda}, \citenamefont {Aarts},\ and\ \citenamefont {Bermudez}}]{Varona24}%
  \BibitemOpen
  \bibfield  {author} {\bibinfo {author} {\bibfnamefont {S.}~\bibnamefont {Varona}}, \bibinfo {author} {\bibfnamefont {S.}~\bibnamefont {Saner}}, \bibinfo {author} {\bibfnamefont {O.~B.}\ \bibnamefont {az\u avan}}, \bibinfo {author} {\bibfnamefont {G.}~\bibnamefont {Araneda}}, \bibinfo {author} {\bibfnamefont {G.}~\bibnamefont {Aarts}},\ and\ \bibinfo {author} {\bibfnamefont {A.}~\bibnamefont {Bermudez}},\ }\href@noop {} {\bibinfo {title} {Towards quantum computing feynman diagrams in hybrid qubit-oscillator devices}} (\bibinfo {year} {2024}),\ \Eprint {https://arxiv.org/abs/arXiv:2411.05092} {arXiv:2411.05092} \BibitemShut {NoStop}%
\bibitem [{\citenamefont {Bell}(2025)}]{Bell_Github}%
  \BibitemOpen
  \bibfield  {author} {\bibinfo {author} {\bibfnamefont {L.}~\bibnamefont {Bell}},\ }\href@noop {} {\bibinfo {title} {{Trotter Scaling Bounds}}} (\bibinfo {year} {2025}),\ \bibinfo {note} {\url{https://github.com/luke-d-bell/trotter-scaling-bounds}}\BibitemShut {NoStop}%
\bibitem [{Note9()}]{Note9}%
  \BibitemOpen
  \bibinfo {note} {This can be proved by using the so-called Hadamard lemma (Campbell identity) $e^{A}Be^{-A} = B + [A,B] + \protect \frac {1}{2!}[A,[A,B]] + \protect \frac {1}{3!}[A,[A,[A,B]]] + \protect \dots $ Note that a nonunitary transformation is applied to transform a Hermitian operator $\protect \hat {x}$ to a non-Hermitian one $\protect \hat {a}^\dagger $.}\BibitemShut {Stop}%
\bibitem [{Note10()}]{Note10}%
  \BibitemOpen
  \bibinfo {note} {Jensen's inequality can be used to show $g_{04} < 0$, that is, $(e_0 + e_4)/2 < e_2$ as follows: $\forall \alpha \protect \neq 0$, $(e_0 + e_4)/2 = [e^{-\alpha ^2 (E_\protect \text {s} - 4)^2/2} + e^{-\alpha ^2 (E_\protect \text {s} + 12)^2/2}]/2 < e^{-\alpha ^2 [(E_\protect \text {s} - 4)^2 + (E_\protect \text {s} + 12)^2]/4} \leq e^{-\alpha ^2 [(E_\protect \text {s} + 4)^2]/2} = e_2$}\BibitemShut {NoStop}%
\bibitem [{\citenamefont {Ciceri}\ \emph {et~al.}(2025)\citenamefont {Ciceri}, \citenamefont {Cottrell}, \citenamefont {Freeland}, \citenamefont {Fry}, \citenamefont {Hirai}, \citenamefont {Intallura}, \citenamefont {Kang}, \citenamefont {Lee}, \citenamefont {Mitra}, \citenamefont {Ohno}, \citenamefont {Pemmaraju}, \citenamefont {Proissl}, \citenamefont {Quanz}, \citenamefont {Rajan}, \citenamefont {Shimada},\ and\ \citenamefont {Yograj}}]{Ciceri2025}%
  \BibitemOpen
  \bibfield  {author} {\bibinfo {author} {\bibfnamefont {A.}~\bibnamefont {Ciceri}}, \bibinfo {author} {\bibfnamefont {A.}~\bibnamefont {Cottrell}}, \bibinfo {author} {\bibfnamefont {J.}~\bibnamefont {Freeland}}, \bibinfo {author} {\bibfnamefont {D.}~\bibnamefont {Fry}}, \bibinfo {author} {\bibfnamefont {H.}~\bibnamefont {Hirai}}, \bibinfo {author} {\bibfnamefont {P.}~\bibnamefont {Intallura}}, \bibinfo {author} {\bibfnamefont {H.}~\bibnamefont {Kang}}, \bibinfo {author} {\bibfnamefont {C.-K.}\ \bibnamefont {Lee}}, \bibinfo {author} {\bibfnamefont {A.}~\bibnamefont {Mitra}}, \bibinfo {author} {\bibfnamefont {K.}~\bibnamefont {Ohno}}, \bibinfo {author} {\bibfnamefont {D.}~\bibnamefont {Pemmaraju}}, \bibinfo {author} {\bibfnamefont {M.}~\bibnamefont {Proissl}}, \bibinfo {author} {\bibfnamefont {B.}~\bibnamefont {Quanz}}, \bibinfo {author} {\bibfnamefont {D.}~\bibnamefont {Rajan}}, \bibinfo {author} {\bibfnamefont {N.}~\bibnamefont {Shimada}},\ and\ \bibinfo {author} {\bibfnamefont {K.}~\bibnamefont {Yograj}},\
  }\href@noop {} {\bibinfo {title} {Enhanced fill probability estimates in institutional algorithmic bond trading using statistical learning algorithms with quantum computers}} (\bibinfo {year} {2025}),\ \Eprint {https://arxiv.org/abs/2509.17715} {arXiv:2509.17715 [quant-ph]} \BibitemShut {NoStop}%
\bibitem [{\citenamefont {Childs}\ \emph {et~al.}(2018)\citenamefont {Childs}, \citenamefont {Maslov}, \citenamefont {Nam}, \citenamefont {Ross},\ and\ \citenamefont {Su}}]{Childs2018}%
  \BibitemOpen
  \bibfield  {author} {\bibinfo {author} {\bibfnamefont {A.~M.}\ \bibnamefont {Childs}}, \bibinfo {author} {\bibfnamefont {D.}~\bibnamefont {Maslov}}, \bibinfo {author} {\bibfnamefont {Y.}~\bibnamefont {Nam}}, \bibinfo {author} {\bibfnamefont {N.~J.}\ \bibnamefont {Ross}},\ and\ \bibinfo {author} {\bibfnamefont {Y.}~\bibnamefont {Su}},\ }\bibfield  {title} {\bibinfo {title} {Toward the first quantum simulation with quantum speedup},\ }\href {https://doi.org/10.1073/pnas.1801723115} {\bibfield  {journal} {\bibinfo  {journal} {Proc. Natl. Acad. Sci.}\ }\textbf {\bibinfo {volume} {115}},\ \bibinfo {pages} {9456} (\bibinfo {year} {2018})}\BibitemShut {NoStop}%
\bibitem [{\citenamefont {Gieres}(2000)}]{Gieres2000}%
  \BibitemOpen
  \bibfield  {author} {\bibinfo {author} {\bibfnamefont {F.}~\bibnamefont {Gieres}},\ }\bibfield  {title} {\bibinfo {title} {Mathematical surprises and {D}irac's formalism in quantum mechanics},\ }\href {https://doi.org/10.1088/0034-4885/63/12/201} {\bibfield  {journal} {\bibinfo  {journal} {Rep. Prog. Phys.}\ }\textbf {\bibinfo {volume} {63}},\ \bibinfo {pages} {1893} (\bibinfo {year} {2000})}\BibitemShut {NoStop}%
\bibitem [{\citenamefont {Weyl}(1931)}]{Weyl1931}%
  \BibitemOpen
  \bibfield  {author} {\bibinfo {author} {\bibfnamefont {H.}~\bibnamefont {Weyl}},\ }\href@noop {} {\emph {\bibinfo {title} {The Theory of Groups and Quantum Mechanics}}},\ \bibinfo {edition} {2nd}\ ed.\ (\bibinfo  {publisher} {Dover},\ \bibinfo {year} {1931})\BibitemShut {NoStop}%
\bibitem [{\citenamefont {Chen}\ and\ \citenamefont {Wei}(2020)}]{Chen2020}%
  \BibitemOpen
  \bibfield  {author} {\bibinfo {author} {\bibfnamefont {Y.}~\bibnamefont {Chen}}\ and\ \bibinfo {author} {\bibfnamefont {T.-C.}\ \bibnamefont {Wei}},\ }\bibfield  {title} {\bibinfo {title} {Quantum algorithm for spectral projection by measuring an ancilla iteratively},\ }\href {https://doi.org/10.1103/PhysRevA.101.032339} {\bibfield  {journal} {\bibinfo  {journal} {Phys. Rev. A}\ }\textbf {\bibinfo {volume} {101}},\ \bibinfo {pages} {032339} (\bibinfo {year} {2020})}\BibitemShut {NoStop}%
\end{thebibliography}
%

\end{document}